\documentclass[10pt,aps, twocolumn,floats,floatfix,
               showpacs,prd,superscriptaddress,
               longbibliography,
               nofootinbib]{revtex4-2}

\usepackage[normalem]{ulem}
\usepackage[utf8]{inputenc}
\usepackage{graphicx,epsfig}
\usepackage{epstopdf}
\usepackage{amssymb,amsmath,amsthm,amsfonts}
\usepackage{bm}
\usepackage[caption=false]{subfig}
\usepackage[usenames,dvipsnames]{xcolor}
\usepackage{url}
\usepackage[inline]{enumitem}
\usepackage{dsfont}
\usepackage{placeins}
\usepackage[T1]{fontenc}
\usepackage{mathrsfs}
\usepackage{anyfontsize}
\usepackage{xspace}
\usepackage{tensor}
\usepackage{accents}
\usepackage{tabularx}
\usepackage{multirow}
\usepackage{booktabs}
\usepackage{comment}
\usepackage{CJKutf8}

\usepackage[linktocpage,breaklinks]{hyperref}
\hypersetup{colorlinks=true,
            citecolor=NavyBlue,
            linkcolor=NavyBlue,
            urlcolor=NavyBlue}

\def\newacronym#1#2#3{\gdef#1{\gdef#1{#2\xspace}#3 (#2)\xspace}}
\def\bh#1{black hole#1 (BH#1)\gdef\bh{BH}}
\newacronym{\GR}{GR}{General Relativity}
\newacronym{\GW}{GW}{gravitational waves}
\newacronym{\QNM}{QNM}{quasi-normal mode}
\newacronym{\LVK}{LVK}{LIGO-Virgo-KAGRA}
\newacronym{\BL}{BL}{Boyer-Lindquist}
\newacronym{\WKB}{WKB}{Wentzel–Kramers–Brillouin}
\newacronym{\RN}{RN}{Reissner-Nordstr\"{o}m}

\def\dif{\textrm{d}}
\def\p{\partial}

\def\ww{{\rm{w}}}
\def\rp{r_{+}}
\def\rmm{r_{-}}
\def\rg{r_{\rm{g}}}
\def\omegaR{\omega_{\rm{R}}}
\def\omegaI{\omega_{\rm{I}}}

\begin{document}

\title{
Charged black holes embedded in matter with anisotropic pressure:\\ Horizon Structure and Quasinormal Mode Spectra
}

\author{D. N. Garzon}
\email{garzon3@illinois.edu}
\affiliation{The Grainger College of Engineering,
Department of Physics \& Illinois Center for Advanced Studies of the Universe, University of Illinois Urbana-Champaign, Urbana, Illinois 61801, USA}
\affiliation{Center for AstroPhysical Surveys, National Center for Supercomputing Applications, University of Illinois Urbana-Champaign, Urbana, IL, 61801, USA}

\author{Jiayi Zhang (\begin{CJK*}{UTF8}{gbsn}张嘉懿\end{CJK*})}
\email{jz124@illinois.edu}
\affiliation{The Grainger College of Engineering,
Department of Physics \& Illinois Center for Advanced Studies of the Universe, University of Illinois Urbana-Champaign, Urbana, Illinois 61801, USA}
\affiliation{Department of Astronomy, University of Illinois Urbana-Champaign, Urbana, Illinois 61801, USA}

\author{Elena Kopteva}
\email{koptieva@illinois.edu}
\affiliation{The Grainger College of Engineering,
Department of Physics \& Illinois Center for Advanced Studies of the Universe, University of Illinois Urbana-Champaign, Urbana, Illinois 61801, USA}

\author{Helvi Witek}\email{hwitek@illinois.edu}
\affiliation{The Grainger College of Engineering,
Department of Physics \& Illinois Center for Advanced Studies of the Universe, University of Illinois Urbana-Champaign, Urbana, Illinois 61801, USA}
\affiliation{Center for AstroPhysical Surveys, National Center for Supercomputing Applications, University of Illinois Urbana-Champaign, Urbana, IL, 61801, USA}

\begin{abstract}
In realistic settings, black holes are expected to be embedded in astrophysical environments.
These environments,
including possible dark matter distributions, can modify observable properties of black holes and leave imprints on their quasinormal mode spectra.
In this work, we model the environment as matter with anisotropic pressure,
and we consider a charged black hole embedded in it.
The resulting spacetime is described by the Kiselev metric.
We first analyze its horizon structure.
We then investigate the quasinormal modes of a massless charged scalar field propagating on this background.
For this purpose, we develop a nontrivial extension of Leaver's continued fraction method to incorporate the effects of the surrounding matter,
and we combine this framework with automatic differentiation techniques.
We also compare our results with those obtained with the sixth-order Wentzel--Kramers--Brillouin approximation
and find agreement between the two methods for the studied cases.

We find that the surrounding matter modifies the oscillation frequencies and decay rates. We also identify the emergence of long-lived modes and regions of avoided crossing like behavior.  Our results demonstrate the importance of incorporating surrounding matter when modeling realistic black holes.
The numerical framework developed in this work provides a tool for studying quasinormal modes in non-vacuum spacetimes and can be extended to a broad class of black-hole geometries embedded in matter fields.
\end{abstract}


\maketitle
\tableofcontents

\section{Introduction}


In real astrophysical scenarios,
\bh{s} are not isolated, they are embedded in environments composed of surrounding matter or energy.
While the majority of binary \bh{} models or studies on \bh{} perturbations have focused on vacuum solutions,
precision gravitational-wave observations~\cite{LIGOScientific:2026wfs,LIGOScientific:2025slb, LIGOScientific:2021aug, LIGOScientific:2016aoc} now motivate understanding the imprints of these environmental effects.

How does surrounding matter affect the observables that we can link to \bh{s}? 
To answer this question, an environmental distribution and corresponding spacetime geometry must be known.
Anisotropic fluids provide a phenomenological yet controlled way to model the influence of a \bh{} environment.
For example,
it has been shown that the presence of anisotropic matter could alter the shape and size of the \bh{} shadow, which can potentially be observed by future Event Horizon Telescope observations~\cite{Badia:2020pnh, Ahmed:2025boj, Abdujabbarov:2015pqp}.
The presence of ultralight dark matter may yield similar modifications of the \bh{} shadow~\cite{Davoudiasl:2019nlo,Cunha:2019ikd,Creci:2020mfg,Acevedo-Munoz:2025ueh}.
A rotating \bh{} surrounded by an anisotropic fluid may act as a particle accelerator~\cite{AhmedRizwan:2020sza}. Other phenomena studied in this context include
superradiance and \bh{} (in-) stabilities~\cite{Cuadros-Melgar:2021sjy},
the Penrose process~\cite{Kim:2019hfp},
quasinormal modes~\cite{Cuadros-Melgar:2020shz, C:2024cnk},
or the photon sphere~\cite{Benavides-Gallego:2018odl, Ivashchuk:2025sjc}.
The environment may also measurably alter
the \QNM{} spectrum~\cite{Cardoso:2013fwa,Pezzella:2024tkf}
or
the phase of gravitational waves of extreme-mass-ratio inspirals that will be measured with LISA~\cite{Alloqulov:2025tdy,Gliorio:2025cbh}.

In this work, we model a single \bh{} immersed in an environment using the Kiselev metric~\cite{Kiselev:2002dx}.
The Kiselev metric is an exact, static, and spherically symmetric solution of Einstein's equations with an (effective) energy-momentum tensor for a fluid with anisotropic pressure.
The effective energy-momentum tensor is a superposition of matter or energy contributions that interact only gravitationally.
Thus, the Kiselev spacetime provides a flexible framework to incorporate matter or energy in \bh{} spacetimes;
special cases include, for example,
the
Schwarzschild, \RN{} 
and \RN{}--de Sitter or --anti-de Sitter spacetimes.
It has also been shown that the anisotropic stress tensor can be decomposed into a perfect fluid plus an electromagnetic field~\cite{Qu:2023rsv}.

\begin{figure}[htbp!]
    \centering   \includegraphics[width=0.23\textwidth]{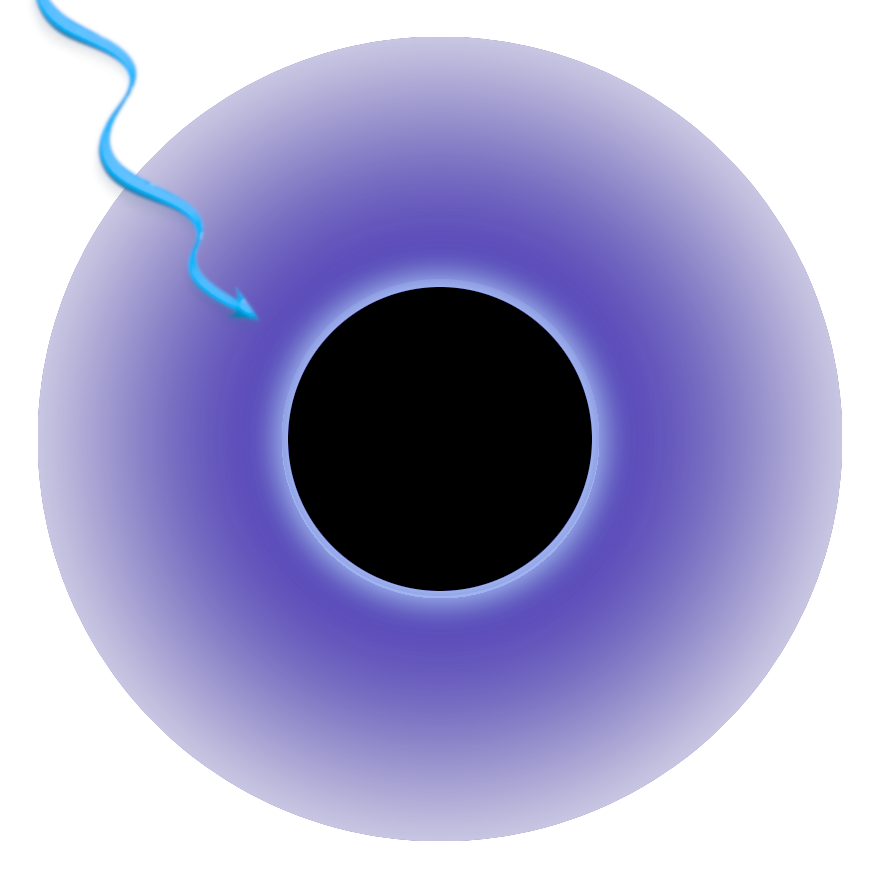}
    \caption{Schematic representation of an anisotropic fluid surrounding a charged black hole, perturbed by an incoming scalar field. The purple region represents the anisotropic fluid distribution, the black circle at the center represents the charged black hole, and the blue wavy arrow indicates the perturbing field.}
    \label{fig:anisotropic_bh_field}
\end{figure}

Here, we focus on a two-component Kiselev model that represents a charged \bh{} embedded in a fluid with anisotropic pressure. The fluid is characterized by a linear equation of state with coefficient $\textrm{w}=-1/3$~\cite{Al-Badawi:2025kbi}.
We display a schematic representation of the setup in Fig.~\ref{fig:anisotropic_bh_field}.
This (effective) fluid can mimic
dark matter distributions~\cite{Kiselev:2003ah},
a condensate of cosmic strings~\cite{Letelier:1979ej, Vilenkin:2000jqa, Vilenkin:1984ib},
or a Barriola--Vilenkin global monopole~\cite{Barriola:1989hx, Harari:1990cz}.
In the latter case, the geometry is locally Schwarzschild-like but globally exhibits a deficit solid angle, yielding a conical spacetime of mathematical interest.

There is extensive observational evidence for the existence of dark matter~\cite{Freese:2008cz}, which constitutes more than $80\%$ of all gravitating matter.
Some authors have explored connections between fluids with anisotropic pressure described by the Kiselev metric and galaxy rotation curves~\cite{Kiselev:2003ah, Rahaman:2011kr}.
This connection was used to explain the asymptotic flatness of some rotation curves~\cite{Kiselev:2003ah}.
Fluid halo models with an anisotropic equation of state have been fitted to observed rotation curve data~\cite{Kuncewicz:2025pma} and used to derive weak-field circular velocities in ``reduced'' Kiselev spacetimes~\cite{Qu:2023rsv}.

The fluid with anisotropic pressure may also be interpreted as a condensate of non-interacting, radially oriented, one-dimensional cosmic strings.
The effective energy-momentum tensor can be derived as an ensemble average over the strings~\cite{Letelier:1979ej} and it describes field theories that generalize action-at-a-distance interactions~\cite{Kalb:1974yc, Letelier:1977tb}.
It was shown that the spacetime around an infinite straight string remains locally flat but is globally conical~\cite{Vilenkin:1981zs}.
This geometric effect may lead to characteristic observational signatures.

After analyzing the horizon structure of the two-component Kiselev spacetime, we determine its response to external perturbations as illustrated in Fig.~\ref{fig:anisotropic_bh_field}.
This response is characterized by the \QNM{} spectrum~\cite{Vishveshwara:1970zz, Kokkotas:1999bd, Berti:2009kk, Berti:2025hly, Konoplya:2011qq}, which is computed within perturbation theory.
\QNM{s} encode key information about the underlying spacetime geometry, making them powerful tools for \bh{} spectroscopy
and studies of their (mode) stability.
Most of the \QNM{} literature focuses on the spectral properties of vacuum \bh{s}, a clear idealization.
In realistic astrophysical settings, massive \bh{s} at galactic centers interact with dense environments, including accretion disks, dark matter structures, and halos, all of which shape galactic and \bh{} dynamics.
Mergers involving such systems generate detectable gravitational waves.
Furthermore, interstellar plasma obscures low-frequency electromagnetic signals, making gravitational waves a unique probe of these environments.
Understanding how environmental effects influence \QNM{s} is therefore essential for accurately interpreting gravitational-wave signals in non-idealized settings, as the \QNM{} frequencies depend sensitively on the \bh{} parameters and serve as fingerprints of the spacetime \cite{Al-Badawi:2025kbi, Varghese:2014xaa, Zhang:2006hh, Ma:2006by, Chen:2005qh, Cuadros-Melgar:2020shz, Cuadros-Melgar:2021sjy}.

In this work, we consider a massless, charged scalar field propagating on
the two-component Kiselev spacetime
and determine its \QNM{} spectrum.
For this purpose, we develop a nontrivial extension of Leaver's continued fraction method~\cite{Leaver:1985ax, Leaver:1990zz}, a traditional method to compute \QNM{s}, to accommodate the complexities introduced by anisotropic-fluid backgrounds. 
We derive the recurrence relation and implement a numerical solver based on automatic differentiation.
The latter is a new technique that enables us to compute the \QNM{} spectrum in parameter regions (e.g., near-extremality) where traditional root finders fail.
In addition, we employ a sixth-order \WKB{} approximation to calculate the \QNM{s},
providing a comparative perspective. In our setup, the \WKB{} effective potential is explicitly frequency dependent, and the treatment developed here applies generally to \WKB{} potentials with this property. We find good agreement between \QNM{s} computed with the \WKB{} approximation and with our extension of Leaver's method.

Although we focus on the analysis of a fluid with anisotropic pressure around a \bh{}, the methods developed here can be applied to a broad class of non-vacuum \bh{} spacetimes. Our results reveal novel features in the \QNM{} spectrum of charged \bh{s} surrounded by an anisotropic fluid, including the presence of long-lived modes, asymmetric structures in the spectrum, and regions of avoided crossing like behavior. These findings contribute to the broader effort to model realistic \bh{} environments.

This paper is organized as follows. In Sec.~\ref{sec:kiselev_general}, we introduce the Kiselev spacetime and the two-component model we consider, and we discuss its physical and geometrical properties.
In Sec.~\ref{sec:field_dynamics}, we present the dynamics of a charged scalar field in this background, reducing the Klein--Gordon equation to a Schr\"odinger-like form and deriving the effective potential.
In Sec.~\ref{sec:qnm_analysis}, we develop a nontrivial extension of Leaver's continued fraction method, including both the full recurrence relation for this background and a numerical implementation based on automatic differentiation.
Here, we also describe the sixth-order \WKB{} method that we employ for comparison.
In Sec.~\ref{sec:numerical_results}, we apply this framework to compute the \QNM{} spectrum and discuss the physical implications of our findings.
Finally, in Sec.~\ref{sec:conclusion}, we conclude with a summary of our contributions.
Throughout this paper, we use geometric units, $c = 8 \pi G = 1$,
and the metric signature $(-+++)$. Symbols and their definitions are summarized in~App.~\ref{app:notation}.

\section{Black holes embedded in an anisotropic fluid}
\label{sec:kiselev_general}

We introduce the spacetime used throughout this work. We begin with the general Kiselev metric and then specialize to the two-component model studied in this paper.
We show that certain components of this metric can be mapped onto an effective energy-momentum tensor $T_{\mu\nu}$ via Einstein's equations,
\begin{align}
\label{eq:EinsteinEquations}
G_{\mu\nu} = T_{\mu\nu} \,,
\end{align}
where $G_{\mu\nu}$ is the Einstein tensor constructed from the metric $g_{\mu\nu}$,
and
we recall our unit choice $c=1=8\pi G$.

We then examine the dominant energy condition for the effective energy-momentum tensor to identify the physically viable parameter range of our model. We conclude the section with a discussion of the horizon structure of this spacetime.

\subsection{The Kiselev metric}
\label{ssec:KiselevMetric}
Kiselev spacetimes~\cite{Kiselev:2002dx} provide a static, spherically symmetric generalization of the Schwarzschild solution that can include matter surrounding the \bh, as well as additional charges or energy in the spacetime.

We choose the Kiselev metric for three main reasons:
(i) it is an exact solution of Einstein's equations with an effective energy-momentum tensor;
(ii) it encompasses a broad class of matter or energy models relevant to cosmology and astrophysics; and
(iii) its symmetries make \QNM{} calculations tractable for a first analysis.

The Kiselev metric,
in Schwarzschild coordinates $(t,r,\theta,\phi)$, is given by
\begin{align}
\label{eq:KiselevMetricGeneral}
\dif s^{2} & = -f(r) \dif t^{2} + f(r)^{-1} \dif r^{2} + r^2 \dif\sigma^{2}
\,,
\end{align}
where
$\dif\sigma^{2}=\dif\theta^2 + \sin^2\theta \dif\phi^{2}$ is the metric on the unit 2-sphere.
The metric function, $f(r)$, is given by
\begin{align}
\label{eq:KiselevMetricFunctionGeneral}
f(r) & = 1 - \frac{\rg}{r} - h(r)
\,,
\end{align}
where $\rg=2M$ is the Schwarzschild radius 
and $M$ is the \bh{} mass parameter.
The function $h(r)$ encodes 
a linear superposition of 
additional,
non-interacting
matter or energy contributions,
and can be written as
\begin{align}
\label{eq:KiselevExtraTermsGeneral}
h(r) & = \sum_{i} \frac{K_i}{r^{3\ww_i+1}}
\,,
\end{align}
where the parameter pairs $(K_{i},\ww_{i})$ describe the $i$-th component;
both parameters may, in general, take arbitrary signs.
Their physical interpretation depends on the matter or energy source that they are describing.
For example, choosing a one-component Kiselev model
and setting
$K_{1}= \Lambda/3$ and $\ww_{1}=-1$
in Eq.~\eqref{eq:KiselevExtraTermsGeneral} yields
the Schwarzschild--(anti-)de Sitter spacetime with the cosmological constant $\Lambda$,
while setting
$K_{1}= - Q^{2}$ and $\ww_{1}=1/3$
selects the
\RN{} \bh{}
with charge $Q$.
A two-component Kiselev model with parameters
$(K_{1}=\Lambda/3,\ww_{1}=-1)$ and $(K_{2}=-Q^{2},\ww_{2}=1/3)$
gives the \RN{}--(anti-)de Sitter metric,
and so on.
The value $\ww_i=0$ is degenerate with the Schwarzschild term and therefore does not define an independent component.
Note that throughout this work, references to the mass and charge of the \bh{} refer to the corresponding mass and charge parameters in the spacetime metric.
We refer to our accompanying paper~\cite{Kopteva:AnisotropicBH}
for a detailed re-derivation of the Kiselev metric,
discussion of its properties, and
an extended list of applications.

In the present paper, we want to model a non-isolated black hole, so we consider a two-component Kiselev model 
representing a charged \bh{} surrounded by some anisotropic fluid.
We set one component in Eq.~\eqref{eq:KiselevExtraTermsGeneral}
to
\begin{align}
\label{eq:2CompKiselev-Charge}
K_{1} & = - Q^{2}
\,,\quad
\ww_{1}   = \frac{1}{3}
\,,
\end{align}
where $Q$ is the charge parameter of the \bh{}.
The associated electromagnetic vector potential due to the central static charge is
\begin{align}
\label{eq:VectorPotential}
A_{\mu} & = \left(-\frac{Q}{r},0,0,0\right)
\,.
\end{align}
The second component is chosen to be
\begin{align}
\label{eq:2CompKiselev-K}
K_{2} & = K
\,,\quad
\ww_{2}   = -\frac{1}{3}
\,,
\end{align}
where $K$ controls the strength of the fluid contribution.

To simplify our analysis, we work with dimensionless coordinates and variables
that are rescaled by the appropriate power of the Schwarzschild radius, $\rg$,
\begin{align}
\label{eq:units}
r_{\rm{new}} & = \frac{r}{\rg}
\,, \quad
t_{\rm{new}} = \frac{t}{\rg}
\,, \quad
\dif s^{2}_{\rm{new}} = \frac{\dif s^{2}}{\rg^2}
\,,\\
Q_{\rm{new}} & = \frac{Q}{\rg}
\,,\quad
r_{\rm{g,new}} = \frac{\rg}{\rg}=1
\,.\nonumber
\end{align}
We henceforth omit the subscript ``new''
for brevity. Note that the fluid parameter $K$ is dimensionless.
Inserting our choices of components
in Eqs.~\eqref{eq:2CompKiselev-Charge} and~\eqref{eq:2CompKiselev-K}
into Eq.~\eqref{eq:KiselevExtraTermsGeneral} gives
\begin{align}
\label{eq:2CompKiselev-hfunction}
h(r) & = h_{1}(r)+ h_{2}(r)
       = - \frac{Q^{2}}{r^2} + K
\,.
\end{align}
Inserting it into
the metric function in Eq.~\eqref{eq:KiselevMetricFunctionGeneral},
and incorporating the units,
Eq.~\eqref{eq:units},
yields the metric function,
\begin{align}
\label{eq:2CompKiselev-MetricFunction}
f(r) & = 1 - K - \frac{1}{r} + \frac{Q^2}{r^2}
\,.
\end{align}
This is the two-component Kiselev model
considered in this paper.
Note that the metric function in Eq.~\eqref{eq:2CompKiselev-MetricFunction} reduces to that of the \RN{} solution for $K=0$.
When $Q=0$, it reduces to a Schwarzschild-like geometry, but with a solid angle deficit~\cite{Barriola:1989hx}.
Asymptotically, the metric function behaves as
\begin{align}\label{eq:f-asymptotic}
\lim_{r\to\infty} f(r) & =1-K,
\end{align}
and the spatial slices exhibit a solid angle deficit: with proper radial distance defined by $\mathrm{d}r_{\mathrm{prop}}=\mathrm{d}r/\sqrt{1-K}$, the area of a sphere at fixed $r_{\mathrm{prop}}$ is smaller than in flat space by a factor $1-K$.
All curvature invariants vanish as $r\to\infty$, so the asymptotic region is locally flat. However, globally it is conical due to the solid angle deficit~\cite{Vilenkin:2000jqa}. Such spacetimes are known as quasi-asymptotically flat~\cite{Nucamendi:1996ac}.

\begin{figure}[b]
\centering
\includegraphics[width=0.47\textwidth,
trim=10mm 10mm 0 10mm,clip]{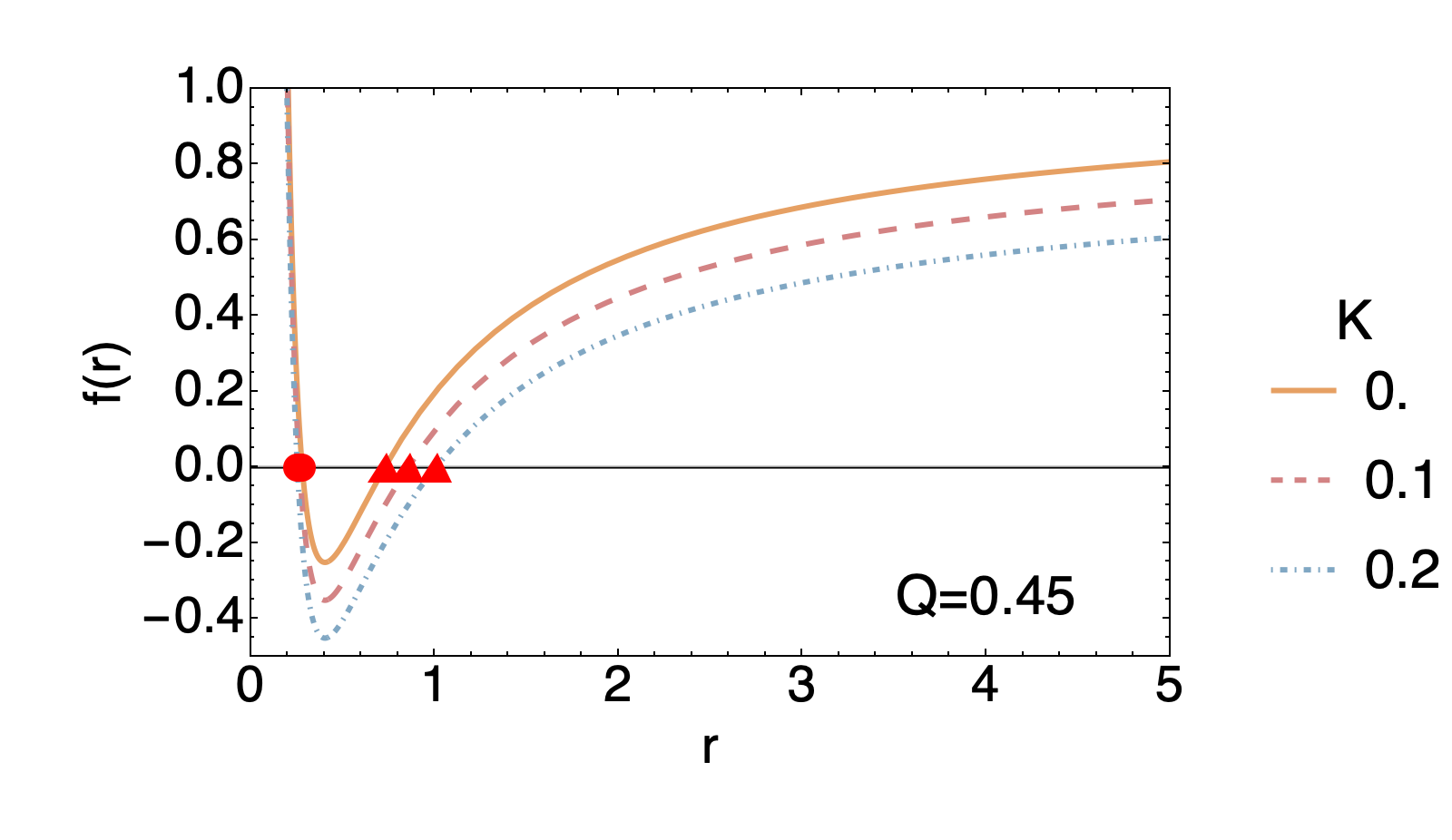}
\includegraphics[width=0.47\textwidth,trim=10mm 15mm 0 10mm,clip]{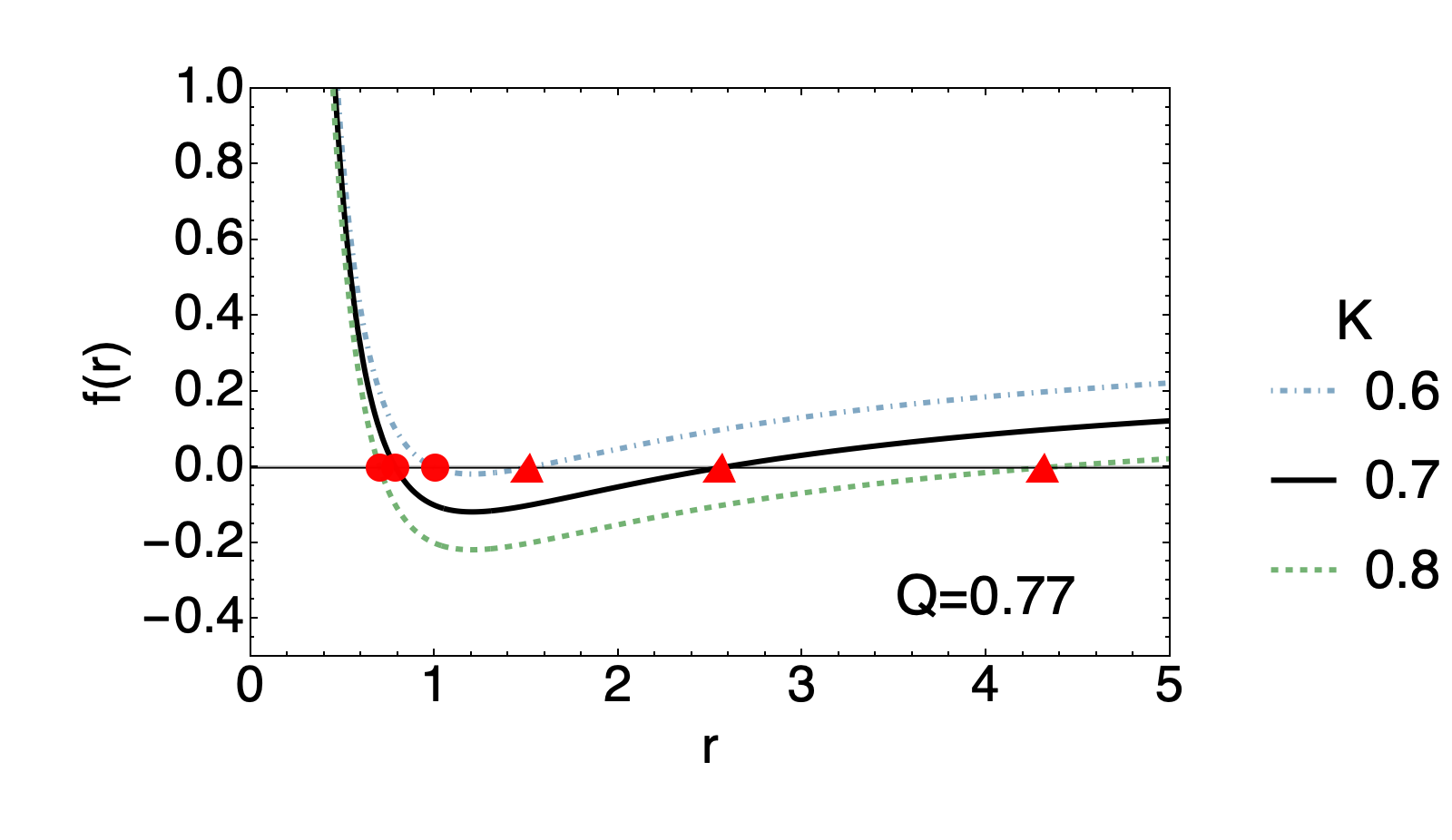}
\caption{Radial profile of the metric function in Eq.~\eqref{eq:2CompKiselev-MetricFunction} describing a \bh{} of charge $Q=0.45$ (top panel) or
$Q=0.77$ (bottom panel),
embedded in a fluid with varying fluid parameter $K$.
We indicate the location of the Cauchy horizon (red dots) and of the event horizon (red triangles).
}
\label{fig:affa4cases_fvsr}
\end{figure}

In Fig.~\ref{fig:affa4cases_fvsr}
we plot the radial profile of the metric function, Eq.~\eqref{eq:2CompKiselev-MetricFunction},
for different values of the fluid parameter $K$ and two exemplary values of the \bh{} charge $Q$.
We indicate the locations of the Cauchy (red dots)
and event (red triangles) horizons where $f(r)=0$.
We see that the 
locations of the horizons shift outwards as 
the fluid parameter $K$ increases.
While this change is small for the inner horizon,
the radius of the event horizon can increase to a few Schwarzschild radii as
$K\to1$.

The sign of the metric function
determines the type of a spacetime region, 
which can be classified as a 
so-called ``R-region'' or ``T-region'';
see, e.g., Refs.~\cite{2001GReGr..33.2259N,Misner:1973prb,Frolov:1998wf}.
An R-region is defined as the region in which
the basis vector $\p_{t}$ is timelike
and, thus, this region is accessible to (ordinary) stationary observers.
It corresponds to a positive metric function, $f(r)>0$.
A T-region is defined as the region in which the basis vector $\p_{t}$ is spacelike, corresponding to $f(r)<0$,
and no static worldlines exist.
The T-region cannot communicate back to the R-region and, thus, is inaccessible to stationary observers.
Fig.~\ref{fig:affa4cases_fvsr} visualizes the different regions for the two-component Kiselev model in Eq.~\eqref{eq:2CompKiselev-MetricFunction}.
Inside the Cauchy horizon is an R-region. Between the Cauchy and the event horizons, the metric function is negative and we have a T-region.
For radii larger than the event horizon
(red triangles), the metric function is positive and the spacetime exhibits an R-region.
In the remainder of this work, we will study perturbations in this physically accessible region.

The two-component Kiselev metric in Eq.~\eqref{eq:2CompKiselev-MetricFunction} gives rise to an effective energy-momentum tensor. In the next section, we derive and analyze it to determine which parameters $(Q,K)$ satisfy the dominant energy condition,
before characterizing the horizon structure of our model.

\subsection{The effective energy-momentum tensor}\label{ssec:TmnEff}

We now derive the effective energy-momentum tensor that is determined by Einstein's equations in Eq.~\eqref{eq:EinsteinEquations}. Note that for a static, spherically symmetric metric of the form in Eq.~\eqref{eq:KiselevMetricGeneral}, the Einstein tensor is diagonal. The same is therefore true of the effective energy-momentum tensor.
We insert the Kiselev metric in Eq.~\eqref{eq:KiselevMetricGeneral} with the general metric function given in Eq.~\eqref{eq:KiselevMetricFunctionGeneral} into the Einstein tensor,
and we obtain
\begin{subequations}
\label{eq:TmnEffKiselevGeneral}
\begin{align}
T^{t}{}_{t} = T^{r}{}_{r}
   & = - \frac{1}{r^2}\left(h + r h' \right)\,,
\\
T^{\theta}{}_{\theta} = T^{\phi}{}_{\phi}
   & = - \frac{1}{2r}\left(2 h' + r h'' \right)
\,,
\end{align}
\end{subequations}
where the prime denotes differentiation with respect to $r$,
and $h(r)$ is given in Eq.~\eqref{eq:KiselevExtraTermsGeneral}.

Because the function $h(r)$ is a superposition of non-interacting energy or matter sources,
the effective energy-momentum tensor is also a superposition of the
individual energy-momentum contributions.

The effective energy-momentum tensor
associated with
the two-component Kiselev model in Eq.~\eqref{eq:2CompKiselev-hfunction} becomes
\begin{align}
\label{eq:2CompKiselev-TmnEff}
T^{\mu}{}_{\nu} & = T_{1}^{\mu}{}_{\nu} + T_{2}^{\mu}{}_{\nu}
\,,
\end{align}
with
\begin{subequations}
\begin{align}
\label{eq:2CompKiselev-TmnEffQ}
T_{1}^{\mu}{}_{\nu} & =  {\rm{diag}}
    \left( -\frac{Q^{2}}{r^{4}}, -\frac{Q^{2}}{r^{4}}, \frac{Q^{2}}{r^{4}}, \frac{Q^{2}}{r^{4}}\right)
\,,\\
\label{eq:2CompKiselev-TmnEffK}
T_{2}^{\mu}{}_{\nu} & = {\rm{diag}}
    \left( - \frac{K}{r^2}, -\frac{K}{r^2}, 0, 0 \right)
\,.
\end{align}
\end{subequations}

In Kiselev's work~\cite{Kiselev:2002dx}, the matter or energy distributions surrounding the \bh{}
are modeled as a superposition of non-interacting anisotropic fluids with
total density $\rho=\sum\rho_{i}$,
radial pressure $p_{r}=\sum p_{r,i}$
and transverse pressure $p_{\perp}=\sum p_{\perp,i}$.
The energy-momentum tensor of this anisotropic fluid (see Fig.~\ref{fig:anisotropic_bh_field}) is~\footnote{Kiselev adopts the same symmetry assumptions but on the matter sector first (staticity, spherical symmetry, no radial energy flux), and then arrives at a metric of the form in Eq.~\eqref{eq:KiselevMetricGeneral}. Therefore, in our case, Eqs.~\eqref{eq:TmnEffKiselevGeneral} should be viewed as the effective $T^{\mu}{}_{\nu}$ of such a superposed anisotropic fluid in Kiselev's sense.}
\begin{align}
\label{eq:TmnAnisotropicFluid}
T^{\mu}{}_{\nu} & = {\rm{diag}}\left(
    - \rho, p_{r}, p_{\perp}, p_{\perp}
    \right)
\,.
\end{align}

Each of the matter or energy components is characterized by an equation of state of the form
\begin{align}
\label{eq:EoSKiselevComponents}
\bar{p}_{i} & = \ww_{i}\, \rho_{i}
\,,
\end{align}
where $\bar{p}_{i}$ is the average pressure
\begin{align}
\label{eq:AveragePressure}
\bar{p}_{i} & =
\frac{1}{3}\left(p_{r,i} + 2 p_{\perp,i}\right)
\,.
\end{align}
This form of the equation of state, written in terms of the average pressure, generalizes the isotropic-fluid relation in which the radial and transverse pressures coincide~\cite{Kiselev:2002dx}. The advantage of this formulation is that it parametrizes the matter source of interest with a single parameter.

We can map the effective energy-momentum tensor of the two-component Kiselev model
considered in this work
to that of an anisotropic fluid.
If we map the effective energy-momentum tensor associated with the \bh{} charge, Eq.~\eqref{eq:2CompKiselev-TmnEffQ}, to Eq.~\eqref{eq:TmnAnisotropicFluid},
we can identify
\begin{align}
\label{eq:2CompKiselev-DensityPressureQ}
\rho_{1} & = \frac{Q^{2}}{r^{4}}
\,,\quad
p_{r,1} = - p_{\perp,1} = - \frac{Q^{2}}{r^{4}}
\,.
\end{align}
If we map the effective energy-momentum tensor of the second component, Eq.~\eqref{eq:2CompKiselev-TmnEffK}, to Eq.~\eqref{eq:TmnAnisotropicFluid},
we find
\begin{align}
\label{eq:2CompKiselev-DensityPressureK}
\rho_{2} & = \frac{K}{r^2}
\,,\quad
p_{r,2} = - \frac{K}{r^2}
\,,\quad
p_{\perp,2} = 0
\,.
\end{align}
Combining both components, we obtain
\begin{align}
\label{eq:2CompKiselev-DensityPressureQK}
\rho & = - p_{r}
    = \frac{Q^{2}}{r^4}+ \frac{K}{r^2}
\,,\quad
p_{\perp} = \frac{Q^{2}}{r^4}
\,.
\end{align}
One can check that both components satisfy the generalized equation of state, Eq.~\eqref{eq:EoSKiselevComponents},
\begin{align}
\label{eq:2CompKiselev-EOSQK}
\bar{p}_{1} & = \ww_{1} \rho_{1}
\,,\quad
\bar{p}_{2} = \ww_{2} \rho_{2}
\,,
\end{align}
with $\ww_{1}=1/3$ and $\ww_{2}=-1/3$,
consistent with
their definitions in
Eqs.~\eqref{eq:2CompKiselev-Charge} and~\eqref{eq:2CompKiselev-K}.
Note that unlike ordinary matter which typically satisfies $p \geq 0$,
or dark-energy-type
contributions
with $p < -\rho/3$ to drive cosmic acceleration,
the second component
occupies an intermediate regime with characteristic gravitational effects.

We impose the dominant energy condition as a conservative criterion for the physical viability of the parameter space of $(Q,K)$.
Applied to Eq.~\eqref{eq:TmnAnisotropicFluid},
it implies
\begin{align}
\label{eq:DECAnisotropicFluid}
\rho \geq |p_{r}|
\,,\quad{\rm{and}}\quad
\rho \geq |p_{\perp}|
\,.
\end{align}
For our two-component Kiselev model,
the first condition is trivially satisfied
as follows from Eq.~\eqref{eq:2CompKiselev-DensityPressureQK}.
The second condition becomes
\begin{align}
\label{eq:tmp2CompKiselev-DEC}
\frac{Q^2}{r^4} + \frac{K}{r^2} & \geq
    \left| \frac{Q^2}{r^4} \right|
\,,
\end{align}
and, thus, requires
\begin{align}
\label{eq:2CompKiselev-DEC}
K & \geq 0
\,.
\end{align}
Therefore, the dominant energy condition is satisfied
if $K \geq 0$,
ensuring that the energy density remains greater than or equal to the pressure magnitudes. We adopt this as the physically admissible range of the fluid parameter.

\subsection{Horizon Structure}\label{ssec:horizon_structure}

We now return to the properties of the \bh{}.

We analyze the horizon structure and classify the possible configurations in the $(Q,K)$ parameter space.
The horizons are determined by the roots of the metric function in Eq.~\eqref{eq:2CompKiselev-MetricFunction}, i.e.,
they are solutions of
\begin{align}
\label{eq:2CompKiselev-HorizonEquation}
f(r) & = 1 - K - \frac{1}{r} + \frac{Q^2}{r^2}
= 0
\,.
\end{align}
Depending on the values of the \bh{} charge parameter $Q$ and the fluid parameter $K$, the spacetime may exhibit zero, one, or two horizons.
We recall that $K\geq0$ from the dominant energy condition
in Eq.~\eqref{eq:2CompKiselev-DEC},
and the radial coordinate is non-negative, $r\geq0$.
Under these constraints,
we identify four regions in the $(Q,K)$ parameter space,
namely
$K=0$, $0\leq K<1$, $K=1$ and $K>1$.
We analyze each of them in the following.

\subsubsection{\texorpdfstring{$K = 0$}{K = 0}}
\label{sssec:RNHorizons}
When the fluid parameter $K$ vanishes, the metric function in Eq.~\eqref{eq:2CompKiselev-MetricFunction}
reduces to that of the
\RN{} solution,
\begin{align}
\label{eq:RN-MetricFunction}
f(r) & = 1 - \frac{1}{r} + \frac{Q^2}{r^2}
\,.
\end{align}
This metric admits two horizons,
\begin{align}
\label{eq:RN-HorizonsSubExtremal}
r_{\rm{RN},\pm} & = \frac{1}{2}
    \left(1 \pm \sqrt{1-4Q^{2}}
    \right)
\,,
\end{align}
as long as
the \bh{} charge satisfies
$0<Q^{2}<1/4$.
Here, $r_{\rm{RN},+}$ corresponds to the event horizon and $r_{\rm{RN},-}$ to the Cauchy horizon~\footnote{
Recent mathematical results indicate that in \bh{} formation through collapse, the Cauchy horizon is a (weak) null singularity~\cite{Dafermos:2017dbw,VandeMoortel:2025qzf}.
}.
The value $Q^{2}=1/4$ corresponds to the extremal \RN{} solution
in which the event and Cauchy horizons coincide.
For charges larger than the extremal value, $Q^{2}>1/4$,
no horizons exist and the spacetime admits a naked singularity.

\subsubsection{\texorpdfstring{$0\leq K < 1$}{0<=K<1}} \label{subsubsection:leq_K_<_1}
We now focus on the horizon structure of the two-component Kiselev metric for a fluid parameter in the range $0\leq K<1$
(which includes the case $K=0$).
By solving Eq.~\eqref{eq:2CompKiselev-HorizonEquation}, we find
\begin{align}
\label{eq:2CompKiselev-HorizonsSubExtremal}
r_{\pm} & = \frac{1}{2(1-K)} \left(
1 \pm \sqrt{ 1 - 4 (1-K) Q^2}
\right)
\,.
\end{align}

Analogous to the \RN{} solution,
the two-component Kiselev
model admits an extremal solution when
\begin{align}
\label{eq:2CompKiselev-HorizonExtremal}
(1-K)Q^{2} & = \frac{1}{4}
\,.
\end{align}
Then, the Cauchy and event horizons merge into a single, degenerate horizon located at
\begin{equation*}
r_{\pm}= \frac{1}{2(1-K)}.
\end{equation*}
The exterior domain is an R-region, i.e., the physical region accessible to stationary observers (cf.~Sec.~\ref{ssec:KiselevMetric}).
The extremality condition in Eq.~\eqref{eq:2CompKiselev-HorizonExtremal} reduces to that of the \RN{} metric for $K=0$.

If the parameters $(Q,K)$
exceed the extremal value,
\begin{align}
\label{eq:2CompKiselev-NakedSingularity}
(1-K) Q^{2} & > \frac{1}{4}
\,,
\end{align}
no horizons exist and the curvature singularity at $r=0$ becomes naked.

If the parameters remain below the extremal value,
\begin{align}
\label{eq:2CompKiselev-ParsSubExtremal}
(1-K) Q^{2} & < \frac{1}{4}
\,,
\end{align}
the spacetime is subextremal
and admits the two real roots, which correspond to the event and Cauchy horizons determined by Eq.~\eqref{eq:2CompKiselev-HorizonsSubExtremal}.
The domain outside the event horizon, $r_{+}$, is again an R-region.
Depending on the value of the \bh{} charge parameter, we can distinguish two
(subextremal)
cases:
\begin{enumerate}[label=Case \arabic*, leftmargin=3.3em]
\item \label{case:two-roots-Qle}
First, we consider a
charge parameter
below the extremal value of the \RN{} solution, i.e., $Q^{2}<1/4$.
Then, any value of the fluid parameter in the range $0\leq K<1$
satisfies the subextremal condition in Eq.~\eqref{eq:2CompKiselev-ParsSubExtremal},
and is allowed.

\item \label{case:two-roots-Qgt}
The condition in Eq.~\eqref{eq:2CompKiselev-ParsSubExtremal} allows for \bh{} charge parameters $Q^{2}>1/4$
for non-vanishing fluid parameter $K$,
as long as
$ 1 - \tfrac{1}{4Q^{2}} <K < 1$.

\end{enumerate}

\begin{figure}[ht]
\centering
\includegraphics[width=.47\textwidth,trim=0cm 2cm 1cm 1.5cm,clip]{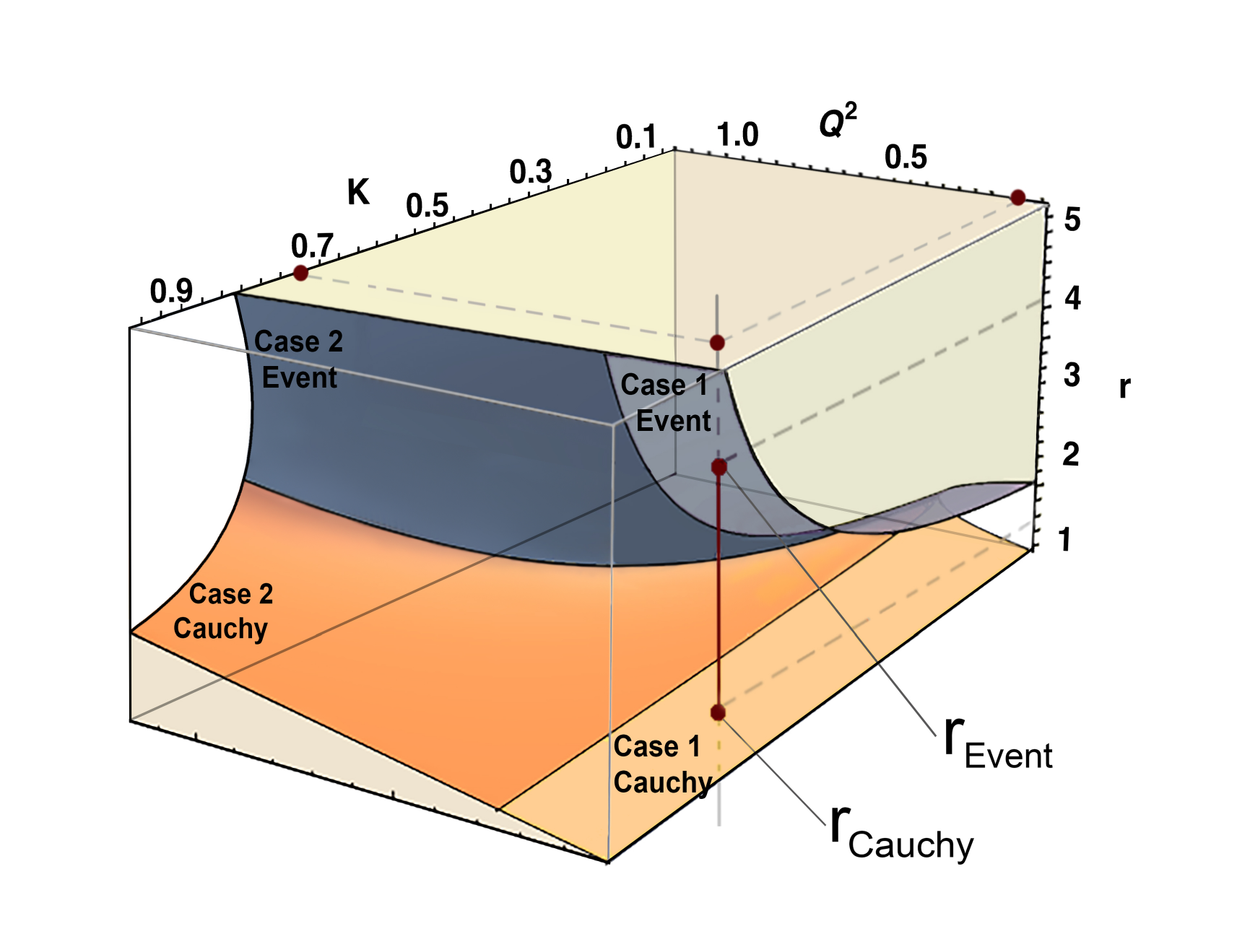}
\caption{Visualization of the roots of the metric function in Eq.~\eqref{eq:2CompKiselev-MetricFunction}
as a function of the \bh{} charge $Q^{2}$ and fluid parameter $0\leq K<1$.
The surface is partitioned into four regions; blue and orange denote Event
and Cauchy horizons, respectively, for \ref{case:two-roots-Qle}
and \ref{case:two-roots-Qgt}.
The enclosed volume corresponds
to the region where $f(r)>0$.
}
\label{fig:parameterSpaceRootsf}
\end{figure}

Fig.~\ref{fig:parameterSpaceRootsf} presents a three-dimensional visualization of the
metric function, $f(r)$, in Eq.~\eqref{eq:2CompKiselev-MetricFunction}
in the $(Q^2,0\leq K<1)$ parameter space for
\ref{case:two-roots-Qle} and \ref{case:two-roots-Qgt}.
The surface represents the
horizons
determined by Eq.~\eqref{eq:2CompKiselev-HorizonsSubExtremal},
while the enclosed volume represents the physical region where $f(r)>0$.
As an example,
we chose a pair $(Q^{2},K)$, indicated by the dots on the corresponding axes.
The planes $Q^2=\mathrm{const}$ and $K=\mathrm{const}$ intersect along a line that cuts through the surface in two points.
The projections of these intersection points onto the $r$-axis give the horizon locations for this parameter choice.


\ref{case:two-roots-Qle} represents a physical parameter space and one may wonder if the parameter range in \ref{case:two-roots-Qgt} is physical.
In particular,
if we reverse the rescaling of Eq.~\eqref{eq:units}
and reinstate the mass units in Eq.~\eqref{eq:2CompKiselev-ParsSubExtremal},
it appears that \ref{case:two-roots-Qgt}
allows for
charges
$Q^{2}>M^{2}$
above the extremal bound of the \RN{} metric.
To address this concern, we inspect the asymptotic behavior of the metric
in Eq.~\eqref{eq:KiselevMetricGeneral} with Eq.~\eqref{eq:2CompKiselev-MetricFunction}.
At spatial infinity,
\begin{align*}
g_{tt} \xrightarrow[r\to\infty]{} -(1-K),
\end{align*}
so the Killing vector $\xi^\mu=(\partial_t)^\mu$ is not unit timelike there. Indeed,
\begin{align*}
\xi^\mu\xi_\mu \xrightarrow[r\to\infty]{} -(1-K).
\end{align*}
However, physical quantities such as the mass, the surface gravity $\kappa$, and the electrostatic potential $\Phi$ must be defined with respect to a unit time translation at infinity. If we rescale the Killing vector by a constant factor as
\begin{align*}
\xi^\mu \mapsto a\,\xi^\mu,
\end{align*}
these quantities rescale accordingly. Therefore, it is convenient to introduce the normalized Killing field~\cite{Brown:1994gs,Nucamendi:1996ac}
\begin{align}
\label{eq:AsymptoticKV}
\hat{\xi}^\mu=\frac{1}{\sqrt{1-K}}\,(\partial_t)^\mu,
\qquad
\hat{\xi}^\mu \hat{\xi}_\mu \xrightarrow[r\to\infty]{} -1.
\end{align}
With this choice, the physical mass becomes
\begin{align*}
M_{\text{phys}}=\frac{M}{\sqrt{1-K}},
\end{align*}
whereas the electric charge does not change. Defined by the Gauss law at infinity, as the flux of the Faraday tensor, it remains
\begin{align*}
Q_{\text{phys}}=Q.
\end{align*}
Then the extremality bound takes the standard form
\begin{align*}
Q_{\text{phys}}^2\le M_{\text{phys}}^2
\qquad\Longleftrightarrow\qquad
(1-K)Q^2\le M^2,
\end{align*}
in agreement with Eq.~\eqref{eq:2CompKiselev-HorizonExtremal}. Thus, \ref{case:two-roots-Qgt} is not excluded on physical grounds.

\subsubsection{\texorpdfstring{$K = 1$}{K = 1}}

When the fluid parameter is set to $K=1$, the metric function becomes
\begin{align}
\label{eq:K1-MetricFunction}
f(r) & = - \frac{1}{r} + \frac{Q^2}{r^2}
\,,
\end{align}
and the spacetime admits a single
Killing horizon at
\begin{align}
\label{eq:K1-Roots}
r_{K=1} & = Q^{2}
\,.
\end{align}
In the region $0<r<Q^2$, one has $f>0$, so this domain corresponds to an $R$-region, within which the timelike curvature singularity at $r=0$ is exposed to a static observer, i.e., it is locally naked.
For $r>Q^2$ one finds $f<0$, so the exterior is a $T$-region in which the radial coordinate plays the role of a timelike variable.
The horizon in Eq.~\eqref{eq:K1-Roots} is therefore the boundary between the R-region and the T-region that extends to infinity.
In this respect, it acts like a cosmological horizon, marking the boundary of the static region that a stationary observer can see, rather than an inner (Cauchy) horizon associated with loss of predictability.
As $r \to \infty$, the metric function approaches zero from below, and the norm of the Killing vector also tends to zero. Therefore, the spacetime does not admit a unit timelike normalization of the Killing vector at infinity.
The case $K=1$ therefore does not describe a \bh{} with a static exterior, but represents a degenerate limit of the quasi-asymptotically flat spacetimes with a deficit solid angle~\cite{Barriola:1989hx,Nucamendi:1996ac}.

\subsubsection{\texorpdfstring{$K>1$}{K>1}}

Finally, we consider the fluid parameter range $K>1$. Solving
Eq.~\eqref{eq:2CompKiselev-HorizonEquation} we find two distinct roots. Since
$1-K<0$, it is convenient to write them as
\begin{align}
\label{eq:2CompKiselev-HorizonKgt1}
r_{1,2} & = - \frac{1}{2 |1-K|} \left(
1 \pm \sqrt{ 1 + 4 Q^{2} |1-K| }
\right)
\,.
\end{align}
The root with the plus sign is always negative,
\begin{align*}
r_1 < 0,
\end{align*}
and therefore lies outside the physical range of the radial coordinate, $r\geq0$.
For $Q\neq0$, one has
\begin{align*}
\sqrt{1+4Q^2|1-K|}>1,
\end{align*}
so the quantity in parentheses is negative and therefore $r_2>0$. Thus, for $K>1$ and $Q\neq0$, the spacetime admits a single positive root. In the special case $Q=0$, this root collapses to $r_2=0$, so no horizon exists at finite radius.
The single root $r_{2}$ corresponds to a simple Killing horizon, and the exterior domain is a T-region
where the Killing vector $\xi^{\mu}=(\p_{t})^{\mu}$ is spacelike.
As in the case $K=1$, this horizon does not represent a Cauchy horizon, but rather bounds a static region in a manner qualitatively analogous to cosmological horizons.

For the remainder of this paper,
we focus on the physically viable subextremal \bh{s} with charge parameter $Q^{2}< 1/4$ and the fluid parameter range $0\leq K<1$
corresponding to~\ref{case:two-roots-Qle} in Sec.~\ref{subsubsection:leq_K_<_1}.


\section{Scalar Field Dynamics}\label{sec:field_dynamics}

\begin{figure*}[tbh]
\centering
\includegraphics[height=3.5cm,clip,trim=0 0 118 0]{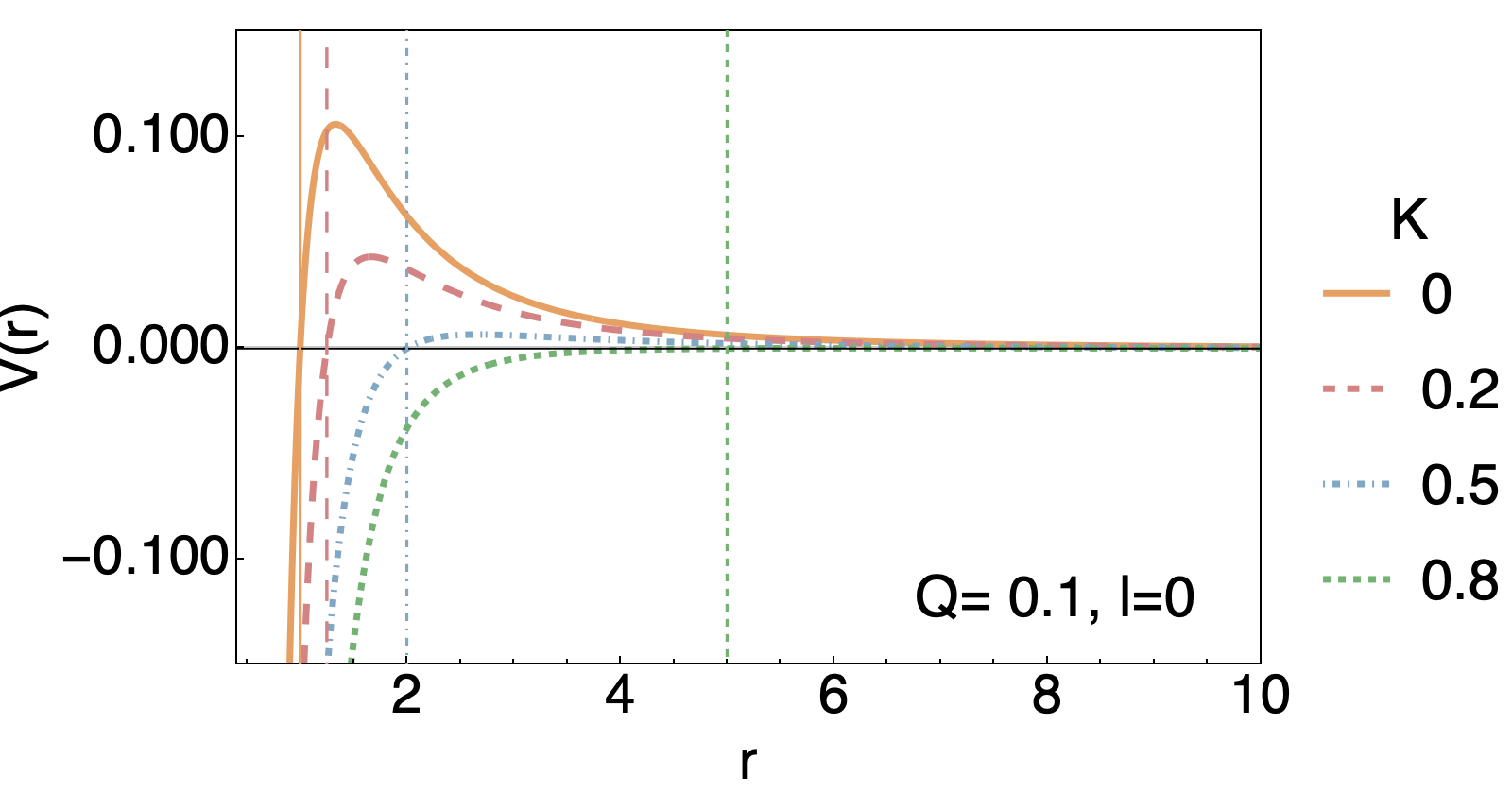}
\includegraphics[height=3.5cm,clip,trim=0 0 118 0]{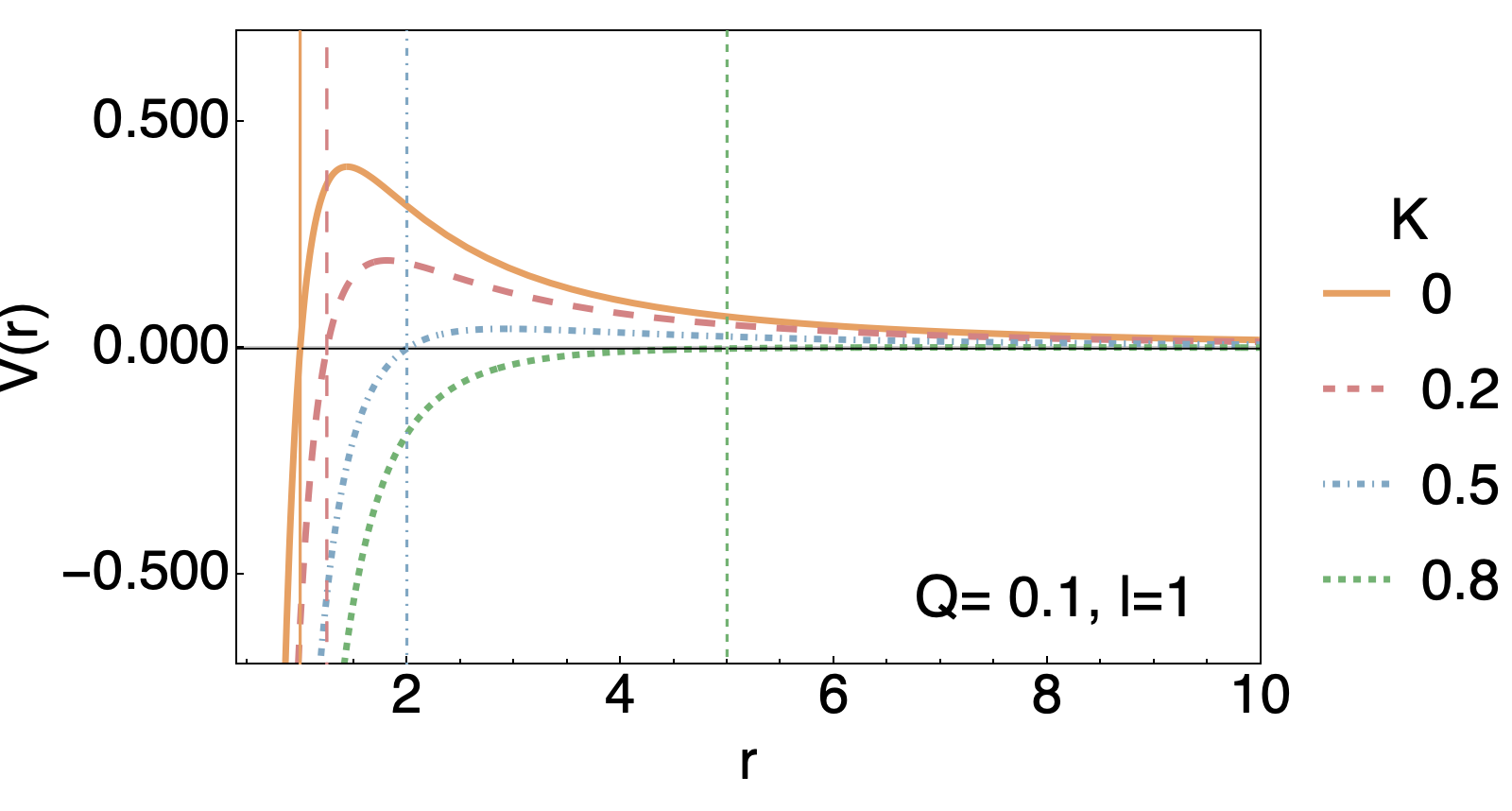}
\includegraphics[height=3.5cm,clip]{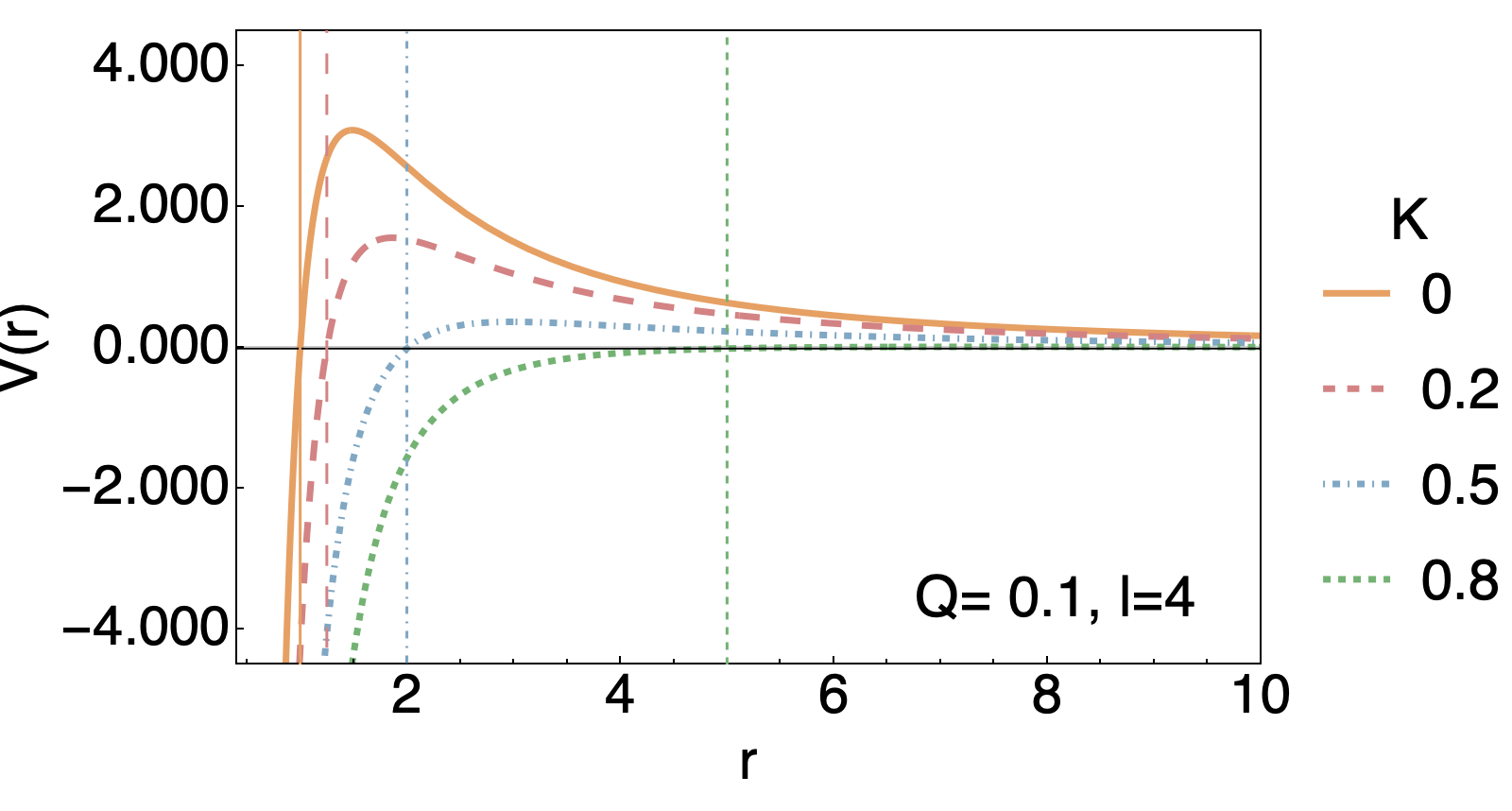}
\\
\includegraphics[height=3.5cm,clip,trim=0 0 118 0]{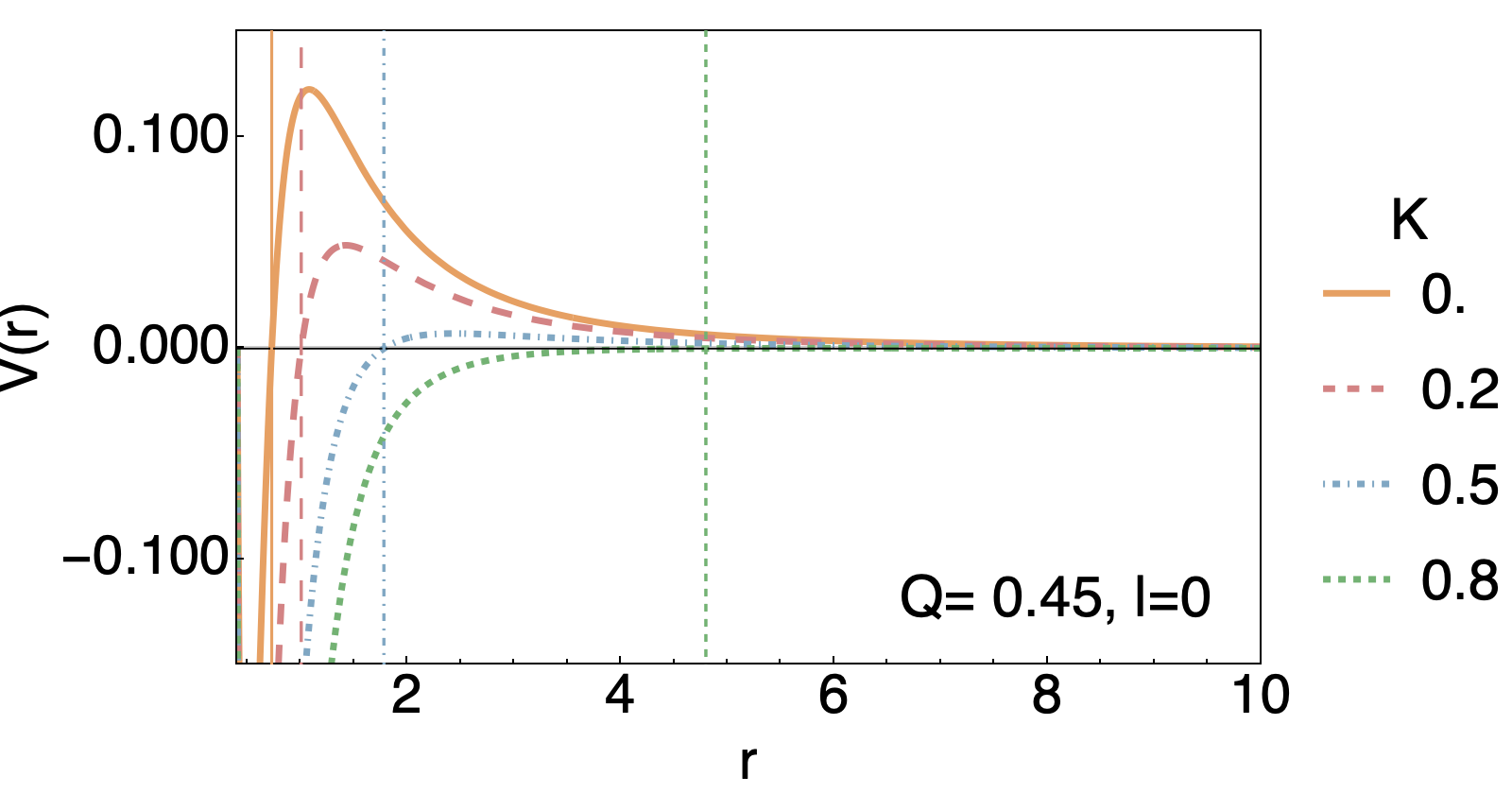}
\includegraphics[height=3.5cm,clip,trim=0 0 118 0]{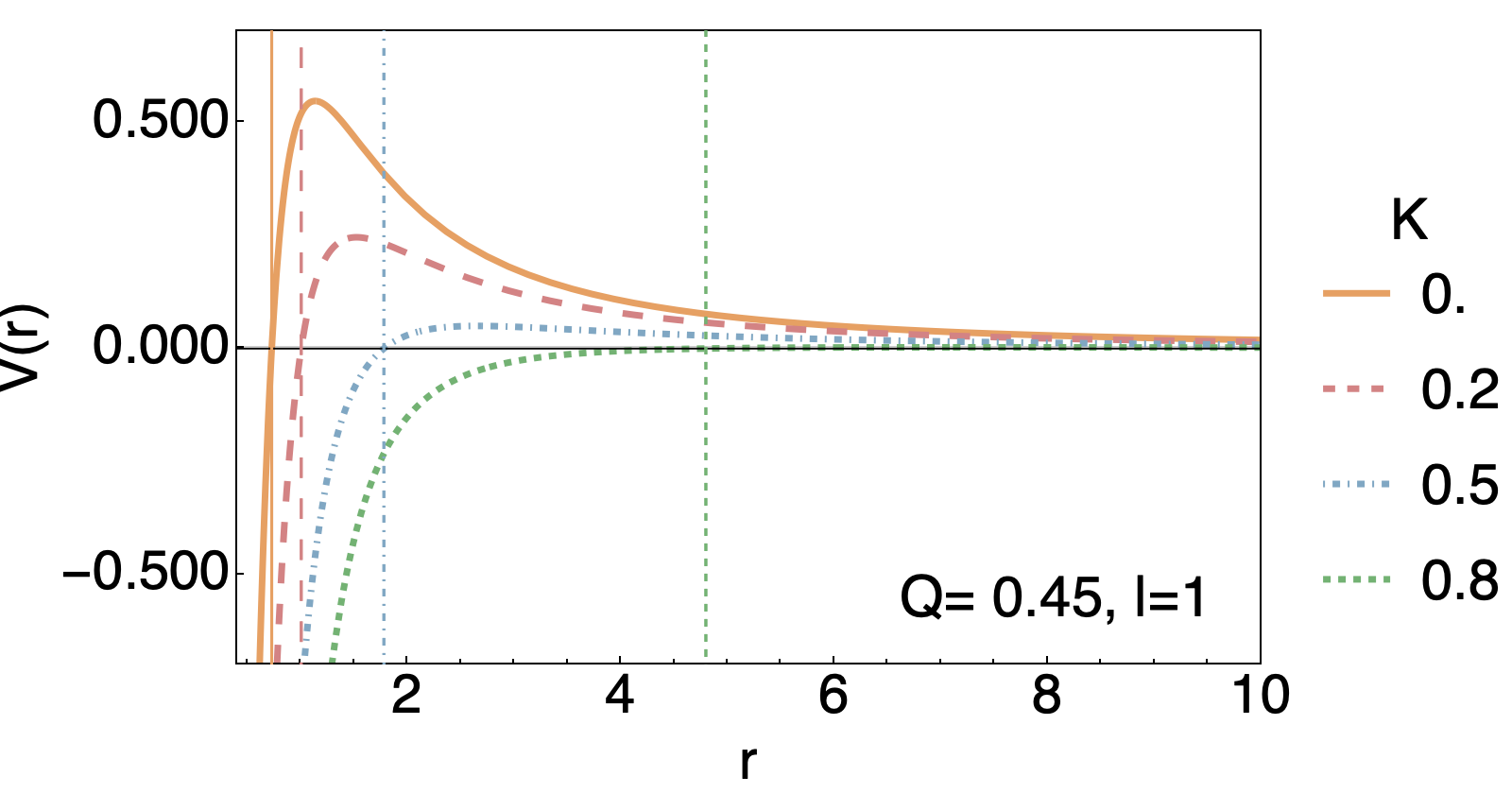}
\includegraphics[height=3.5cm,clip]{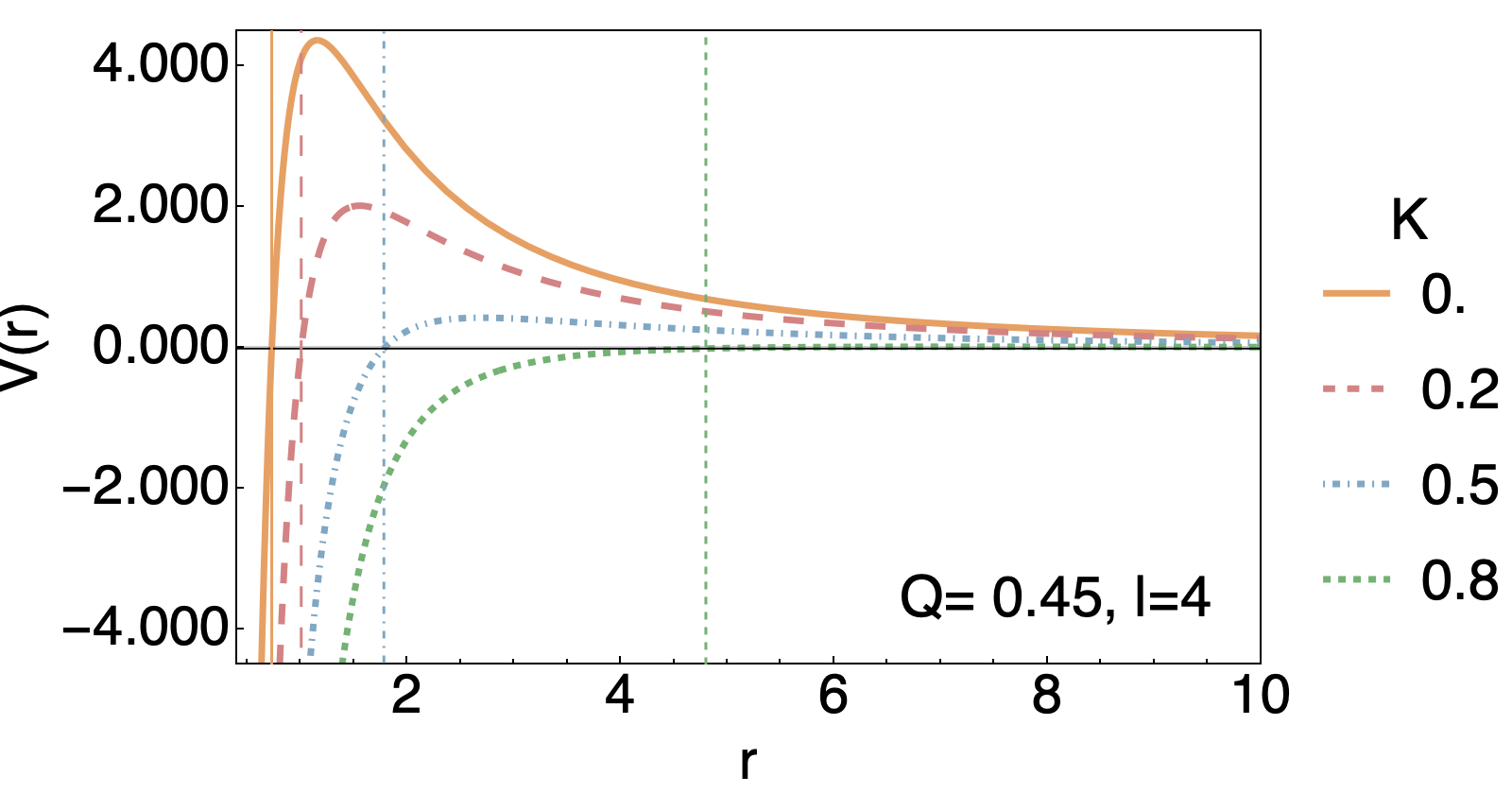}
\caption{Effective potential, Eq.~\eqref{eq:2CompKiselev-Veff-polynomial},
for the two-component Kiselev model with \bh{} charge $Q=0.1$ (top panel) and $Q=0.45$ (bottom panel), and fluid parameters $0\leq K<1$ as indicated in the legend.
We consider multipoles
$l=0$ (left panel),
$l=1$ (middle panel),
and $l=4$ (right panel).
The location of the event horizon for given $(Q,K)$ is indicated by vertical lines in the same color coding as the corresponding fluid parameter $K$.
}
\label{fig:v_combined}
\end{figure*}


In this section, we examine the scattering of a test massless scalar field, $\psi$, with charge $q$ in the background of the spherically symmetric, non-vacuum \bh{}
described by the two-component Kiselev spacetime given in Eq.~\eqref{eq:2CompKiselev-MetricFunction}.
The dynamics of the scalar field are governed by the Klein--Gordon equation (see App.~\ref{sec:action_eom}),
\begin{align}
\label{eq:KG-massless}
D^{\mu} D_{\mu} \psi &= 0
\,,
\end{align}
where
$D_{\mu} = \nabla_{\mu} - i q A_{\mu}$
is the covariant derivative that incorporates the electromagnetic interaction between the scalar field and the \bh.
The electromagnetic vector potential $A_{\mu}$ is given in Eq.~\eqref{eq:VectorPotential}.

Following the symmetries of the spacetime,
we take a mode ansatz for the solution $\psi(t, r, \theta, \phi)$ to Eq.~\eqref{eq:KG-massless},
\begin{align}
\label{eq:psidecomp-massless}
    \psi(t, r, \theta, \phi) & =
        \sum_{l,m} \frac{\Psi_{l m}(r)}{r}\, Y_{l m}(\theta, \phi)\, e^{-i \omega_{l m} t}
\,,
\end{align}
where $Y_{l m}(\theta, \phi)$ are spherical harmonics,
$\omega_{l m}$ is the complex frequency for a given $(l, m)$ multipole,
and $\Psi_{l m}(r)$ is the radial part of the scalar field.
Since Eq.~\eqref{eq:KG-massless} is linear,
the multipoles decouple.
We therefore suppress the
labels $(l,m)$ and
simply
write $\Psi(r)$ and $\omega$ in the following.

Using the ansatz~\eqref{eq:psidecomp-massless}
and the orthogonality of the spherical harmonics,
Eq.~\eqref{eq:KG-massless} reduces to a second-order ordinary differential equation for the radial function,
\begin{align}
\label{eq:KG-massless-2CompKiselev}
\frac{\dif^{2} \Psi(r)}{\dif r_{\ast}^{2}}
+ \left((\omega - \Phi(r))^2 - V(r)\right)\Psi(r) &= 0
\,.
\end{align}
Here, we introduce the tortoise radial coordinate $r_{\ast}$, which is implicitly defined by
\begin{align}
\label{eq:tortoise-massless}
\frac{\dif r_{\ast}}{\dif r}
&= \frac{1}{f(r)}
\,,
\end{align}
so that the horizon $r\to r_{+}$ corresponds to $r_{\ast}\to -\infty$
and spatial infinity $r\to\infty$ to $r_{\ast}\to\infty$.
The function
\begin{align}
\label{eq:ElectrostaticPotential}
\Phi(r) &= \frac{qQ}{r}
\,,
\end{align}
is the electrostatic potential
that encodes the electrostatic interaction between the \bh{} charge $Q$ and the scalar field charge $q$.
We refer to the term $(\omega - \Phi(r))$
in Eq.~\eqref{eq:KG-massless-2CompKiselev}
as the electrostatically shifted
frequency, or equivalently as the shifted energy.
The geometric
effective potential is
\begin{align}
\label{eq:Veff}
V(r) &= f(r) \left(\frac{l(l+1)}{r^2} + \frac{f'(r)}{r}\right)
\,,
\end{align}
where  the prime denotes differentiation with respect to $r$.
Notably, $V(r)$ is purely geometric: it depends on the spacetime through the metric function $f(r)$ and on the angular structure through $l$, but it is independent of the scalar field's charge and frequency.

After inserting the metric function, Eq.~\eqref{eq:2CompKiselev-MetricFunction},
the effective potential becomes
\begin{align}
\label{eq:2CompKiselev-Veff-polynomial}
V(r)
&= - \frac{2Q^{4}}{r^{6}}
+ \frac{3 Q^{2}}{r^{5}}
+ \frac{\lambda Q^{2} - 2 (1-K) Q^{2} -1}{r^{4}}\\
&\quad + \frac{1-K -\lambda}{r^{3}}
+ \frac{\lambda (1-K)}{r^{2}}
\,,\nonumber
\end{align}
where $\lambda = l(l+1)$.

From Eq.~\eqref{eq:2CompKiselev-Veff-polynomial}, we immediately obtain the asymptotic limits.
At spatial infinity,
\begin{align}
\label{eq:V_asymptotic_inf}
\lim_{r\to\infty} V(r) &= 0
\,,
\end{align}
and at the event horizon,
\begin{align}
\label{eq:V_asymptotic_hor}
\lim_{r\to r_{+}} V(r) &= 0
\,,
\end{align}
since $f(r_{+})=0$ in Eq.~\eqref{eq:Veff}.

Next, we analyze the extrema of the geometric effective potential.
They are found by solving $V'(r) = 0$,
which, for the polynomial form in Eq.~\eqref{eq:2CompKiselev-Veff-polynomial},
yields the quartic equation
\begin{align}
\label{eq:VeffExtremaPolynEquation}
0 & =
      2\lambda(1-K)\, r^4
    + 3 (1-K-\lambda)\, r^3
\\ & \quad
    + 4 \left( \lambda Q^{2} - 2 (1-K) Q^{2} - 1 \right) r^2
    + 15 Q^2\, r - 12 Q^4
\,.\nonumber
\end{align}
A maximum of $V(r)$ corresponds to a potential barrier, while a minimum would correspond to a potential well.

For the parameter range considered in this work ($0\le K<1$, $Q^2<1/4$, and $l\ge 0$), Eq.~\eqref{eq:VeffExtremaPolynEquation} admits a single real root in the exterior region, $r>r_+$. From Eqs.~\eqref{eq:V_asymptotic_inf} and~\eqref{eq:V_asymptotic_hor}, the potential tends to zero at both boundaries of this region, and Eq.~\eqref{eq:Veff} gives $V(r)>0$ for all $r>r_+$. Thus, the root of Eq.~\eqref{eq:VeffExtremaPolynEquation} is always a maximum of the geometric effective potential.

In Fig.~\ref{fig:v_combined}, we plot the effective potential given in Eq.~\eqref{eq:2CompKiselev-Veff-polynomial} for representative parameters of the two-component Kiselev model.
As background spacetime, we consider \bh{s} with charge parameters $Q=0.1$ (top panel) and $Q=0.45$ (bottom panel) and fluid parameters $K\in\{0,\,0.2,\,0.5,\,0.8\}$.
The location of the event horizon for each given $(Q,K)$ is indicated by vertical lines with the corresponding color coding.
We show the potential for multipoles $l=0$ (left panel), $l=1$ (middle panel) and $l=4$ (right panel).
We note that the magnitude of the potential near the horizon
can be small, depending on the combination of parameters;
for instance, it is $\lesssim\mathcal{O}(10^{-2})$ for the monopole.

For the subextremal charge and fluid parameters considered here, we always find a potential barrier.
The solid orange curves correspond to the RN limit, $K=0$.

For fixed $Q$ and $K$, the barrier height increases with the multipole number, as seen by comparing the left, middle, and right panels. For fixed $l$ and $K$, increasing the charge $Q$ also raises the barrier height in the range shown, as seen by comparing the upper and lower rows.
As the fluid parameter $K$ increases, the barrier height decreases. The barrier becomes lower and broader, and its peak moves farther away from the \bh{}.

To interpret
the profiles of the geometric effective potential
in terms of the radial propagation of the scalar field, it is useful to introduce the effective local wavenumber, $k(r)$, through the relation~\footnote{Note that the square of the wavenumber, $k^{2}$, is identical to the \WKB{} effective potential that we introduce and analyze in Sec.~\ref{subsec:wkb}.}
\begin{align}
\label{eq:LocalWavenumber}
k^{2}(r) &= (\omega - \Phi(r))^{2} - V(r)
\,.
\end{align}
For complex mode frequencies $\omega$, the quantity $k^{2}(r)$ is generally complex.
Thus, the distinction between locally oscillatory and evanescent radial behavior is understood here through the sign of its real part, $\mathrm{Re}\,k^{2}(r)$.
Its classical turning points, $r_{\rm{tp}}$, are then defined by
\begin{equation}
\label{eq:TurningPointCondition}
\mathrm{Re}\,k^{2}(r_{\rm tp}) = 0
\,,
\end{equation}
or, since $V(r)$ is real and using Eq.~\eqref{eq:LocalWavenumber},
\begin{equation}\label{eq:TurningPointCondition2}
\mathrm{Re}\!\left[(\omega - \Phi(r_{\rm tp}))^{2}\right] = V(r_{\rm tp})
\,.
\end{equation}

Because the turning points are determined by comparing the shifted energy with $V(r)$ through Eq.~\eqref{eq:TurningPointCondition2}, the shape of $V(r)$ directly controls their location. In particular, the extrema of the geometric potential set the characteristic scale against which the shifted energy term is measured. The turning points separate the locally oscillatory and evanescent radial regions and play a central role in the \WKB{} analysis of quasinormal modes (see Sec.~\ref{sec:qnm_analysis}).

The features visible in Fig.~\ref{fig:v_combined} reflect the interplay between the shifted energy term and the geometric potential in Eq.~\eqref{eq:LocalWavenumber}.
The square of the effective local wavenumber at the horizon reduces to
\begin{equation}
k^{2}(\rp) = \left(\omega - \Phi(\rp) \right)^2
\,,
\end{equation}
which follows from Eq.~\eqref{eq:V_asymptotic_hor}.
Near the maximum of the geometric effective potential, if it is sufficiently high, the real part of the squared local wavenumber, $\mathrm{Re}\,k^{2}(r)$, may become negative, thus signaling an evanescent region.
Farther out, the geometric effective potential decreases, and the field may become oscillatory again.

In addition, when the equation
\begin{equation}
\label{eq:ShiftedFrequency}
\mathrm{Re}(\omega)-\Phi(r)=0
\end{equation}
admits a solution outside the event horizon, the real part of the shifted energy vanishes at that location while $V(r)$ remains positive.
This occurs at
\begin{equation}\label{eq:second_turning_point}
r_{\Phi}=\frac{qQ}{\mathrm{Re}(\omega)},
\end{equation}
provided that $qQ$ and $\mathrm{Re}(\omega)$ have the same sign and $r_{\Phi}>r_{+}$.
In that case, a second evanescent region can arise outside the geometric barrier.
If no such exterior point exists, this second evanescent region does not occur.
Thus, the resulting radial
profile of
$\mathrm{Re}\,k^2(r)$ may contain a finite oscillatory cavity, bounded on the inside by the geometric barrier and on the outside by the radius $r_{\Phi}$, defined in Eq.~\eqref{eq:second_turning_point}, at which the real part of the shifted energy vanishes.
In this sense, the trapping is produced jointly by the geometric potential and the electromagnetic interaction.
This mechanism does not require a local minimum of $V(r)$.
If the geometric potential additionally admits a local minimum, the trapping region may become wider, favoring a richer spectrum of long-lived resonances.

We stress, however, that this is only a qualitative interpretation of the possible radial propagation
determined by the effective local wavenumber.
Determining whether such configurations actually give rise to quasi-bound states, and whether these states become unstable, requires solving the full boundary-value problem.
This calculation lies beyond the scope of the present paper, but quasi-bound states have been found in similar physical systems~\cite{Hod:2012px,
Herdeiro:2013pia,
Dolan:2015dha,
Dias:2018zjg,
Senjaya:2023rna,
Sanchis-Gual:2025uvm,
Qin:2026axh}.



\section{Quasinormal Modes Calculation}
\label{sec:qnm_analysis}

In this section, we describe our methodology for calculating the \QNM{} spectrum of a massless, charged scalar field scattering off the \bh{} described by the metric in Eq.~\eqref{eq:2CompKiselev-MetricFunction}.
In particular, we solve the boundary value problem
posed by Eq.~\eqref{eq:KG-massless-2CompKiselev} using Leaver's continued fraction method and the \WKB{} approximation.
In addition to standard root-finding algorithms,
we employ automatic differentiation
to numerically solve the continued fraction equation in challenging regions of the parameter space.
We describe how this new method is constructed and applied to \QNM{} calculations.

\subsection{Boundary Conditions}
\label{ssec:BoundaryConditions}

We first discuss the boundary conditions satisfied by the radial function, $\Psi(r)$;
cf. Eq.~\eqref{eq:psidecomp-massless}.
At spatial infinity, the solution represents a purely outgoing wave.
For mathematical clarity, we write the outgoing boundary condition in terms of the tortoise radial coordinate defined in Eq.~\eqref{eq:tortoise-massless},

\begin{align}
\label{eq:psi_infinity}
\Psi(r_{\ast})
&\sim
A^{\rm out}
\exp{\left[
i\omega r_{\ast}
-
\frac{i qQ}{1-K}\ln r_{\ast}
\right]}
\,,\,\,\,
r_{\ast} \to \infty
\,,
\end{align}
where $A^{\rm{out}}$ denotes the amplitude of the outgoing wave.
Note that, despite the solid angle deficit mentioned in Sec.~\ref{ssec:KiselevMetric}, the effective potential vanishes at spatial infinity and the outgoing boundary condition introduced in Eq.~\eqref{eq:psi_infinity} remains applicable.

At the event horizon, the
physical solution is a purely ingoing wave,
ensuring that no information escapes from the \bh{}.
This condition is written as
\begin{align}
\label{eq:psi_horizon}
\Psi(r_{\ast}) & \sim A^{\rm{in}}
\exp{\left[-i r_{\ast} \left( \omega - \frac{q Q}{\rp} \right)\right]}
\,, \quad
r_{\ast} \to -\infty
\,,
\end{align}
where $A^{\rm{in}}$ denotes the amplitude of the ingoing wave
and $\rp$ is the location of the event horizon in Schwarzschild coordinates given in Eq.~\eqref{eq:2CompKiselev-HorizonsSubExtremal}.

The differential equation in Eq.~\eqref{eq:KG-massless-2CompKiselev},
together with the boundary conditions in Eqs.~\eqref{eq:psi_infinity} and~\eqref{eq:psi_horizon},
defines an eigenvalue problem.
Its solution is a discrete set of complex frequencies $\omega=\omegaR+i\omegaI$, where
\begin{equation}
\label{eq:omega-real-imag}
    \omegaR\equiv\mathrm{Re}(\omega),
    \quad
    \omegaI\equiv\mathrm{Im}(\omega).
\end{equation}
We use the shorthand notation in Eq.~\eqref{eq:omega-real-imag} in places to keep expressions compact.

The following subsections describe how the frequencies are computed numerically with the continued fraction method
and
the \WKB{} approximation.


\subsection{Continued Fraction Method}
\label{ssec:ContinuedFractionMethod}

Leaver's continued fraction method~\cite{Leaver:1985ax}
is a powerful tool for computing \QNM{}
frequencies of perturbations around \bh{s}.
In this section, we extend
the continued fraction method
to perturbations around a charged \bh{} surrounded by an anisotropic fluid, i.e., the two-component Kiselev metric introduced in Sec.~\ref{sec:kiselev_general}.
Leaver's approach employs the
Frobenius method \cite{Arfken:2005mmp}, in which the solution of the second-order differential equation is expanded as a power series around a singular point.
The series expansion is then inserted into the differential equation to obtain a recurrence relation for
the series coefficients.
The recurrence relation defines an eigenvalue problem,
and its eigenvalues are the complex \QNM{} frequencies that we seek.

To apply Leaver's method,
we first rewrite the differential equation in Eq.~\eqref{eq:KG-massless-2CompKiselev}
in a form
that makes the singular points explicit.
We rescale the radial function,
\begin{align}
\label{eq:wavefunction_transformation}
\Psi(r) & = r R(r)
\,,
\end{align}
and we rewrite the metric function in Eq.~\eqref{eq:2CompKiselev-MetricFunction}
as
\begin{align}
\label{eq:MetricFctInDelta}
f(r) & = \frac{\Delta(r)}{r^{2}}
\,,
\end{align}
where we introduce the polynomial
\begin{align}
\label{eq:DeltaPolynomial}
\Delta(r) & = (1-K) (r-\rp) (r-\rmm)
\,.
\end{align}
Here, $r_{\pm}$ denote the horizons given by Eq.~\eqref{eq:2CompKiselev-HorizonsSubExtremal}.

Substituting the transformations in Eqs.~\eqref{eq:wavefunction_transformation} and~\eqref{eq:MetricFctInDelta}
into Eq.~\eqref{eq:KG-massless-2CompKiselev}
gives a second-order differential equation for the rescaled radial function, $R(r)$,
\begin{align}
\label{eq:final_transformed_eq}
0 & = \frac{\Delta^{2}}{r} R''
    + \frac{\Delta \Delta'}{r} R'
\\ & \quad
+ \left[  \frac{\Delta \Delta'}{r^2}
    - \frac{2\Delta^2}{r^3}
    + r^{3} \left[
        \left(\omega - \Phi(r)\right)^2
        - V(r) \right]
 \right] R
\,,\nonumber
\end{align}
where
the prime again denotes differentiation with respect to $r$.
The effective potential becomes
\begin{align}
\label{eq:VeffInDelta}
V(r) & = \frac{\Delta}{r^2}
\left( \frac{l(l+1)}{r^2}
    +\frac{\Delta'}{r^3}
    -\frac{2\Delta}{r^4}
    \right)
\,.
\end{align}
From Eq.~\eqref{eq:final_transformed_eq},
we see that the singular points are
$r = r_{\pm}$,
$r = 0$,
and $r \to \infty$,
and we proceed with the series ansatz.

\subsubsection{Ansatz as power series}


The solution of the differential equation in Eq.~\eqref{eq:final_transformed_eq} can be written as a power series around one of the singular points.
Following Leaver~\cite{Leaver:1985ax},
we
expand
it around the event horizon, $\rp$,
and capture
the remaining singular points in the prefactors.
Then, we take the ansatz~\cite{Baber:1935TC},
\begin{align}
\label{eq:ansatz_power_series}
R(r) & = \exp\left[\frac{i\omega r}{1-K}\right]
    \left(r - \rmm \right)^\epsilon
    \sum_{n=0}^\infty a_n \left(\frac{r - \rp}{r - \rmm}\right)^{n + \Gamma}
\,,
\end{align}
where
$r_{\pm}$ are given in Eq.~\eqref{eq:2CompKiselev-HorizonsSubExtremal},
$n=0,1,2,\ldots$ denotes
the summation index,
and the exponents $\Gamma$ and $\epsilon$ follow from the boundary conditions in Sec.~\ref{ssec:BoundaryConditions}.
The sequence of coefficients,
$\{a_{n}\}$,
is determined by the recurrence relation that we derive below.

We identify the exponent $\Gamma$ from the ingoing boundary condition at the horizon.
Matching the power-law behavior of Eq.~\eqref{eq:ansatz_power_series}
to that in Eq.~\eqref{eq:psi_horizon}, we find
\begin{align}
\label{eq:SeriesExponentGamma}
\Gamma & = - \frac{i \rp}{(1-K)(\rp-\rmm)}
\left(\omega\rp - q Q \right)
\,.
\end{align}
We identify the exponent $\epsilon$ using the outgoing boundary condition.
Matching the behavior of the radial function for large radii, $R(r)\propto r^{\epsilon} e^{i\omega r/(1-K)}$,
from Eq.~\eqref{eq:ansatz_power_series} to the boundary condition in Eq.~\eqref{eq:psi_infinity}, we find
\begin{align}
\label{eq:epsilon_exponent}
\epsilon & = - 1
    - i q Q \left(\rp + \rmm \right)
    + i \omega\left(\rp + \rmm \right)^2
\,.
\end{align}
Here, we used the identity 
\begin{equation*}
  \rp+\rmm=\frac{1}{1-K}
\,.  
\end{equation*}
that follows from Eq.~\eqref{eq:2CompKiselev-HorizonsSubExtremal}. 

\subsubsection{The Recurrence Relation}

We now derive a recurrence relation for the coefficients, $a_{n}$, of the series expansion in Eq.~\eqref{eq:ansatz_power_series}.
We find a three-term recurrence relation that defines an eigenvalue problem, which we solve using the continued fraction method to compute the complex frequencies $\omega$.

We start by substituting the series ansatz of Eq.~\eqref{eq:ansatz_power_series} into the differential equation in Eq.~\eqref{eq:final_transformed_eq}
which yields an expression in
powers of
$\frac{r - \rp}{r - \rmm}$.
Following Leaver's procedure~\cite{Leaver:1990zz},
we collect the terms at each power of
$\frac{r - \rp}{r - \rmm}$
and find a three-term recurrence relation,
\begin{subequations}
\label{eq:3termrecurrence}
\begin{align}
\label{eq:3termrecurrence_n0}
\alpha_0 a_1 + \beta_0 a_0 &= 0
\,,\\
\label{eq:3termrecurrence_general}
\alpha_n a_{n+1} + \beta_n a_n + \gamma_n a_{n-1} &= 0
\,,\text{ for } n \geq 1
\,.
\end{align}
\end{subequations}
The coefficients
$(\alpha_{n},\beta_{n},\gamma_{n})$
in Eqs.~\eqref{eq:3termrecurrence} are given by
\begin{widetext}
\begin{subequations}
\label{eq:3RecurrenceCoefficients}
\begin{align}
\alpha_{n}
&=
(1+n)
\Big[
    \rp \Big(1+n+2i\rp(qQ-\rp\omega)\Big)
  - \rmm \Big(1+n+2i\rp(-qQ+\rp\omega)\Big)
\Big]
\,,\\
\beta_{n}
&=
\rmm^2
\Big[
    \lambda
    + (qQ-2\rp\omega)
      \Big(i+2in+4\rp(qQ-\rp\omega)\Big)
\Big]
\\ &\quad
+ \rmm
\Big[
    1+2n^2
    +2\rp(qQ-\rp\omega)
      \Big(-i+4\rp(qQ-2\rp\omega)\Big)
    +n\Big(2+4i\rp(-qQ+\rp\omega)\Big)
\Big]
\nonumber\\
&\quad
+ \rp
\Big[
    -1-2n^2-3i qQ \rp
    +\rp
      \Big(
        -\lambda
        +4i\rp\omega
        +4\rp(qQ-2\rp\omega)(qQ-\rp\omega)
      \Big)
    +n\Big(-2+2i\rp(-3qQ+4\rp\omega)\Big)
\Big]
\,,\nonumber\\
\gamma_{n}
&=
n^2(-\rmm+\rp)
+
2 i n(\rmm+\rp)
\Big(
    -qQ\rmm
    +2qQ\rp
    +\rmm^2\omega
    -2\rp^2\omega
\Big)
\\ &\quad
-
4\rp(\rmm+\rp)^2
\Big(-qQ+\rp\omega\Big)
\Big(-qQ+(\rmm+\rp)\omega\Big)
\,,\nonumber
\end{align}
\end{subequations}
\end{widetext}
and we recall $\lambda=l(l+1)$. Note that, if a nonminimal common polynomial factor is retained in the reduced
differential equation, the resulting recurrence can appear with more than
three terms; in that case, Gaussian elimination may be used to reduce it
to a three-term recurrence. In the present derivation, we chose a prefactor that gives the three-term recurrence directly.
The recurrence relation is the discrete analogue of the differential equation in Eq.~\eqref{eq:final_transformed_eq}.

As a consistency check, in the Schwarzschild limit $K=0$, $\rmm=0$, $\rp=1$, and $Q=0$, the recurrence coefficients in Eqs.~\eqref{eq:3RecurrenceCoefficients} reduce to those in Leaver's work~\cite{Leaver:1985ax} for scalar
perturbations of a Schwarzschild \bh{.} 

A three-term recurrence relation has two linearly independent solution
sequences for the coefficients $\{a_n\}$. Any solution can be written as a linear combination of these two sequences. When a nonzero linear combination exists whose large-$n$ behavior is asymptotically smaller than that of any linearly independent solution, this subdominant sequence is unique up to
normalization and is called the ``minimal solution''.
Since the argument of the series in
Eq.~\eqref{eq:ansatz_power_series} approaches unity as $r \to \infty$,
convergence at spatial infinity is governed by the large-$n$ behavior of
the coefficients $a_n$. Only the minimal solution has the behavior
required for convergence and  preserves the required outgoing
asymptotics. 
Numerically, frequencies associated with the minimal
solution should converge to a stable\footnote{Here, ``stable'' means
invariant under small changes in the number $N$ of terms retained in the
truncated continued fraction, for sufficiently large $N$.} value as
$N$ becomes sufficiently large.

The three-term recurrence relation in Eq.~\eqref{eq:3termrecurrence} yields a tri-diagonal matrix equation,
\begin{align}
\label{eq:3termrecurrence-matrix}
\begin{pmatrix}
\beta_{0}  & \alpha_{0} & & &  \\
\gamma_{1} & \beta_{1}  & \alpha_{1} & &  \\
 & \ddots & \ddots & \ddots & \\
 & & \gamma_{N-1} & \beta_{N-1} & \alpha_{N-1} \\
 & &  & \gamma_{N} & \beta_{N}
\end{pmatrix}
\begin{pmatrix}
a_0 \\
a_1 \\
\vdots \\
a_{N-1} \\
a_N
\end{pmatrix}
& = 0
\,.
\end{align}
The coefficients $(\alpha_{n},\beta_{n},\gamma_{n})$,
given in Eqs.~\eqref{eq:3RecurrenceCoefficients},
depend on the spacetime's and scalar field's parameters and,
importantly, on the complex frequency $\omega$
that we want to solve for.
Hence, Eq.~\eqref{eq:3termrecurrence-matrix}
poses an eigenvalue problem
whose solutions determine a discrete set of frequencies.
The \QNM{} frequencies correspond to those values of $\omega$ for which the determinant of the matrix in Eq.~\eqref{eq:3termrecurrence-matrix} approaches zero in the limit of $N\to\infty$.
Thus, Eq.~\eqref{eq:3termrecurrence-matrix} becomes a root-finding problem in the complex frequency plane
that needs to be solved numerically.

\subsubsection{Continued Fraction}

We note that the three-term recurrence relation in Eq.~\eqref{eq:3termrecurrence}
is mathematically equivalent to the
matrix equation in Eq.~\eqref{eq:3termrecurrence-matrix},
which, in turn, is equivalent to an infinite continued fraction. Thus, to find the roots
(comprising the \QNM{} frequencies)
of the eigenvalue problem in Eq.~\eqref{eq:3termrecurrence-matrix},
we write it as a continued fraction.
We therefore consider the ratio of successive coefficients,
$a_{n+1}/a_{n}$,
of the minimal solution
and find~\cite{Leaver:1990zz},
\begin{align}
\label{eq:continued_fraction}
\cfrac{a_{n+1}}{a_n} & = \cfrac{
  -\gamma_{n+1}}{\beta_{n+1}
    -\cfrac{\alpha_{n+1}\gamma_{n+2}}
    {\beta_{n+2}
    - \cfrac{\alpha_{n+2}\gamma_{n+3}}{\beta_{n+3} - \dots}}}
\,.
\end{align}

We see that each term depends on the ratio of the next two coefficients and,
thus, creates a recursive chain.
To derive a characteristic equation for the \QNM{} frequencies,
we relate the continued fraction for the first coefficient,
\begin{align}
\label{eq:continued_fraction_n0}
\cfrac{a_{ 1}}{a_0} & = \cfrac{-\gamma_{ 1}}{\beta_{ 1}
    - \cfrac{\alpha_{ 1}\gamma_{ 2}}{\beta_{ 2}
    - \cfrac{\alpha_{ 2}\gamma_{ 3}}{\beta_{ 3} - \dots}}}
\,,
\end{align}
to the ratio obtained from the recurrence relation in Eq.~\eqref{eq:3termrecurrence_n0}
\begin{align}
\label{eq:Ratio_n1n0_recurrence}
\frac{a_1}{a_0} & = -\frac{\beta_0}{\alpha_0}
\,.
\end{align}
Equating Eqs.~\eqref{eq:continued_fraction_n0} and~\eqref{eq:Ratio_n1n0_recurrence},
we find the characteristic equation
\begin{align}
\label{eq:characteristic_equation}
F(\omega) & \equiv \beta_0
    - \cfrac{\alpha_0 \gamma_1}{\beta_1
    - \cfrac{\alpha_1 \gamma_2}{\beta_2
    - \cfrac{\alpha_2 \gamma_3}{\beta_3 - \dots}}} = 0
\,,
\end{align}
whose roots are the \QNM{} frequencies.
The coefficients $(\alpha_{n},\beta_{n},\gamma_{n})$ are functions of the frequency,
c.~f.~Eqs.~\eqref{eq:3RecurrenceCoefficients},
and therefore Eq.~\eqref{eq:characteristic_equation}
is a transcendental equation that
needs to be solved numerically.
In principle, the characteristic equation in Eq.~\eqref{eq:characteristic_equation}
can be solved to find both the fundamental and any of the overtone \QNM{} frequencies
if the continued fraction is truncated at a sufficiently large number $N$ of terms.
In practice, we solve Eq.~\eqref{eq:characteristic_equation}
only for the fundamental \QNM{} frequency,
and we invert it iteratively for overtones $n\geq 1$.
This inversion yields a relation between two continued fractions,
\begin{widetext}
\begin{align}
\label{eq:inverted_equation}
\left[\beta_n - \cfrac{\alpha_{n-1} \gamma_n}{\beta_{n-1} - \cfrac{\alpha_{n-2} \gamma_{n-1}}{\beta_{n-2} - \cdots - \cfrac{\alpha_0 \gamma_1}{\beta_0}}}\right]
& =
\left[\cfrac{\alpha_n \gamma_{n+1}}{\beta_{n+1} - \cfrac{\alpha_{n+1} \gamma_{n+2}}{\beta_{n+2} - \cfrac{\alpha_{n+2} \gamma_{n+3}}{\beta_{n+3} - \cdots}}}\right]
\,,
\end{align}
\end{widetext}
We solve Eqs.~\eqref{eq:characteristic_equation} and~\eqref{eq:inverted_equation} numerically using up to $N=5000$ terms. In the simplest cases, we use $N=700$. We use the conventional MINPACK Powell Hybrid root-finding method for most calculations,
and complement them with automatic differentiation methods in the most challenging parameter ranges.
The results are presented in Sec.~\ref{sec:numerical_results}, and we quantify the numerical error in App.~\ref{sec:error_analysis}.

\subsection{Automatic Differentiation for the Continued Fraction}
\label{ssec:automaticdifferentiation}
We can find the roots of the characteristic equation defined in Eq.~\eqref{eq:characteristic_equation} using traditional root finders.
Traditional solvers are efficient and reliable
for small to intermediate values of the \bh{} charge
$Q$ and the fluid parameter $K$.
However, as the parameters
approach their extremal values, or when the electrostatic coupling, $qQ$,
between the scalar-field charge and the \bh{} charge becomes large, closely spaced
roots may arise.
Then, conventional root-finding algorithms may fail to
distinguish nearby roots or converge to spurious solutions.
Therefore, we reformulate the problem as a minimization task to find the desired frequencies and solve it using the limited-memory Broyden--Fletcher--Goldfarb--Shanno (L-BFGS) algorithm together with automatic differentiation \cite{Griewank:2003ad}.

L-BFGS is a quasi-Newton optimization method. In our case, it constructs an approximation to the inverse Hessian matrix of the function under consideration using information from previous frequency updates and gradient evaluations~\cite{Liu:1989lbfgs}. This approximation is then used to determine an efficient update of the candidate frequency associated with Eq.~\eqref{eq:characteristic_equation}.

Instead of solving the characteristic equation in Eq.~\eqref{eq:characteristic_equation}
directly, we introduce the real, non-negative objective function~\cite{Boyd:2004fnq},
\begin{align}
\label{eq:ObjectiveFunctionML}
J(\omega) = |F(\omega)|^{2} \geq 0
\,,
\end{align}
where $F(\omega)$ is defined in Eq.~\eqref{eq:characteristic_equation}.
The minima of the objective function occur when $J(\omega)=0$,
so finding the
minima of this objective function
is mathematically equivalent to solving the characteristic equation in Eq.~\eqref{eq:characteristic_equation}.
This reformulation converts the root-finding problem into a nonlinear minimization problem for a real, non-negative function,
allowing us to employ gradient-based optimization methods.

The optimization is performed in the two-dimensional real space
$(\omegaR,\omegaI)$
spanned by the real and imaginary parts of the
frequency.
To minimize the objective function, $J(\omega)$,
we use the L-BFGS optimizer with a strong Wolfe line search implemented in PyTorch~\cite{Paszke:2019xhz}.
The optimizer iteratively updates the frequency
using gradient information.
The latter is determined by the gradient of the objective function with respect to the frequencies,
\begin{align}
\label{eq:GradJomega}
\nabla_{\omega} J & = \left(
\frac{\partial J}{\partial \omega_R},
\,
\frac{\partial J}{\partial \omega_I}
\right)
\,.
\end{align}

The gradient in Eq.~\eqref{eq:GradJomega} is evaluated by PyTorch's automatic differentiation engine \textsc{autograd}, which tracks the arithmetic operations defining $J$ as a computational graph and computes the derivatives automatically within floating-point precision.

Starting from an initial guess
for the real and imaginary parts of the frequency, $\omegaR$, $\omegaI$, the optimizer iteratively reduces the residual of $J(\omega)$.
We typically set the gradient tolerance to $10^{-14}$ and the parameter-change tolerance to $10^{-16}$ within the L-BFGS optimizer. In addition to the native convergence criteria of the optimizer, we impose the condition
\begin{equation}
J(\omega)=|F(\omega)|^2 \leq 10^{-16},
\label{eq:root_criterion}
\end{equation}
to verify that the converged solution corresponds to a root of the characteristic equation. The corresponding value of the frequency $\omega$ is then accepted as a root.

We obtain the initial guess for the root search by evaluating the characteristic equation, Eq.~\eqref{eq:characteristic_equation},
in the complex frequency plane.
Regions in which the magnitude of the characteristic equation is small indicate candidate \QNM{} frequencies and are used as initial guesses.
Examples of these low-residual regions are shown in Fig.~\ref{fig:collage_CE_QNM}.

\begin{figure}[h]
\centering
\includegraphics[width=0.47\textwidth,trim=30 30 30 25,clip]{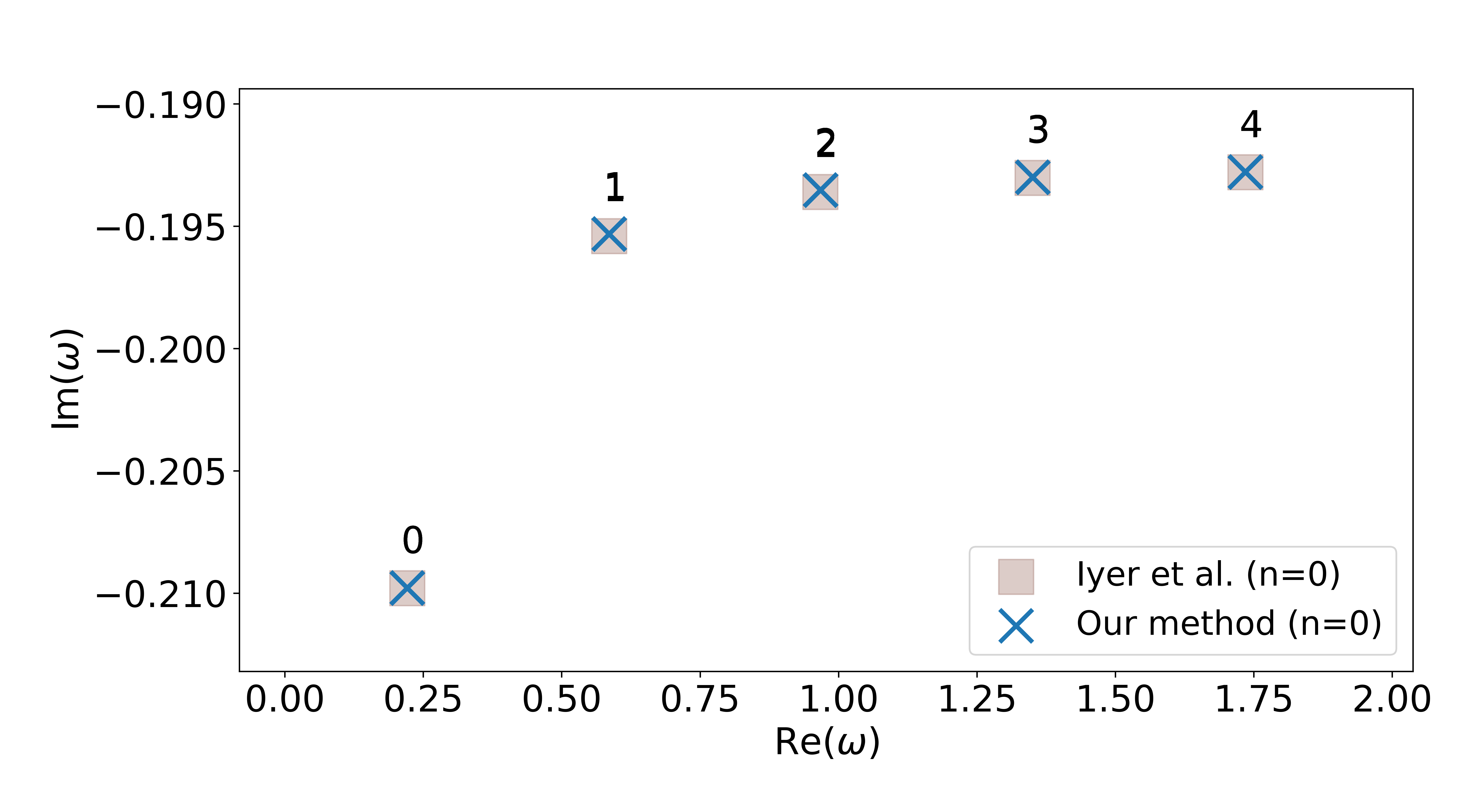}
\caption{
Spectrum of fundamental \QNM{} frequencies
of a massless scalar field propagating
on a Schwarzschild \bh{} background, shown as $\mathrm{Re}(\omega)$ versus
$\mathrm{Im}(\omega)$ for $l = 0, \dots, 4$.
We compare our results (crosses) against the original calculation (squares) by Iyer
\emph{et al.}~\cite{Iyer:1986np}.
}
\label{fig:qnms_comparison}
\end{figure}

We validate the automatic differentiation method
that we apply to \QNM{} computations
by benchmarking our results
against known \QNM{} spectra of scalar perturbations around Schwarzschild and \RN{} \bh{s}~\cite{Iyer:1986np, Richartz:2014lpa}.
In Fig.~\ref{fig:qnms_comparison}, we display
the
spectrum of the fundamental $l=0,\ldots,4$ multipoles of a massless, uncharged scalar field propagating on a Schwarzschild background.
We compare our results against the classic result by Iyer \emph{et al.}~\cite{Iyer:1986np},
and we find excellent agreement within $\lesssim 10^{-6}$.
We compare the numerical errors of the automatic-differentiation approach with those of the traditional root-finding implementation in App.~\ref{sec:error_analysis}, and find that, for sufficiently large truncation level $N$, the errors decrease further and exhibit a weaker dependence on the initial guesses for the roots.

In practice, we perform all calculations on a standard workstation or laptop (Mac M4, $32$ GB RAM).
A typical optimization run takes a few seconds for each root.
While not the only possible approach,
the technique presented in this section is straightforward to implement and
performs well across the full
(allowed)
parameter space.
In particular, it enables us to reliably resolve closely spaced roots that
arise in near-extremal configurations.
To our knowledge, this is the first application
of
automatic differentiation through PyTorch's \textsc{autograd} engine in combination with Leaver's continued fraction method.
This yields a flexible and robust approach that can be applied to different problems in \bh{} spectroscopy. Moreover, because the method is formulated in terms of differentiable tensor operations within PyTorch, it is naturally compatible with GPU-based implementations, which could become advantageous in future applications involving extensive parameter scans.

 \subsection{Wentzel–Kramers–Brillouin Approximation}\label{subsec:wkb}

We compare the \QNM{} frequencies obtained with the continued fraction method to those computed with the \WKB{} approximation for the same parameter sets.
The \WKB{} method is a semi-analytic technique used to approximate solutions to Schr\"{o}dinger-like equations such as Eq.~\eqref{eq:KG-massless-2CompKiselev}.
We write Eq.~\eqref{eq:KG-massless-2CompKiselev} as
\begin{align}
    \frac{\dif^{2}\Psi}{\dif r^{2}_{\ast}} + U(r,\omega) \Psi & = 0
\,,
\end{align}
where
$r_{\ast}$ is
the tortoise coordinate
(cf. Eq.~\eqref{eq:tortoise-massless})
and
\begin{equation} \label{eq:WKBeffpot}
 U(r,\omega)=(\omega-\Phi(r))^2 - V(r)
\,
\end{equation}
is the \WKB{} effective potential
determined by the geometric effective potential $V(r)$ in Eq.~\eqref{eq:Veff} and the electrostatic potential $\Phi$ given in Eq.~\eqref{eq:ElectrostaticPotential}.
We note that the \WKB{} potential is equivalent to the square of the local wavenumber, $k^2(r)$, introduced in Eq.~\eqref{eq:LocalWavenumber}.

Explicitly, the \WKB{} effective potential reads
\begin{align}
\label{eq:WKBEffPotential}
U(r,\omega) & =
\left(\omega - \frac{qQ}{r} \right)^{2}
-f(r)\left(\frac{l(l+1)}{r^2} + \frac{f'(r)}{r}\right)\,,
\end{align}
where the metric function $f(r)$ is given in Eq.~\eqref{eq:2CompKiselev-MetricFunction}.

In the original Schutz--Will and Iyer--Will formulations, see~Refs.~\cite{Iyer:1986np,Schutz:1985km}, the \WKB{} construction for \bh{s} is treated as a one-dimensional scattering problem in the tortoise coordinate
with an effective potential that approaches a constant as
$r_*\to\pm\infty$
and that has a smooth barrier possessing a single nondegenerate extremum.

For the parameters that we consider here,
the potential in Eq.~\eqref{eq:WKBEffPotential} is compatible with these assumptions.
In particular, asymptotically it behaves as
\begin{subequations}
\label{eq:WKBEffPot_Asymptotics}
\begin{align}
\lim_{r\to\infty} U(r,\omega) & = \omega^{2}
\,,\\
\lim_{r\to \rp} U(r,\omega) & = \left(\omega-\frac{qQ}{r_{+}}\right)^2
\,,
\end{align}
\end{subequations}
which are constant, and we recall that $r_{\ast}\to\infty$ as $r\to\infty$.
We also verify numerically that, for the parameter sets to which the \WKB{} method is applied, the potential has a single, smooth, nondegenerate extremum.

In the \WKB{} method, the approximation is built by matching the asymptotic solution onto a local expansion about the extremum of the \WKB{} effective potential.
The \WKB{} formula requires specifying an expansion point $r_0$ in the radial coordinate,
about which the \WKB{} potential is Taylor-expanded.
Because the \QNM{} frequency $\omega$ is
complex, the \WKB{} potential is complex-valued even along the real $r$-axis,
and there is no single natural notion of an extremum there.

We define $r_0$ by extremizing the real part of the \WKB{} potential, $\mathrm{Re}\,U$, consistent with the discussion of the turning points in Eqs.~\eqref{eq:TurningPointCondition}
and~\eqref{eq:TurningPointCondition2}, where the distinction between locally oscillatory and evanescent regions is made using $\mathrm{Re}\,k^2$, or equivalently $\mathrm{Re}\,U$.
For a complex frequency, the real part of the \WKB{} potential is
\begin{align}
\label{eq:ReU}
\mathrm{Re}\,U(r,\omega) & =
    \left(\omegaR - \Phi(r) \right)^2 - \omegaI^2 - V(r)
\,.
\end{align}
Since the imaginary part of the frequency is independent of $r$, it does not affect the extremum condition.
Thus, for each value of $\omegaR$, the location $r_0\equiv r_0(\omegaR)$ of the real extremum is obtained from
\begin{align}
\label{eq:WKB-extremum-real}
\bigl( \mathrm{Re}\,U \bigr)'|_{r=r_0} & = 0
\,.
\end{align}
For brevity, we refer to this solution hereafter as the ``real extremum''.

We employ the sixth-order \WKB{} approximation developed by Konoplya \cite{Konoplya:2003ii}, following the procedure outlined in Ref.~\cite{Konoplya:2002tz} for applying the method to \RN{} spacetimes.
In our analysis, the sixth-order approximation provides an independent computation of the \QNM{} frequencies and yields good quantitative agreement with Leaver's continued fraction method for the examples considered, except for the case when the fluid parameter $K=0$, charge parameter $Q\simeq0.49$ at multipole number $l=4$ and interaction strength $qQ=1$, where the \WKB{} root finder fails.
Higher-order extensions up to the thirteenth order, combined with Pad\'e approximations~\cite{Konoplya:2019hlu}, are available and may be adopted when greater numerical accuracy is required in future work. The sixth-order \WKB{} formula adapted to frequency-dependent potentials reads~\cite{Konoplya:2003ii},
\begin{align}
\label{eq:wkb3}
\frac{i U_0}{\sqrt{2U^{(2)}_0}}
- \Lambda_2 - \Lambda_3 - \Lambda_4
- \Lambda_5 - \Lambda_6
& = \textrm{n} + \frac{1}{2}
\,,
\end{align}
where
\begin{equation}\label{eq:WKB-eff-pot-deriv}
 U_0 \equiv U(r_0,\omega), \quad U_0^{(k)}\equiv \partial^k_{r_{\ast}}U|_{r=r_0}\,,
\end{equation}
\\
\noindent the derivatives $U_0^{(k)}$ are taken with respect to the tortoise coordinate $r_{\ast}$,
$r_0$ is the real extremum defined in Eq.~\eqref{eq:WKB-extremum-real},
and $\textrm{n}$ is the overtone number.
Positive overtone numbers $\textrm{n}=0,1,2,\ldots$ correspond to $\mathrm{Re}(\omega)>0$, while negative
numbers
$\textrm{n}=-1,-2,\,\ldots$ correspond to $\mathrm{Re}(\omega) <0$.

The correction terms $\Lambda_2$ and $\Lambda_3 $ are given by
\footnote{In the original formulation of Ref.~\cite{Iyer:1986np}, the second-order correction term $\Lambda(n)$, which corresponds to our $\Lambda_{2}(n)$, was printed with a factor of $1$ instead of $i$ in the numerator of the common factor. After correcting this typo, our Schwarzschild \QNM{} results agree with Ref.~\cite{Iyer:1986nq}.}

\begin{widetext}
\begin{subequations}
\begin{align}
\Lambda_2(\textrm{n}) & = \frac{i}{\sqrt{2 U^{(2)}_0}}
\Bigg[ \frac{1}{8} \frac{U_0^{(4)}}{U^{(2)}_{0}}
    \left(\frac{1}{4} + b^2(\textrm{n})\right)
- \frac{1}{288} \left(\frac{U_0^{(3)}}{U^{(2)}_0}\right)^2
    \left(7 + 60\,b^2(\textrm{n})\right)
\Bigg]
\,,\\
\Lambda_3(\textrm{n}) & = \frac{b(\textrm{n})}{2 U^{(2)}_0}
\Bigg[ \frac{5}{6912} \left(\frac{U_0^{(3)}}{U^{(2)}_0}\right)^4 \left(77 + 188\, b^2(\textrm{n})\right)
- \frac{1}{384} \frac{\left(U_0^{(3)}\right)^2 U_0^{(4)}}{\left(U^{(2)}_0\right)^3} \left(51 + 100\, b^2(\textrm{n})\right)
\\ & \quad
+ \frac{1}{2304} \left(\frac{U_0^{(4)}}{U^{(2)}_0}\right)^2 \left(67 + 68\, b^2(\textrm{n})\right)
+ \frac{1}{288} \frac{U_0^{(3)}\,U_0^{(5)}}{\left(U^{(2)}_0\right)^2} \left(19 + 28\, b^2(\textrm{n})\right)
- \frac{1}{288} \frac{U_0^{(6)}}{U^{(2)}_0} \left(5 + 4\,b^2(\textrm{n})\right)
\Bigg]
\,,\nonumber
\end{align}
\end{subequations}
\end{widetext}
where
$b(\textrm{n}) \equiv \textrm{n} + \frac{1}{2}$.
The remaining
correction terms
\(\Lambda_4\), \(\Lambda_5\), and \(\Lambda_6\) can be found in the appendix of Ref.~\cite{Konoplya:2003ii}.

We now compute the \QNM{} frequencies using the
sixth-order
\WKB{} formula
in Eq.~\eqref{eq:wkb3}
for fixed sets of system parameters
that include
the \bh{} charge $Q$, fluid parameter $K$,
scalar field charge $q$, and
multipole number $l$.
First, we solve Eq.~\eqref{eq:WKB-extremum-real} numerically on a dense grid.
This procedure produces a discrete data set $(\omegaR,r_{0})$, which we interpolate to obtain a smooth function $r_{0}(\omegaR)$.
Then, we insert the interpolated function $r_{0}(\omegaR)$
into the sixth-order \WKB{} formula in Eq.~\eqref{eq:wkb3},
and analytically continue the resulting expression to complex frequencies.
Finally, we solve Eq.~\eqref{eq:wkb3} for the complex frequencies using the root finder
\texttt{FindRoot}
in
\textit{Mathematica}~\cite{mathematica14}.
For small fluid parameters $K$ and electromagnetic interaction $qQ$, we take the sixth-order \WKB{} spectrum of Schwarzschild spacetime, whose \QNM{} frequencies are well known~\cite{Konoplya:2003ii,Leaver:1985ax}, as the initial guess.
We then use the \QNM{} spectrum obtained for each pair of small $(K,qQ)$ values as the initial guess for the frequencies at the next, slightly larger pair of parameter values.
To validate our results, we employ the frequencies obtained with Leaver’s continued fraction method as initial guesses;
see Secs.~\ref{ssec:ContinuedFractionMethod} and~\ref{ssec:automaticdifferentiation}.
We estimate the error of the root finder with the \texttt{NMinimize} function, and typically find an error of $\lesssim10^{-9}$.

We benchmark our implementation against known \QNM{} results in Schwarzschild and \RN{} spacetimes~\cite{Konoplya:2002ky,Varghese:2008hyu}, and we find excellent agreement;
see e.g. Figs.~1–2 in Ref.~\cite{Varghese:2008hyu},
and Table~I and Figs.~1–2
in Ref.~\cite{Konoplya:2002ky}.

To understand the dependence on the parameters explored in the results section (Sec.\ref{sec:numerical_results}), we perform a sensitivity analysis.
Specifically, we examine how the expression used in the sixth-order \WKB{} calculation depends directly on the parameters $Q$, $K$, $q$, and $l$ around representative configurations; see App.~\ref{sec:append_sensitivity}.
The dependence is strongest for the fluid parameter $K$ and the scalar multipole number $l$, while the direct effects of the \bh{} charge $Q$ and the scalar field charge $q$ are smaller. 
In this comparison, the frequency $\omega$ and the extremum location $r_0$ are held fixed, and therefore the analysis does not describe the full dependence of the calculated \QNM{} frequencies on these parameters.

\section{Results: Quasinormal Modes}\label{sec:numerical_results}

In this section, we present the \QNM{} spectra of (charged) scalar perturbations propagating in the background of the two-component Kiselev spacetime.
We restrict our
analysis to the physical region in parameter space, corresponding to~\ref{case:two-roots-Qle}
discussed in Sec.~\ref{ssec:horizon_structure}.

We compute the \QNM{} spectra using the methods described in Sec.~\ref{sec:qnm_analysis}.
Focusing on the fundamental \QNM{s},
we analyze their
characteristic patterns
as we scan the parameter space
spanned by the \bh{} charge $Q$, the fluid parameter $K$, and the scalar field charge $q$.

To guide the selection of representative examples and parameter ranges for the \QNM{} analysis, we use a sensitivity analysis of the potential in Eq.~\eqref{eq:KG-massless-2CompKiselev}.
We group all terms that do not depend on the frequency in the second term of Eq.~\eqref{eq:KG-massless-2CompKiselev}, including the electrostatic potential, $qQ/r$,
and quantify their dependence on the parameters; see App.~\ref{sec:append_sensitivity} for a description of a general methodology.
We found that near the event horizon, the system is most sensitive to variations in the \bh{} charge $Q$ and the scalar multipole number $l$. Far from the horizon, the fluid parameter $K$ becomes dominant.

In the following, we first examine the \QNM{} spectra of an uncharged scalar field. Then, we turn to the spectra of a charged scalar field and analyze their dependence on the \bh{} charge $Q$ and the electromagnetic interaction, $qQ$. Next, we present representative examples of the \QNM{} spectrum beyond the fundamental mode. Finally, we discuss the influence of the fluid parameter $K$. The frequencies shown in the plots and tables are the coordinate frequencies $\omega$.

\subsection{Uncharged Scalar Field}

\begin{figure*}[t]
\centering
\includegraphics[width=0.98\textwidth]{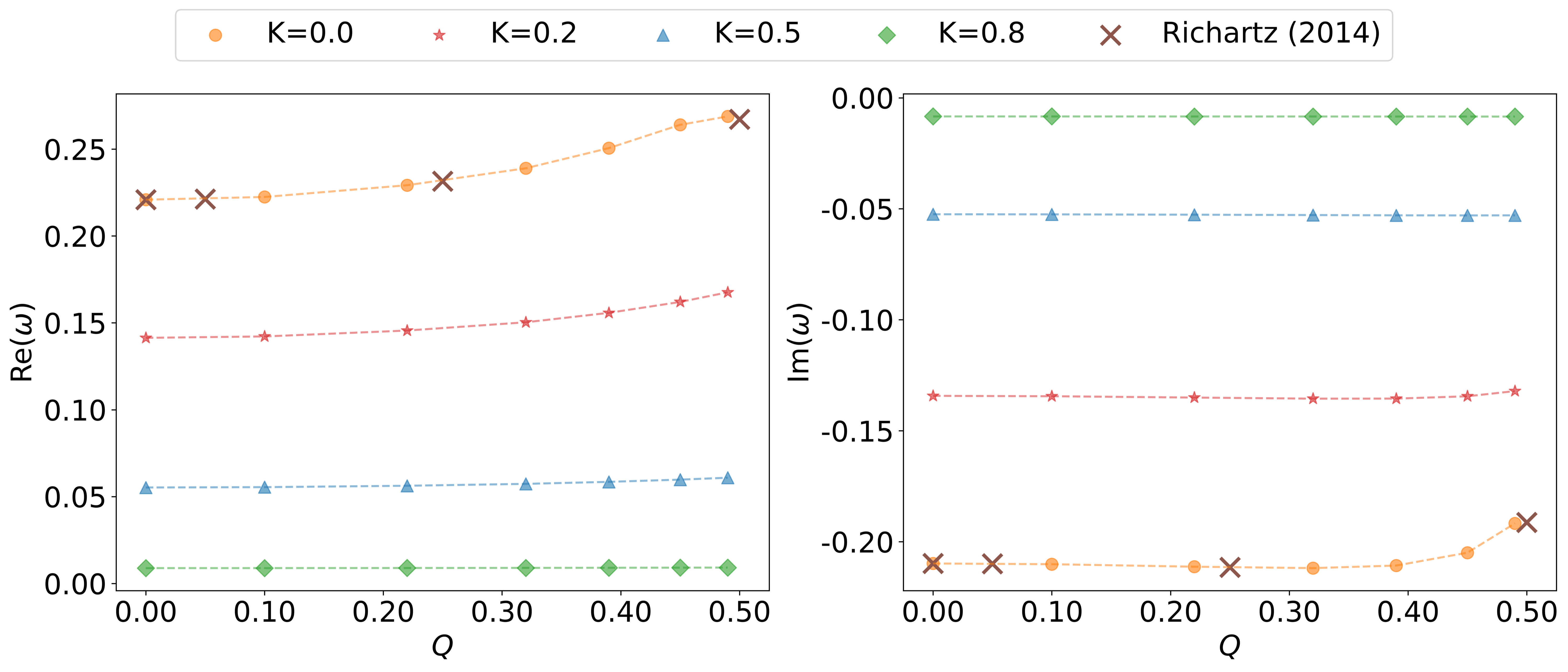}
\caption{Real (left panel) and imaginary (right panel) parts of the fundamental
\QNM{}
frequency $\omega$ of the $l = 0$ multipole of an uncharged scalar field ($q=0$)
as a function of the \bh{} charge $Q$.
The different lines correspond to
$K \in \{0.0, 0.2, 0.5, 0.8\}$.
For $K=0$, corresponding to the \RN{} spacetime, we benchmark our results against those of Ref.~\cite{Richartz:2014lpa}
(indicated as crosses).
}
\label{fig:non_interacting_combined}
\end{figure*}

We begin our analysis with the case of an uncharged scalar field ($q=0$)
propagating in the background of a charged \bh{} surrounded by an anisotropic fluid.
This setup includes the Schwarzschild limit $(Q=0,K=0)$.
Here, we focus on the fundamental
\QNM{} frequencies.

Fig.~\ref{fig:non_interacting_combined} displays the real (left panel) and imaginary (right panel) parts of the monopole ($l=0$)
\QNM{} frequency
as functions of the \bh{} charge $Q$
for fixed values of the fluid parameter $K\in\{0.0,0.2,0.5,0.8\}$.
We summarize the numerical values of the monopole \QNM{} frequencies
in Table~\ref{tab:qnm_values_l0_qQ0_corrected}.
We benchmark our results
in Schwarzschild and \RN{} spacetimes
($K=0$)
against the results of Richartz et al.~\cite{Richartz:2014lpa},
indicated as crosses in Fig.~\ref{fig:non_interacting_combined},
and we find excellent agreement.

In the left panel of Fig.~\ref{fig:non_interacting_combined},
we show the positive branch of the real part of the frequency,
that determines the oscillation frequency.
The negative branch is determined by the frequency's symmetry about zero.
For fixed  values of the fluid parameter
the oscillation frequency
slightly increases with increasing \bh{} charge, $Q$.
This trend is most pronounced when the fluid parameter vanishes.
For a large fluid parameter, $K=0.8$, the oscillation frequency becomes almost insensitive to the \bh{} charge
as can be seen in Fig.~\ref{fig:non_interacting_combined} and Table~\ref{tab:qnm_values_l0_qQ0_corrected}.
Fixing the \bh{} charge,
we find that the oscillation frequency decreases with increasing fluid parameter.

In the right panel of Fig.~\ref{fig:non_interacting_combined},
we show the imaginary part of the frequency
for positive \bh{} charges.
The values for negative charges are the same due to symmetry.
The imaginary part of the frequency
indicates the decay or growth rates of the scalar field.
For all parameters studied here, we find that the imaginary part of the \QNM{} frequency is negative, thus indicating
the decay of the scalar field.
That is, the two-component Kiselev metric is mode stable against uncharged scalar perturbations.

At fixed fluid parameter, the dependence of the decay rate on the
\bh{} charge is weak.
For small fluid parameters its magnitude, $|\omegaI|$, decreases for near extremal \bh{} charge.
This trend is most pronounced when the fluid parameter vanishes,
and is consistent with zero-damped modes of near-extremal \RN{} spacetimes~\cite{Zimmerman:2015trm,Saha:2025nsg}. 
As the fluid parameter increases,
the decay rate remains approximately constant with increasing \bh{} charge.

Fixing the \bh{} charge and varying the fluid parameter,
we find that the scalar's decay rate decreases with increasing fluid parameter.
This behavior indicates that the presence of the fluid yields long-lived scalar modes.
We refer to Sec.~\ref{sec:dependence_fluid} for a discussion of the impact of the fluid parameter on the \QNM{} spectra.

In summary, both the oscillation frequency and the decay rate of the monopole \QNM{} frequencies
decrease as the fluid parameter increases.

\begin{table*}[htbp]
\centering
\caption{
Real and imaginary parts of the fundamental, monopole ($l=0$) \QNM{} frequency, $\omega$, of an uncharged, massless scalar field ($q=0$),
computed with Leaver's method.
Recall that, following the rescaling introduced in Eqs.~\eqref{eq:units}, the symbols $\omega$ and $Q$ denote the dimensionless frequency, $\rg\omega$, and \bh{} charge, $Q/\rg$, and we employ units with $\rg=1$.
}
\label{tab:qnm_values_l0_qQ0_corrected}
\begin{tabular}{ccccccccc}
\toprule
\multirow{2}{*}{$Q$}
& \multicolumn{2}{c}{$K = 0.0$}
& \multicolumn{2}{c}{$K = 0.2$}
& \multicolumn{2}{c}{$K = 0.5$}
& \multicolumn{2}{c}{$K = 0.8$} \\
\cmidrule(lr){2-3} \cmidrule(lr){4-5} \cmidrule(lr){6-7} \cmidrule(lr){8-9}
& $\mathrm{Re}(\omega)$ & $\mathrm{Im}(\omega)$
& $\mathrm{Re}(\omega)$ & $\mathrm{Im}(\omega)$
& $\mathrm{Re}(\omega)$ & $\mathrm{Im}(\omega)$
& $\mathrm{Re}(\omega)$ & $\mathrm{Im}(\omega)$ \\
\midrule
0.000032 & 0.220910 & -0.209791 & 0.141382 & -0.134267 & 0.055227 & -0.052448 & 0.008836 & -0.008392 \\
0.100000 & 0.222479 & -0.210122 & 0.142183 & -0.134437 & 0.055422 & -0.052490 & 0.008849 & -0.008394 \\
0.223607 & 0.229238 & -0.211242 & 0.145581 & -0.135039 & 0.056230 & -0.052646 & 0.008899 & -0.008405 \\
0.316228 & 0.238995 & -0.211877 & 0.150333 & -0.135529 & 0.057310 & -0.052810 & 0.008964 & -0.008417 \\
0.387298 & 0.250597 & -0.210722 & 0.155770 & -0.135501 & 0.058478 & -0.052926 & 0.009030 & -0.008429 \\
0.447214 & 0.264003 & -0.204941 & 0.162027 & -0.134460 & 0.059749 & -0.052969 & 0.009099 & -0.008440 \\
0.489898 & 0.268820 & -0.191746 & 0.167569 & -0.132119 & 0.060848 & -0.052930 & 0.009155 & -0.008448 \\
\bottomrule
\end{tabular}
\end{table*}

In addition, we compute the \QNM{} spectra of the $l=1$ and $l=4$ multipoles.
The results are summarized in Table~\ref{tab:qnm_values_l1_qQ0_corrected} for $l=1$, and in Table~\ref{tab:qnm_values_l4_qQ0} and Fig.~\ref{fig:wkb_leaver_l=4qQ=0} for $l=4$ in App.~\ref{app:wkb_comparison}.
We observe that the oscillation frequency strongly depends on the multipole number and increases with increasing multipole.
In contrast, the decay rate decreases only slightly
with increasing multipole number.


Furthermore, we compare the results obtained with Leaver's continued fraction method with those obtained using the sixth-order \WKB{} approximation
for the $l=4$ multipole;
see
Table~\ref{tab:qnm_values_l4_qQ0} and Fig.~\ref{fig:wkb_leaver_l=4qQ=0}
in App.~\ref{app:wkb_comparison}.
The two methods show excellent agreement in the parameter range considered, with relative differences below $1.7\times10^{-6}$ in the real part and $8.1\times10^{-6}$ in the imaginary part of the frequency.

\subsection{Charged Scalar Field }

Next, we compute the \QNM{} frequencies of a charged scalar field propagating in the background of a charged \bh{} surrounded by an anisotropic fluid.
We first vary the \bh{} charge while keeping the electrostatic interaction, $qQ$, fixed.
In the second part of the analysis, we fix the \bh{} charge and vary the electrostatic interaction.

\subsubsection{Dependence on the black-hole charge at fixed electrostatic interaction}

\begin{figure*}[htbp]
    \centering
    \includegraphics[width=0.98\textwidth]{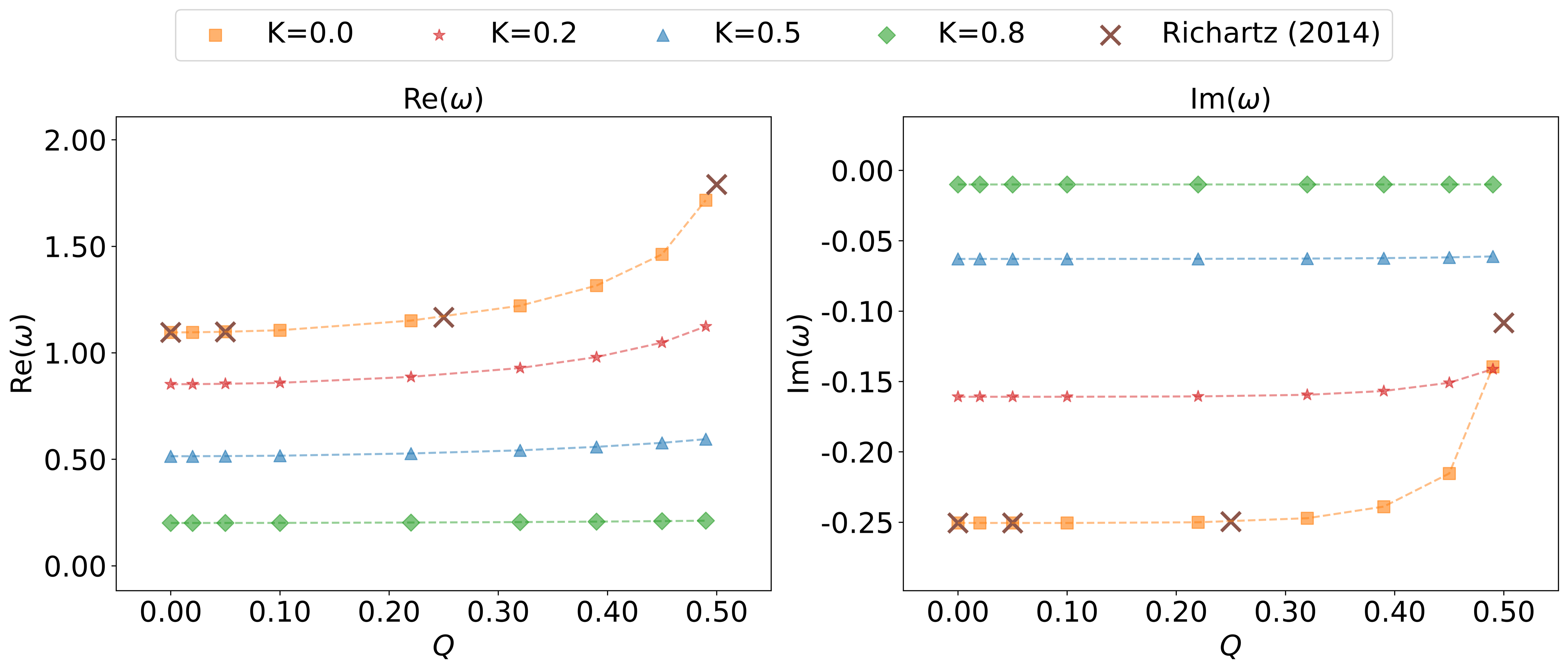}
\caption{
Real (left panel) and imaginary (right panel) parts of the fundamental,
monopole ($l=0$)
\QNM{} frequency $\omega$ of a charged scalar field
as functions of the \bh{} charge $Q$.
The electrostatic interaction is set to $qQ=1$.
The different lines correspond to
$K \in \{0.0, 0.2, 0.5, 0.8\}$.
We show benchmark values for $K=0$, taken from Ref.~\cite{Richartz:2014lpa}, as crosses. Note the rightmost benchmark point is evaluated  beyond
the largest charge \(Q=0.489898\) included in our numerical scan.}

    \label{fig:interaction_re_im_vs_q}
\end{figure*}

\begin{table*}[hbtp]
\centering
\caption{
Real and imaginary parts of the fundamental, monopole ($l=0$) \QNM{} frequency, $\omega$,
of a charged scalar field,
computed with Leaver's method.
We fix the electrostatic interaction to $qQ=1$,
yielding a scalar charge of $q=1/Q$.
Recall that, following the rescaling introduced in Eqs.~\eqref{eq:units}, the symbols $\omega$ and $Q$ denote the dimensionless frequency, $\rg\omega$, and \bh{} charge, $Q/\rg$, in units with $\rg=1$.
}
\label{tab:qnm_values_KQ_methods_qQ1}

\begin{tabular}{ccccccccc}
\toprule
\multirow{2}{*}{$Q$}
& \multicolumn{2}{c}{$K = 0.0$}
& \multicolumn{2}{c}{$K = 0.2$}
& \multicolumn{2}{c}{$K = 0.5$}
& \multicolumn{2}{c}{$K = 0.8$} \\
\cmidrule(lr){2-3} \cmidrule(lr){4-5} \cmidrule(lr){6-7} \cmidrule(lr){8-9}
& $\mathrm{Re}(\omega)$ & $\mathrm{Im}(\omega)$
& $\mathrm{Re}(\omega)$ & $\mathrm{Im}(\omega)$
& $\mathrm{Re}(\omega)$ & $\mathrm{Im}(\omega)$
& $\mathrm{Re}(\omega)$ & $\mathrm{Im}(\omega)$ \\
\midrule
0.000032 & 1.095607 & -0.250515 & 0.852449 & -0.160865 & 0.514145 & -0.062897 & 0.200980 & -0.010021 \\
0.022361 & 1.096107 & -0.250518 & 0.852768 & -0.160865 & 0.514269 & -0.062897 & 0.201000 & -0.010021 \\
0.054772 & 1.098621 & -0.250531 & 0.854366 & -0.160867 & 0.514892 & -0.062896 & 0.201100 & -0.010021 \\
0.100000 & 1.105789 & -0.250549 & 0.858909 & -0.160864 & 0.516652 & -0.062892 & 0.201380 & -0.010021 \\
0.223607 & 1.150947 & -0.250058 & 0.886967 & -0.160609 & 0.527208 & -0.062837 & 0.203016 & -0.010019 \\
0.316228 & 1.220813 & -0.247174 & 0.928326 & -0.159474 & 0.541763 & -0.062665 & 0.205139 & -0.010015 \\
0.387298 & 1.315278 & -0.238943 & 0.979761 & -0.156767 & 0.558173 & -0.062340 & 0.207355 & -0.010008 \\
0.447214 & 1.462140 & -0.215409 & 1.047636 & -0.150926 & 0.576953 & -0.061794 & 0.209671 & -0.009999 \\
0.489898 & 1.716349 & -0.139573 & 1.123658 & -0.141199 & 0.594188 & -0.061131 & 0.211604 & -0.009989 \\
\bottomrule
\end{tabular}
\end{table*}

Here, we analyze the dependence of the
\QNM{} frequencies of a charged scalar field
on the \bh{} charge
and fluid parameter,
while keeping the electrostatic interaction fixed at $qQ=1$.
In this section, we focus on the fundamental, monopole mode, and we present
the higher multipoles $l=1,4$ in App.~\ref{app:wkb_comparison}.
The results are shown in Table~\ref{tab:qnm_values_KQ_methods_qQ1}
where we list the numerical values of the frequency,
and in Fig.~\ref{fig:interaction_re_im_vs_q}
where we display its
real (left panel) and imaginary (right panel) parts.
We vary the \bh{} charge
from
$Q=0.000032$
to 
$Q=0.489898$
and we consider fluid parameters in the range
$K\in\{0.0,0.2,0.5,0.8\}$.
The scalar field charge is determined by the electrostatic interaction,
and here we fix it to $q = 1/Q$.
For vanishing fluid parameter ($K=0$),
which corresponds to \RN{} spacetimes, we benchmark our results against those by Richartz et al.~\cite{Richartz:2014lpa}
and find excellent agreement.

In the left panel of Fig.~\ref{fig:interaction_re_im_vs_q}, we display the oscillation frequency,
$\mathrm{Re}(\omega)$,
as a function of the \bh{} charge.
For fixed values of the fluid parameter,
the oscillation frequency increases with increasing \bh{} charge.
This trend is most pronounced when the fluid parameter vanishes, and the effect decreases as the fluid parameter increases.
When $K\gtrsim0.8$, the oscillation frequency is almost constant for all \bh{} charges.
Next, we consider the dependence on the fluid parameter while keeping the \bh{} charge fixed.
We find that the oscillation frequency decreases as the fluid parameter increases.

In the right panel of Fig.~\ref{fig:interaction_re_im_vs_q}, we present the imaginary part of the frequency, $\mathrm{Im}(\omega)$,
as a function of the \bh{} charge.
The imaginary part determines the decay rate of the \QNM{}.
We first fix the fluid parameter and analyze the dependence on the \bh{} charge.
The imaginary part of the frequency is almost constant for small and intermediate values of the \bh{} charge.
As the charge approaches its extremal value, $Q\to0.5$, the magnitude of the frequency's imaginary part decreases.
This trend is most pronounced for a vanishing fluid parameter.

For nonzero fluid parameter, $K\neq0$,
the dependence on the \bh{} charge washes out and 
the magnitude of the frequency's imaginary part remains almost constant with increasing \bh{} charge.
For a fixed \bh{} charge, the magnitude of ${\rm{Im}}(\omega)$ decreases with increasing fluid parameter.
In the limiting regime
$K\to1$, both the oscillation frequency and decay rate tend to zero.
The reduction in the decay rate gives rise to long-lived modes.

In addition, we list the \QNM{} frequencies of the $l=1$ and $l=4$ multipoles in Tables~\ref{tab:qnm_values_l1_qQ1_interaction} and \ref{tab:qnm_values_l4_qQ1},
respectively,
in App.~\ref{app:wkb_comparison}.
With increasing multipole number, the oscillation frequency increases significantly
while the decay rate slightly decreases.


In App.~\ref{app:wkb_comparison}, we compare the results obtained with Leaver's continued fraction method to those that we obtained with the sixth-order \WKB{} approximation
for the $l=4$ multipole;
see
Table~\ref{tab:qnm_values_l4_qQ1} and Fig.~\ref{fig:wkb_leaver_l=4qQ=1}.
For the parameter sets where the \WKB{} calculation succeeds, the relative differences range from approximately $1.2\times 10^{-4}$ to $6.5 \times 10^{-4}$ in the real part and from $3\times 10^{-3}$ to $4.5\times 10^{-3}$ in the imaginary part.
The sixth-order \WKB{} approximation fails only for $K=0$ and $Q\approx0.49$, where the \bh{} charge approaches its extremal value.

\subsubsection{Dependence on the
electrostatic interaction at fixed black-hole charge}

\begin{figure*}[htbp]
    \centering
\includegraphics[width=0.9\textwidth]{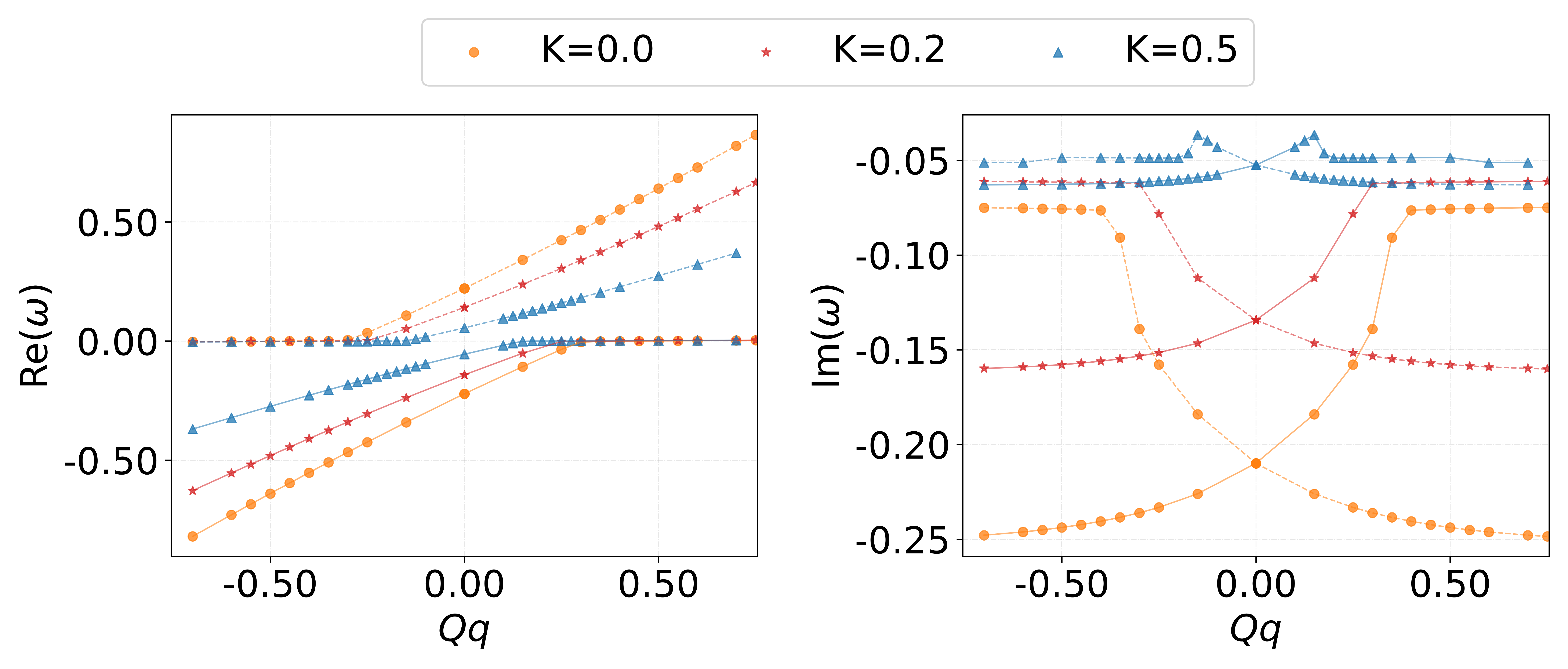}
    \caption{
Real (left panel) and imaginary (right panel) parts of the fundamental,
monopole ($l=0$)
\QNM{} frequency $\omega$
of the charged, massless scalar field
as functions of the electrostatic interaction, $qQ$.
We fix the \bh{} charge to $Q=0.05$.
The different lines correspond
to
$K \in \{0.0, 0.2, 0.5\}$.
Solid lines indicate the negative branch, while dashed lines indicate the positive branch.
Here, ``positive'' and ``negative''
refer to those branches that have
${\rm Re}(\omega)>0$ and ${\rm Re}(\omega)<0$,
respectively,
in the limit $K\to0$.
}
\label{fig:qq_neg_combined}
\end{figure*}

Next, we investigate the dependence of the \QNM{} frequencies on the electrostatic interaction, $qQ$,
while keeping the \bh{} charge fixed at $Q=0.05$.
We focus on the fundamental, monopole mode
throughout this section.

The results are shown in Fig.~\ref{fig:qq_neg_combined}, where we display the real (left panel) and imaginary (right panel) parts of the frequency
as a function of the electrostatic interaction for fluid parameters in the range $K\in\{0.0,0.2,0.5\}$.

We first analyze the real part of the
frequency shown in the left panel of Fig.~\ref{fig:qq_neg_combined}.
As we scan the parameter space in the range $qQ\in[-0.7,0.7]$,
we find two branches of frequencies
that are antisymmetric,
$\omega\to -\omega^{\ast}$,
under the transformation $qQ\to -qQ$.
We find that one of the branches admits only positive oscillation frequencies, ${\rm{Re}}(\omega)\geq0$,
while the other branch admits only negative frequencies, ${\rm{Re}}(\omega)\leq0$.
We, therefore, refer to them as ``positive'' and ``negative'' branches. We indicate them by dashed and solid lines, respectively, in Fig.~\ref{fig:qq_neg_combined}.

The frequencies' magnitudes increase approximately linearly with that 
of the electrostatic interaction.
The curves' slope is largest for electro-vacuum, $K=0$, and decreases as the fluid parameter increases, $K>0$.
We find that the oscillation frequency vanishes when the electrostatic potential is below (positive branch) or above (negative branch) a critical value.
That is, for large electrostatic interactions, the scalar perturbation ceases to oscillate and decays monotonically in time.
This behavior is analogous to that of a damped harmonic oscillator
that transitions from the underdamped region in parameter space
to an overdamped region.
In our problem, the electrostatic potential appears to play the role of the damping parameter,
yielding modes with near-zero oscillation frequency above a critical value.
We verify these ``zero-frequency'' modes numerically
using $N=5,000$ terms in the continued fraction method~\footnote{Previous studies~\cite{Richartz:2014lpa} describe the mode as disappearing above a critical
electrostatic potential.
Here, we find that the branch persists with decaying modes as shown in Fig.~\ref{fig:qq_neg_combined}. Given the numerical challenges associated with this regime, additional computations would be valuable for further verification.
}.

We see that the two branches approach each other as the interaction term goes to zero,
$qQ\to 0$,
but they never intersect.
Similar ``avoided regions'' have been observed in \bh{} \QNM{} spectra in different spacetimes, where they are referred to as ``repulsion of modes''~\cite{Yang:2025dbn, Dias:2022oqm, Motohashi:2024fwt,Lo:2025njp}.
This spectral behavior is not unique to \bh{} \QNM{};
it typically arises when two modes are coupled, and it has also been observed in the context of condensed matter or quantum mechanical systems
where it is called ``avoided-crossing'' regions~\cite{Yarkony:1996zz, Zener:1932,Landau:1932}. Establishing a nonzero minimum spectral separation and a corresponding exchange of mode character,  characteristic signatures of an avoided-crossing region, requires further investigation in future work.

Next, we analyze the imaginary part of the \QNM{} frequency,
as a function of the electrostatic interaction, as shown in the right panel of Fig.~\ref{fig:qq_neg_combined}.
For all parameters that we consider, the imaginary part 
remains negative and, thus, indicates the mode's decay rate.
For the positive and negative branches, the dependence of the rate on the electrostatic interaction is symmetric about $qQ=0$.



Therefore, we focus only on the positive branch in the following discussion and recover the negative branch by symmetry. When the fluid parameter vanishes, the decay rate is largest for an electrostatic interaction of $qQ\sim 0.7$, and it decreases as the interaction decreases. The decay rate is smallest near $qQ\sim -0.7$.

As the fluid parameter $K$ increases the decay rate decreases over the sampled $qQ$ range. For small, non zero fluid parameters, $K\simeq0.2$, the overall trend is similar to that of the $K=0$ curve, although with smaller decay rate. For intermediate values, $K\simeq0.5$, the decay rate depends non monotonically on the interaction.

\subsection{Frequency Spectrum} \label{subsubsec:freq_spectrum}

\begin{figure*}[!t]
  \centering
  \includegraphics[
    width=0.975\textwidth,
    trim=0pt 0pt 0pt 30pt,clip
  ]{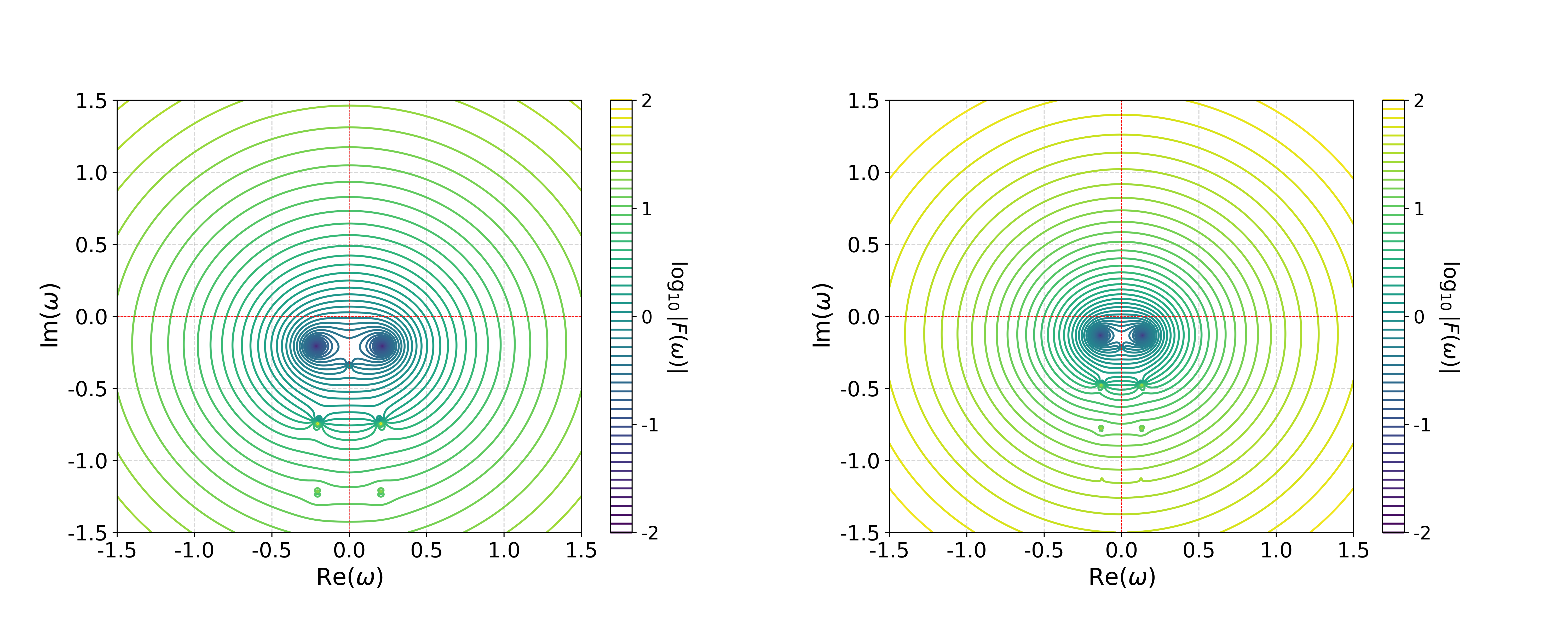}
  \centering
  \includegraphics[
    width=0.975\textwidth,
    trim=0pt 0pt 0pt 30pt,clip
  ]{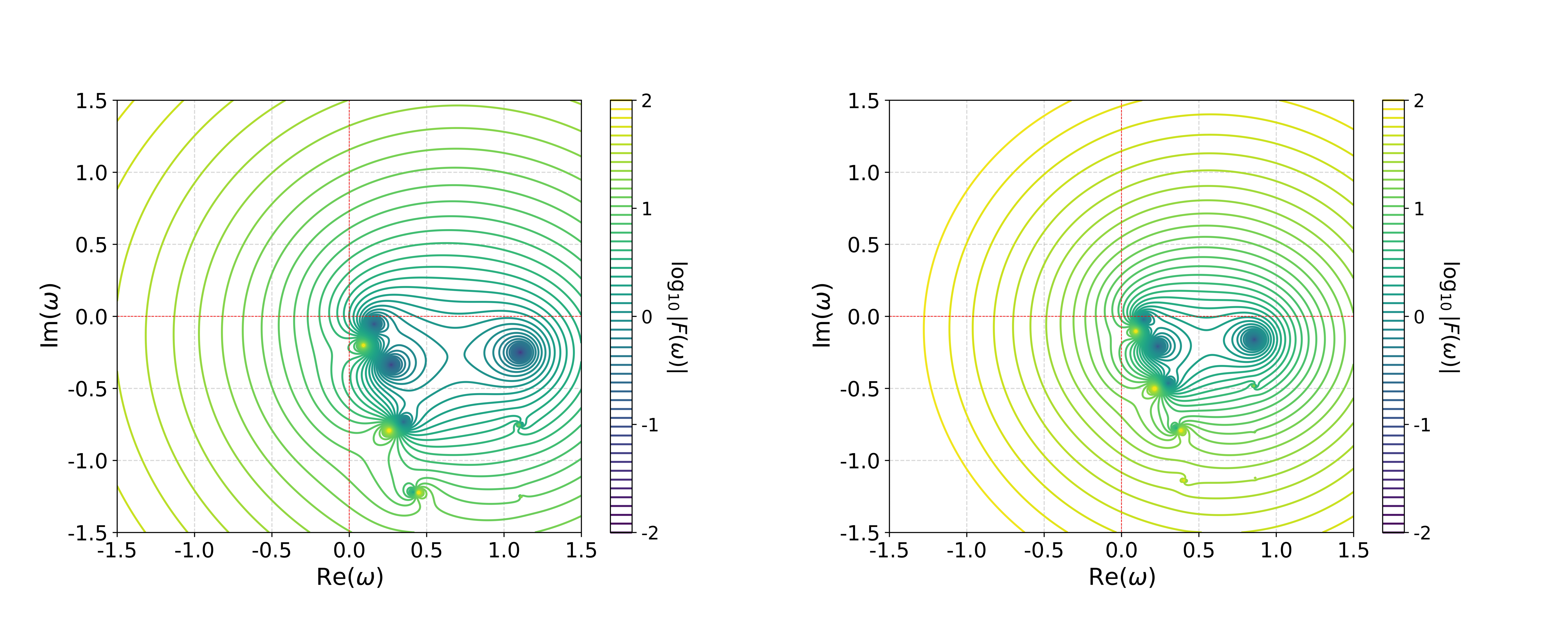}
  \caption{
  Contour plots of the magnitude of the characteristic equation, $|F(\omega)|$, in Eq.~\eqref{eq:characteristic_equation}
  in the complex frequency plane.
  We evaluate the characteristic equation for the scalar's monopole using Leaver's method.
  The color bar shows $\log_{10}\lvert F(\omega) \rvert$.
  The dark regions indicate the minima that correspond to the candidate \QNM{} frequencies.
  We label the root(s) with the smallest magnitude of the imaginary part of the frequency as the fundamental mode, while other roots indicate overtones.
  For all panels, we consider a \bh{} with charge $Q=0.1$,
  and we vary the electrostatic interaction, $qQ$, and the fluid parameter $K$.
    Top left:       $qQ=0$, $K=0.0$;
    Top right:      $qQ=0$, $K=0.2$;
    Bottom left:    $qQ=1$, $K=0.0$;
    Bottom right:   $qQ=1$, $K=0.2$.
  }
  \label{fig:collage_CE_QNM}
\end{figure*}

Here, we examine representative examples of the
\QNM{} frequency spectrum of the monopole,
including the frequencies of the fundamental mode and overtones.
Therefore, we evaluate the magnitude $|F(\omega)|$ of the characteristic equation~\eqref{eq:characteristic_equation}
in the complex frequency plane.

Representative results are shown in the contour plots in Fig.~\ref{fig:collage_CE_QNM}
for a \bh{} with charge $Q=0.1$. 
We present results for uncharged scalars, $qQ=0$, in the top panels
and for charged scalars with interaction strength $qQ=1$ in the bottom panels.
We consider fluid parameters $K=0$
(left panels) and $K=0.2$
(right panels).
The resulting contour maps in Fig.~\ref{fig:collage_CE_QNM}
display $\log_{10}|F(\omega)|$.
The dark purple regions indicate the smallest residuals,
where the characteristic equation is close to zero
and, thus, indicate the candidate
\QNM{} frequencies.
Note the color scale is clipped at \(\log_{10}|F|=-2\), so the contour plot does not by itself resolve the exact roots.

In the top panels of Fig.~\ref{fig:collage_CE_QNM},
in which we set $qQ=0$,
the spectra are symmetric with respect to the imaginary axis (i.e., about $\mathrm{Re}(\omega) = 0$).
This
behavior is expected when the interaction between \bh{} and scalar field charges is zero.
Next, we compare the spectrum obtained around a \bh{} surrounded by an anisotropic fluid with $K=0.2$ (top right panel) with that of a
\RN{} black hole ($K=0$, top left panel).
While the spectrum remains symmetric about the imaginary axis,
we see that the fluid reduces
both the oscillation frequency and the decay rate of the displayed overtones.
This is a consequence of the decrease of the potential barrier in the presence of the fluid;
see Fig.~\ref{fig:v_combined}
and discussion in the following section~\ref{sec:dependence_fluid}.

In the bottom panels of Fig.~\ref{fig:collage_CE_QNM}, we show the \QNM{} spectrum for nonzero
interaction between the \bh{} and scalar field charges, which we set to $qQ=1$.
Due to this nonzero interaction, the spectrum becomes asymmetric,
i.e., the oscillation frequencies are no longer symmetric about the imaginary axis and are shifted towards the positive real part of the frequency.
We also see that the decay rates, indicated by the imaginary part of the frequency, are spaced differently for the two branches.


Comparing the bottom left and bottom right panels of Fig.~\ref{fig:collage_CE_QNM} reveals that increasing the fluid parameter $K$ reduces
both the oscillation frequency and the decay rate of the displayed overtones. We see a similar pattern in the branches of the $l=1$ and $l=4$ modes for all values of the fluid parameter $K$ considered here.

Finally, we find no \QNM{s} with a positive imaginary part in the parameter range explored here.
This provides evidence that nonextremal charged \bh{s} surrounded by the anisotropic fluid remain (mode) stable under massless, charged scalar field perturbations.

\subsection{Dependence on the Fluid Parameter} \label{sec:dependence_fluid}
We now discuss how the fluid parameter affects the \QNM{} spectrum for both uncharged and charged scalar fields. Figs.~\ref{fig:non_interacting_combined} and~\ref{fig:interaction_re_im_vs_q} show the real and imaginary parts of the fundamental frequency as functions of the \bh{} charge $Q$ for fluid parameter values $K\in\{0,0.2,0.5,0.8\}$. Increasing the fluid parameter reduces both the oscillation frequency and the decay rate. This behavior is related to the way the fluid parameter reshapes the metric function and affects the scattering region in three different ways.

First, from Eq.~\eqref{eq:2CompKiselev-MetricFunction}, the asymptotic value of the metric function $f(r)$ is given by Eq.~\eqref{eq:f-asymptotic}, which we restate here for convenience,
\begin{equation}
    \lim_{r\to\infty}f(r)=1-K.
\end{equation}

Second, the event horizon in Eq.~\eqref{eq:2CompKiselev-HorizonsSubExtremal} moves outward as the fluid parameter increases because it scales as $r_{+}\sim(1-K)^{-1}$.
Third, the maximum of the geometric potential $V(r)$ in Eq.~\eqref{eq:Veff} also moves outward as the fluid parameter increases. The potential barrier becomes lower and broader while remaining outside the horizon, as visible in Fig.~\ref{fig:v_combined}. Consequently, as the fluid parameter increases, the scattering region moves farther away from the \bh{}, and the potential barrier seen by the field becomes lower and flatter.

We now discuss how the features listed above lead to the observed decrease in the oscillation frequency and the decay rate as the fluid parameter increases. We use the \WKB{} potential $U(r,\omega)$ in Eq.~\eqref{eq:WKBEffPotential}, together with the leading-order \WKB{} condition from Eq.~\eqref{eq:wkb3}, where all correction terms are neglected.
We also use the location of the real extremum $r_{0}$
defined in Eq.~\eqref{eq:WKB-extremum-real}.
In the following, we
denote quantities evaluated at the extremum by a subscript $0$.
Then, the electrostatic potential, the geometric potential in Eq.~\eqref{eq:2CompKiselev-Veff-polynomial}, and the \WKB{} potential evaluated at $r_{0}$ are given by
\begin{equation}
    \Phi_{0}=\frac{qQ}{r_{0}},\qquad V_{0}=V(r_{0}),\qquad U_{0}=U(r_{0},\omega),
\end{equation}
and the metric function is $f_{0}=f(r_{0})$.

At the real extremum, Eq.~\eqref{eq:WKB-extremum-real} gives
\begin{equation}
\label{eq:U-extremum-discussion}
    2\left(\omegaR-\Phi_{0}\right)\Phi'(r_{0})=-V'(r_{0}).
\end{equation}
To leading order, the \WKB{} condition in Eq.~\eqref{eq:wkb3} gives two coupled relations
\begin{subequations}
\label{eq:Eqs-balance-K}
\begin{align}
\label{eq:Re-balance-K}
    \left(\omegaR - \Phi_0\right)^{2}
    &\simeq
    V_0+\omegaI^{2}
\,, \\
\label{eq:Im-balance-K}
    2\left(\omegaR - \Phi_0\right)\omegaI
    &\simeq
    -\left(n+\tfrac{1}{2}\right)\sqrt{2\,U_{0}^{(2)}}
\,.
\end{align}
\end{subequations}
From Eqs.~\eqref{eq:Eqs-balance-K}, the leading-order estimates for
the frequencies,
derived in more detail in App.~\ref{app:leading_order_estimates}, read
\begin{equation}
\label{eq:Im-omega-order-K}
    |\omegaI|
    =
    \mathcal{O}\!\left[
        \sqrt{f_0}
        \left(
        \frac{|qQ|}{r_0}
        +
        \frac{\sqrt{1-K}}{r_0}
        +
        \frac{1}{r_0^{3/2}}
        \right)
    \right],
\end{equation}
and
\begin{equation}
\label{eq:Re-omega-order-K}
    \omegaR
    =
    \frac{qQ}{r_0}
    +
    \mathcal{O}\!\left[
        \sqrt{f_0}
        \left(
        \frac{\sqrt{\lambda}}{r_0}
        +
        \frac{|qQ|}{r_0}
        +
        \frac{\sqrt{1-K}}{r_0}
        +
        \frac{1}{r_0^{3/2}}
        \right)
    \right].
\end{equation}

Both estimates are suppressed as the scattering region moves outward and the value of the metric function $f_0$ decreases with increasing fluid parameter.
The electrostatic contribution is also suppressed as the extremum $r_{0}$ moves outward.
This behavior leads to the observed decrease in both the oscillation frequency and the decay rate as the fluid parameter increases.

For an uncharged scalar field, the scaling with the fluid parameter can be made explicit. In this case, Eq.~\eqref{eq:U-extremum-discussion} reduces to $V'(r_{0})=0$, and the leading-order \WKB{} real extremum scales as; see App.~\ref{app:leading_order_estimates},
\begin{equation}
    r_0=\mathcal{O}\!\left((1-K)^{-1}\right).
\end{equation}
Thus, using $f_0=\mathcal{O}(1-K)$, we obtain
\begin{equation}
    |\omega_I|
    =
    \mathcal{O}\!\left((1-K)^2\right)
\end{equation}
and
\begin{equation}
    \omega_R
    =
    \begin{cases}
        \mathcal{O}\!\left((1-K)^2\right),
        & \lambda=0,
        \\[4pt]
        \mathcal{O}\!\left((1-K)^{3/2}\right),
        & \lambda\neq0,
    \end{cases}
\end{equation}
where we recall $\lambda=l(l+1)$.

For nonzero electromagnetic coupling, Fig.~\ref{fig:interaction_re_im_vs_q} shows that,
at fixed \(Q\), $|$Im$(\omega)|$ decreases monotonically
with increasing fluid parameter \(K\).

Related studies of \bh{s} embedded in matter environments confirm that environmental parameters can significantly reshape the \QNM{} spectrum, although the detailed behavior remains model-dependent~\cite{Pezzella:2024tkf,Tan:2025usr,Das:2023ess,Jusufi:2019ltj}.

Note also that, in our model, the physical surface gravity $\kappa$ vanishes in the limiting regime, $K \to 1$, as shown in App.~\ref{sec:append_surface_grav}.
Since the Hawking temperature is proportional to the surface gravity,
$T_H = \frac{\kappa}{2\pi}$,
the horizon becomes increasingly cold in this limit.
In different models, such as near-extremal \bh{s}, the vanishing surface gravity is associated with the emergence of zero-damped
\QNM{s}~\cite{Joykutty:2021fgj, Zimmerman:2015trm, Saha:2025nsg,Richartz:2015saa, Yang:2013uba}.

\section{Conclusion}\label{sec:conclusion}

Astrophysical \bh{s} are not isolated objects.
They are surrounded by complex environments that can leave observable imprints on their dynamics.
In this work, we have modeled an environment through an anisotropic fluid described within Kiselev's framework,
and we have analyzed the \bh{'s} response to perturbations by computing the \QNM{} spectra of a charged scalar field.

In particular, we have considered a two-component Kiselev model
(see~Eq.~\eqref{eq:2CompKiselev-MetricFunction})
comprising a charged \bh{} surrounded by an anisotropic fluid that may represent dark matter or a cosmic string network.
We have analyzed the metric's horizon structure 
and we have shown that 
in the charged subextremal parameter range it contains two distinct horizons: an inner Cauchy
horizon and an outer event horizon.


To investigate the mode stability of the spacetime and its response to perturbations,
we have computed the \QNM{} spectra of a massless, charged scalar field propagating on the two-component Kiselev metric.

Therefore, we have developed a nontrivial extension of Leaver's continued fraction method that accounts for the anisotropic fluid around the \bh{}.
This method results in a characteristic equation whose roots are the complex \QNM{} frequencies that we sought.
In addition to standard root finders, we have implemented a new algorithm using automatic differentiation in the \textsc{PyTorch} framework.
This new algorithm has proven crucial in challenging regions of parameter space, such as near-extremal spacetimes, where conventional root finders failed.
To our knowledge, this is the first application of automatic differentiation techniques to \bh{} perturbation theory.

We have corroborated our results using a sixth-order \WKB{} approximation extended to consistently account for a frequency-dependent \WKB{} effective potential.
We found that the \WKB{} approximation reproduces the results obtained using Leaver's continued-fraction method with good accuracy for the cases explored in this work.

We have analyzed how the scalar-field \QNM{} spectra depend on the \bh{} charge, the fluid parameter, and the interaction between the \bh{} and scalar-field charges. 
In summary, we find the following results:
\begin{itemize}[label={-},leftmargin=7pt,itemsep=0.0em]
\item For all parameters in the physically allowed range that we have considered,
we found no evidence of \QNM{} frequencies with a positive imaginary part.
That is, we only found decaying \QNM{s},
which strongly indicates mode stability of the two-component Kiselev metric under massless, charged scalar perturbations.

\item For an uncharged scalar field, increasing the \bh{} charge
increases the oscillation frequency of the fundamental mode. The
decay rate has a weak, nonmonotonic, and fluid parameter-dependent behavior. 
Increasing the fluid parameter at a fixed \bh{} charge decreases both the oscillation frequency and the decay rate. The surrounding fluid therefore yields longer-lived scalar modes.

\item For a charged scalar field at fixed electrostatic interaction strength, increasing the \bh{} charge increases the oscillation frequency.
This effect becomes weaker as the fluid parameter increases.
The surrounding fluid therefore suppresses the influence of the electromagnetic interaction on the \QNM{} spectrum.

\item For a charged scalar field at fixed \bh{} charge, the electrostatic
interaction changes both the oscillation frequency and the decay rate, and we find two branches of \QNM frequencies.
For the positive branch, the magnitude of the oscillation frequency grows
as the interaction increases.
The slope of the curve weakens for larger values of the fluid parameter. 
The decay rate is weakest near the most negative interaction in the sampled interval.
The negative branch follows by symmetry. 
For values close to vanishing interaction, we also
observe avoided-crossing like behavior, where the two branches of \QNM frequencies approach each
other but do not intersect.

\item For both charged and uncharged scalar fields, increasing the angular multipole number increases the oscillation frequency and slightly decreases the decay rate.
These trends persist as the \bh{} charge and the fluid parameter are varied.

\item We have also analyzed representative portions of the \QNM{} spectrum,
including  overtones. For uncharged
scalar fields, the spectrum is symmetric about the imaginary axis. This symmetry is broken when the electromagnetic interaction is nonzero. For both uncharged and charged fields, increasing the fluid parameter reduces both the oscillation and the decay rate for the  overtones.

\end{itemize}
Taken together, our results support mode stability in the parameter range studied and show that the surrounding fluid tends
to suppress both the oscillation and the decay rate of the scalar modes and to reduce the electromagnetic imprint on the frequencies.

As part of future work, it would be valuable to investigate massive scalar perturbations and gravitational perturbations, and to analyze the limit of the fluid parameter, $K \to 1$, in greater detail.
A geometric analysis of the near-horizon region in this limit may clarify the origin of the long-lived modes and their connection to the near-horizon geometry.

%

The good agreement between the sixth-order \WKB{} approximation and Leaver's method supports the applicability and numerical reliability of the \WKB{} approximation for the cases considered. In future work, higher-order formulations, such as the thirteenth-order \WKB{} approximation supplemented by Pad\'e resummation, could provide even greater numerical accuracy. If appropriately extended to the frequency-dependent effective potential considered here, this approach could serve as an independent and highly accurate method for computing the \QNM{} spectrum.

\section{Acknowledgments}
 
We thank
H.~O.~Silva,
M.~Richartz,
E.~Berti, A.~Zhidenko
and E.~Han
for insightful discussions and comments.
We thank O.~Stashko
for his valuable comments on the manuscript.
We thank M.~Richartz for sharing the calculations from Ref.~\cite{Richartz:2014lpa}.
We thank R.~A.~Konoplya for sharing sixth-order \WKB{} \textit{Mathematica} code.
The authors acknowledge support provided by the National Science Foundation under NSF Award No.~PHY-2409726 and
No.~OAC-2411068.
D.~N.~Garzon acknowledges support from the AAUW Doctoral Fellowship,  the Illinois Center for Advanced Studies of the Universe (ICASU) Graduate Fellowship at the University of Illinois Urbana-Champaign and the Forum on Graduate Student Affairs of the American Physical Society.
J.Z. acknowledges financial support from the American Physical Society Division of Gravitational Physics (APS DGRAV) and the Department of Physics at the University of Illinois Urbana-Champaign.
We acknowledge the Texas Advanced Computing Center (TACC) at the University of Texas at Austin for providing HPC resources on Frontera via allocation PHY22041.
This work used
Expanse at the San Diego Supercomputer Center
through allocation
PHY240323
from the Advanced Cyberinfrastructure Coordination Ecosystem: Services and Support (ACCESS) program, which is supported by U.S. National Science Foundation grants Nos.~2138259, 2138286, 2138307, 2137603, and 2138296.
We acknowledge the Illinois Computes project which is supported by the University of Illinois Urbana-Champaign and the University of Illinois System.
Calculations were performed using Mathematica~\cite{mathematica14},
together with the open-source libraries PyTorch~\cite{paszke2019pytorch},
NumPy~\cite{harris2020array}, and SciPy~\cite{2020SciPy-NMeth}.

\appendix
\section{Notation}\label{app:notation}

For convenience, we summarize the notation used throughout the manuscript in Table~\ref{tab:notation}.

\begin{table*}[ht]
\centering
\footnotesize
\setlength{\tabcolsep}{3pt}
\begin{tabular}{|c|c|c|}
\hline
\textbf{Symbol} & \textbf{Definition} & \textbf{Note} \\ \hline
$ds^2$ & \parbox[t]{150pt}{Line element of the Kiselev spacetime: $\dif s^2 = -f(r)\,\dif t^2 + f(r)^{-1} \dif r^2 + r^2 \left( \dif\theta^2 + \sin^2\theta\,\dif\phi^2\right)$} & \parbox[t]{150pt}{Static, spherically symmetric metric in (dimensionless) Schwarzschild coordinates $(t,r,\theta,\phi)$, rescaled by $\rg=2M$.} \\ \hline
$M$ & \parbox[t]{150pt}{Mass parameter of the \bh{}} & \parbox[t]{150pt}{Sets the gravitational radius $\rg = 2M$.} \\ \hline
$Q$ & \parbox[t]{150pt}{Electric charge parameter of the \bh{}} & \parbox[t]{150pt}{--} \\ \hline
$K$ & \parbox[t]{150pt}{Fluid parameter characterizing the anisotropic fluid} & \parbox[t]{150pt}{Considered in the range $0 \leq K < 1$ (\ref{case:two-roots-Qle}).} \\ \hline
$\rmm$ & \parbox[t]{150pt}{Cauchy horizon} & \parbox[t]{150pt}{Inner horizon in the charged subextremal regime (\ref{case:two-roots-Qle}).} \\ \hline
$\rp$ & \parbox[t]{150pt}{Event horizon} & \parbox[t]{150pt}{Outer horizon in the subextremal regime (\ref{case:two-roots-Qle}); $\rmm < \rp$.} \\ \hline
$r_{*}$ & \parbox[t]{150pt}{Tortoise coordinate} & \parbox[t]{150pt}{Defined by $\dif r_{*}/\dif r = 1/f(r)$.} \\ \hline
$T^{\mu}{}_{\nu}$ & \parbox[t]{150pt}{Effective energy-momentum tensor of the anisotropic fluid: $T^{\mu}{}_{\nu} = \mathrm{diag}(-\rho,\,p_r,\,p_{\perp},\,p_{\perp})$} & \parbox[t]{150pt}{--} \\ \hline
$\rho$ & \parbox[t]{150pt}{Energy density of the anisotropic fluid} & \parbox[t]{150pt}{Dominant energy condition applies.} \\ \hline
$p_r,\,p_{\perp}$ & \parbox[t]{150pt}{Radial and tangential pressures} & \parbox[t]{150pt}{In general, for an anisotropic fluid, $p_r \neq p_{\perp}$.} \\ \hline
$\psi$ & \parbox[t]{150pt}{Scalar field with ansatz $\psi(t, r, \theta, \phi) = \sum_{l,m} \frac{\Psi_{l m}(r)}{r}\, Y_{l m}(\theta, \phi)\, e^{-i \omega_{l m} t}$} & \parbox[t]{150pt}{Mode decomposition in spherical harmonics $Y_{lm}$ with mode numbers $l \geq 0$ and $-l \leq m \leq l$.} \\ \hline
$\lambda$ & \parbox[t]{150pt}{Angular multipole factor $\lambda = l(l+1)$, with $l=0,1,2,\ldots$} & \parbox[t]{150pt}{$\lambda=0$ for the monopole mode $l=0$ and $\lambda>0$ for $l\ge 1$.} \\ \hline
$\omega$ & \parbox[t]{150pt}{Complex \QNM{} frequency $\omega = \mathrm{Re}(\omega) + i\,\mathrm{Im}(\omega)$} & \parbox[t]{150pt}{--} \\ \hline
$q$ & \parbox[t]{150pt}{Charge parameter of the scalar field} & \parbox[t]{150pt}{--} \\ \hline
$\Phi(r)$ & \parbox[t]{150pt}{Electrostatic potential $\Phi(r) = qQ/r$} & \parbox[t]{150pt}{Determines the electrostatic interaction.} \\ \hline
$V(r)$ & \parbox[t]{150pt}{Geometric effective potential: $V(r)=f(r)\left(\dfrac{l(l+1)}{r^2}+\dfrac{f'(r)}{r}\right)$} & \parbox[t]{150pt}{--} \\ \hline
$U(r,\omega)$ & \parbox[t]{150pt}{\WKB{} effective potential: $U(r,\omega)=(\omega-\Phi(r))^2-V(r)$} & \parbox[t]{150pt}{--} \\ \hline
$\Lambda_i$ & \parbox[t]{150pt}{\WKB{} correction terms ($i=2,3,4,5,6$)} & \parbox[t]{150pt}{Higher-order correction terms in the \WKB{} expansion.} \\ \hline
\multicolumn{3}{|l|}{\textbf{Units:} Geometric units $c=1$, $8\pi G=1$;}\\
\multicolumn{3}{|l|}{Coordinates and parameters are made dimensionless by appropriate rescaling with the gravitational radius $\rg=2M=1$.}\\ \hline
\end{tabular}
\caption{Notation and definitions adopted throughout the manuscript.}
\label{tab:notation}
\end{table*}

\section{Surface gravity}\label{sec:append_surface_grav}

In this section, we derive the surface gravity $\kappa$ of the black hole in the Kiselev spacetime.
This quantity appears in the discussion of the horizon structure in Sec.~\ref{ssec:horizon_structure} and in the effect of the fluid parameter on the \QNM{} spectrum in Sec.~\ref{sec:dependence_fluid}.

In a static, spherically symmetric spacetime, the surface gravity $\kappa$ on the \bh{} horizon is defined as
\begin{align}
\kappa & = \left. \frac{1}{2} \frac{\dif f}{\dif r} \right|_{r=\rp}
\,,
\end{align}
where $\rp$ is the location of the event horizon.
After inserting the metric function in Eq.~\eqref{eq:2CompKiselev-MetricFunction},
we find
\begin{align}
\label{eq:kappa_K}
\kappa & = \frac{\rp - 2Q^2}{2 \rp^3}
\\ & =
    \frac{2(1-K)^2 \left[1-4(1-K)Q^2 + \sqrt{1-4(1-K)Q^2} \right]}{\left( 1+\sqrt{1-4(1-K)Q^2}  \right)^{3} }
\,,\nonumber
\end{align}
where we have used Eq.~\eqref{eq:2CompKiselev-HorizonsSubExtremal} in the last line.

Finally, we reconsider the physical surface gravity using the normalized asymptotic Killing field introduced in Eq.~\eqref{eq:AsymptoticKV}.
Then, the physical surface gravity is
\begin{align}
\label{eq:physical_kappa}
\kappa_{\rm phys}=\frac{\kappa}{\sqrt{1-K}}.
\end{align}

Next, we consider the surface gravity in the limiting regime, $K\to1$.
Using Eqs.~\eqref{eq:kappa_K} and~\eqref{eq:physical_kappa},
we obtain
\begin{align}
    \lim_{K \to 1} \kappa_{\rm phys} &= 0.
\end{align}
These results confirm that the physical surface gravity
vanishes in the limiting regime.

\section{Action and Equations of Motion}\label{sec:action_eom}

In this work, we consider Einstein-Maxwell theory
with
a minimally coupled, massless, complex scalar field $\psi$ with charge $q$, which is treated as a test field on a fixed \bh{} background.
With the dimensionless variables and electromagnetic normalization adopted
in this work, the action takes the form
\begin{align}
\label{eq:action_app}
S = \int\dif^{4}x \sqrt{-g}
\left[\frac{1}{2} R
    - \frac{1}{2} F^{\mu\nu} F_{\mu\nu}
    - (D_{\mu}\psi)^{\ast} D^{\mu}\psi
    \right]
+ S_{\rm{M}}
\,,
\end{align}
where
$S_{\rm{M}}$
denotes the action of the additional effective matter associated with the anisotropic fluid,
$(\cdot)^{\ast}$ denotes the complex conjugate,
and the Maxwell tensor in terms of the vector potential, $A_{\mu}$, is
\begin{align}
F_{\mu\nu} & = \nabla_{\mu} A_{\nu} - \nabla_{\nu} A_{\mu}
\,.
\end{align}
The gauge-covariant derivative is given by
\begin{align}
\label{appeq:GaugeCovDeriv}
D_{\mu} \psi & = \left(\nabla_{\mu} - i q A_{\mu} \right) \psi
\,,
\end{align}
where $\nabla_\nu$ is the Levi--Civita covariant derivative compatible with the metric $g_{\mu\nu}$.

Varying the action with respect to the complex conjugate of the scalar field, $\psi^{\ast}$, yields the charged Klein--Gordon equation
\begin{align}
\label{appeq:scalar_eom}
D^{\mu} D_{\mu} \psi & = 0
\,.
\end{align}
We expand the
gauge-covariant d'Alembertian
in terms of the Levi--Civita covariant derivative and the vector potential by inserting Eq.~\eqref{appeq:GaugeCovDeriv} in Eq.~\eqref{appeq:scalar_eom}.
We find
\begin{align}
\label{eq:D2_expand_app}
D^{\mu} D_{\mu} \psi & = \left(
    \nabla^{\mu} \nabla_{\mu}
    -2 i q\,A^{\mu} \nabla_{\mu}
    - i q\,\nabla_{\mu} A^{\mu}
    - q^2 A^{\mu} A_{\mu}
\right)\psi
\,.
\end{align}
We choose the Lorenz gauge
\begin{align}
\label{eq:lorenz_app}
\nabla_{\mu} A^{\mu} & = 0
\,.
\end{align}
With this choice, Eq.~\eqref{eq:D2_expand_app} becomes
\begin{align}
\label{eq:D2_lorenz_app}
D^{\mu} D_{\mu} \psi & = \left(
    \nabla^{\mu} \nabla_{\mu}
    -2 i q\,A^{\mu} \nabla_{\mu}
    - q^2 A^{\mu} A_{\mu}
\right)\psi
\,.
\end{align}

In this work, we consider a single \bh{} with charge parameter $Q$.
The corresponding vector potential is given in Eq.~\eqref{eq:VectorPotential},
and we display it again for completeness,
\begin{align}
\label{appeq:VectorPotential}
A_{\mu} & = \left(-\frac{Q}{r},0,0,0\right)
\,.
\end{align}
Here, we consider the Kiselev metric, Eq.~\eqref{eq:KiselevMetricGeneral},
with components $g_{00}=-f(r)$ and $g^{00}=-1/f(r)$.
Thus, the terms in the Klein-Gordon equation~\eqref{eq:D2_lorenz_app}
reduce to
\begin{align}
\label{eq:A0_and_identities_app}
A^0 & =g^{00}A_0 = \frac{Q}{f(r)\,r}
\,, \quad
A^\nu\nabla_\nu = A^0\nabla_0
\,,\\
A_{\nu}A^{\nu} & = g^{00}A_0^{\,2}=-\frac{Q^2}{f(r)\,r^2}
\,.\nonumber
\end{align}
Then, the scalar field equation~\eqref{eq:D2_lorenz_app} becomes
\begin{align}
\label{eq:D2_specialized_app}
D_\nu D^\nu\psi & =
\left(
\nabla_\nu\nabla^\nu
-2iq\,A^{0}\nabla_0
- q^2\,A_\nu A^\nu
\right)\psi
\nonumber \\ & =
\nabla_\nu \nabla^\nu \psi
- 2iq\,\frac{Q}{f(r)\,r}\,\frac{\partial \psi}{\partial t}
+ \frac{q^2\, Q^2}{f(r)\,r^2}\,\psi
\,.
\end{align}

\section{Sensitivity Analysis}\label{sec:append_sensitivity}

For a function $P$ that depends on a set of parameters $\{p\}$, one naturally wants to determine which parameters contribute most strongly to variations of that function.
This question can be addressed with a
local sensitivity analysis~\cite{Arriola:2009uq}, which measures the response (or ``sensitivity'') of the function $P$ to variations in one of the parameters while all other parameters are held fixed.

In this section, we first introduce the general methodology of sensitivity analysis for functions depending on multiple parameters, and then apply it to the sixth-order \WKB{} approximation as an example.

The sensitivity is measured by the partial derivative of $P$ with respect to that parameter, while all others are held fixed. Since the parameters may have different units and scales, it is convenient to compare their influence using the dimensionless sensitivity coefficients
\begin{equation}\label{eq:sensitivity}
S_p = \left| \frac{\partial P}{\partial p} \cdot \frac{p}{P} \right|.
\end{equation}
These derivatives are evaluated at nominal parameter values, chosen to represent central or characteristic values within the range of interest.
The analysis is local in the sense that it quantifies the response of $P$ to small variations of each parameter around these chosen nominal values, rather than the behavior of $P$ over the full parameter space.
Equation~\eqref{eq:sensitivity} normalizes these derivatives and therefore gives dimensionless coefficients that compare the relative contribution of each parameter to local variations in $P$.

\subsection{Application to the sixth-order \WKB{} Approximation}

\begin{figure*}[ht]
  \centering
\includegraphics[width=0.4\textwidth]{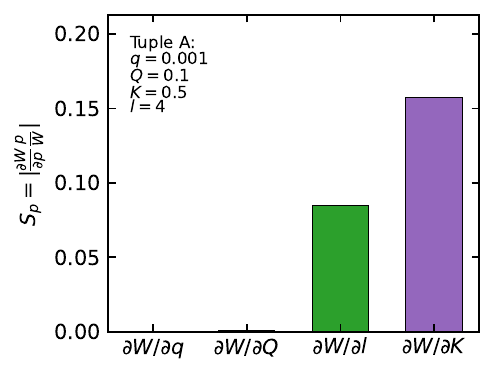}
\includegraphics[width=0.4\textwidth]{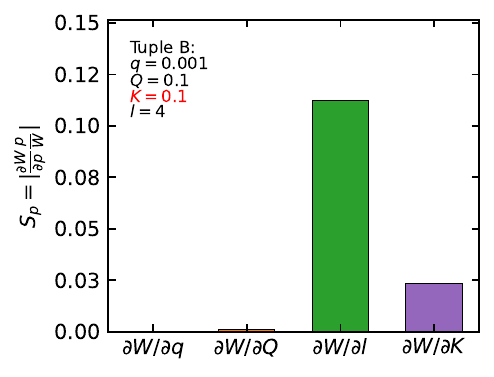}
\includegraphics[width=0.4\textwidth]{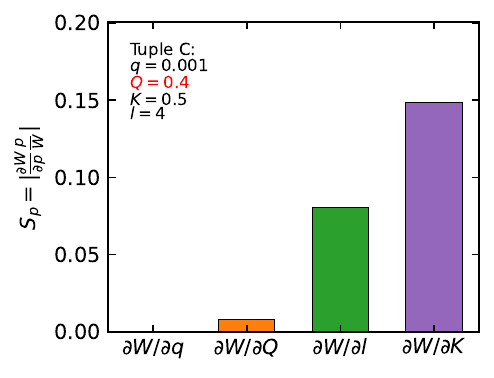}
\includegraphics[width=0.4\textwidth]{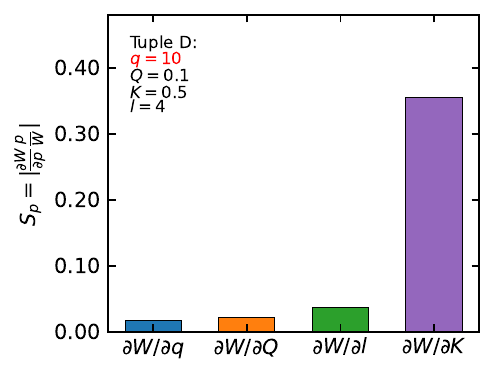}
\caption{Sensitivity coefficients $S_p$
defined in Eq.~\eqref{eq:wkb_sensitivity}
entering the \WKB{} condition in Eq.~\eqref{eq:wkb3}.
In each panel, one parameter is varied, as indicated in red in the upper-left box, while the remaining parameters are fixed at the nominal tuple
$\{q=0.001,Q=0.1, K=0.5,l=4\}$.
The panels correspond to tuple A (upper left), the nominal configuration;
tuple B (upper right), with $K=0.1$;
tuple C (lower left), with $Q=0.4$;
and tuple D (lower right), with $q=10$.
For each tuple, the \QNM{} frequency $\omega$ and the location $r_0$ of the real extremum of the effective potential are recomputed self-consistently for the updated parameter values.}

  \label{fig:sensitivity_bars}
\end{figure*}

In the sixth-order \WKB{} approximation, the quantity of interest is the right-hand side of Eq.~\eqref{eq:wkb3}.
Here, we introduce it as the function
\begin{align}
\label{appeq:WKB-Wfunction}
W & := \frac{i U_0}{\sqrt{2U^{(2)}_0}}
- \Lambda_2 - \Lambda_3 - \Lambda_4
- \Lambda_5 - \Lambda_6
\,,
\end{align}
which is a combination of the \WKB{} effective potential and its derivatives.
We consider the parameter set
\begin{align}
\label{appeq:set}
\{p\} & = \{Q, K, q, l\}
\,.
\end{align}
We select a tuple of representative parameter values as the nominal values, and compute the dimensionless sensitivity coefficients
\begin{align}
\label{eq:wkb_sensitivity}
S_p & = \left| \frac{\partial W}{\partial p} \cdot \frac{p}{W} \right|
\,.
\end{align}
We present the results of the sensitivity analysis in Fig.~\ref{fig:sensitivity_bars}.
In practice, our calculation proceeds as follows:

\begin{enumerate}
\item We begin by choosing a nominal tuple from the parameter set in Eq.~\eqref{appeq:set} that is representative of our \QNM{} computations with the \WKB{} method,
    \begin{equation}
    \label{appeq:Sensitivity-TupleA}
    \{q=0.001, Q=0.1, K=0.5, l=4\}
    \,.
    \end{equation}
    We refer to this nominal tuple as tuple~A in Fig.~\ref{fig:sensitivity_bars}.
    The value $K=0.5$ is taken as the midpoint of the range considered in~\ref{case:two-roots-Qle}.

\item For this nominal tuple, we compute the location $r_0$ of the real extremum of the \WKB{} effective potential defined in Eq.~\eqref{eq:WKB-extremum-real} and the corresponding \QNM{} frequency $\omega$.

\item We then calculate the partial derivatives of the function $W$
in Eq.~\eqref{appeq:WKB-Wfunction}
with respect to $q$, $Q$, $K$, and $l$,
and we compute the sensitivity coefficients using Eq.~\eqref{eq:wkb_sensitivity}. In this step, $r_0$ and $\omega$ are kept fixed at the values obtained in the previous step.

\item  Next, we vary one of the continuous parameters, $K$, $Q$, or $q$, away from its nominal value while keeping the others fixed.
Choosing representative values from this broader variation defines tuples B, C, and D in Fig.~\ref{fig:sensitivity_bars}.
They are given by
\begin{subequations}
\label{appeq:Sensitivity-TupleBCD}
\begin{align}
B: &\, \{q=0.001, Q=0.1, K=0.1, l=4\}
\,,\\
C: &\, \{q=0.001, Q=0.4, K=0.5, l=4\}
\,,\\
D: &\, \{q=10, Q=0.1, K=0.5, l=4\}
\,.
\end{align}
\end{subequations}
For each new tuple, we first recompute the corresponding
location of the real extremum in the potential and the \QNM{} frequency, and then evaluate the sensitivity coefficients using Eq.~\eqref{eq:wkb_sensitivity}.
\end{enumerate}

The multipole number $l$ is kept fixed at its nominal value $l=4$ throughout the parameter scan.
It labels the perturbation mode and takes discrete values rather than representing a continuously tunable physical parameter.
Note that the multipole number affects the validity of the \WKB{} approximation:
it is expected to perform best in the eikonal limit where
the angular multipole number is large compared with the overtone number, $l\gg \textrm{n}$.
We only evaluate the coefficient $S_l$ for each tuple through the derivative $\partial W/\partial l$ to estimate how strongly the \WKB{} condition depends on the angular mode number; we do not vary the mode number.

Fig.~\ref{fig:sensitivity_bars} illustrates representative examples of the sensitivity coefficients obtained by varying $K$, $Q$, and $q$ around the nominal parameter set;
see Eqs.~\eqref{appeq:Sensitivity-TupleA} and~\eqref{appeq:Sensitivity-TupleBCD}.
The full analysis is carried out dynamically as the parameters are changed, so the coefficients are not evaluated only for the single values highlighted in red.
Instead, each panel shows one representative tuple taken from that broader parameter variation.
In each panel, the parameter highlighted in red in the upper-left box is the one varied from its nominal value, while the remaining parameters are kept at the nominal tuple.

Comparing tuple B (top right panel with $K=0.1$) with nominal tuple A (top left panel), the sensitivity coefficient $S_K$ increases with $K$, while the sensitivity coefficient $S_l$ decreases.
When comparing tuple D (bottom right panel with $q=10$) with tuple A, the sensitivity to changes in $q$ increases only slightly and remains the smallest among all parameters.
The sensitivity coefficient $S_Q$ is the second smallest overall. As $Q$ increases from the nominal tuple to $Q=0.4$ in tuple C (bottom left panel), its contribution becomes moderately larger.

Overall, $S_K$ and $S_l$ are the largest explicit local
sensitivity coefficients for the representative tuples considered,
while \(S_Q\) and \(S_q\) are comparatively smaller. This indicates
that the WKB residual \(W\)
depends more strongly on \(K\) and \(l\) than on \(Q\) and \(q\).

\section{Error Analysis}
\label{sec:error_analysis}


In this section, we analyze the truncation error in our solutions for the complex \QNM{} frequencies,
$\omega=\omegaR + i \omegaI$, which we obtain by solving the characteristic equation in Eq.~\eqref{eq:characteristic_equation}.
We solve this equation using conventional root finders as well as automatic differentiation techniques, and we determine the error of both methods.

We compute the truncation error for representative \bh{} and scalar-field parameters,
and set
\begin{align}
\label{eq:error_analysis_parameters}
K & = 0.2,\quad
Q = 0.05,\quad
qQ = 0.2,\quad
l = 0
\,.
\end{align}
We vary the truncation level, i.e., the number $N$ of terms in the series expansion in Eq.~\eqref{eq:ansatz_power_series},
in the range
$N \in \{30, 100,  300, 500, 700, 1100,1500,2000,3000\}$.
The frequency obtained with a given truncation level, $\omega_{N}$, is compared against a reference solution $\omega_{\rm{ref}}$
that we compute with $N_{\rm ref}=6,000$ terms.
We analyze the numerical error of both the real and the imaginary parts of the frequency.

We define the absolute error of a quantity $X$ as
\begin{align}
\label{eq:error_abs}
E_{\rm abs} & =
    \left|X_N-X_{\rm ref}\right|
\,,
\end{align}
where $X$ is a placeholder for $\omegaR$ and $\omegaI$,
$X_{N}$ denotes the quantity computed with truncation level $N$, and $X_{\rm{ref}}$ refers to the reference solution.
The corresponding relative error is defined as
\begin{align}
\label{eq:error_rel}
E_{\rm rel} & =
    \frac{\left|X_N-X_{\rm ref}\right|}
    {\left|X_{\rm ref}\right|}
\,.
\end{align}

\subsection{Error Analysis for conventional root finder}

\begin{figure*}[ht]
  \centering
\includegraphics[width=0.75\textwidth]{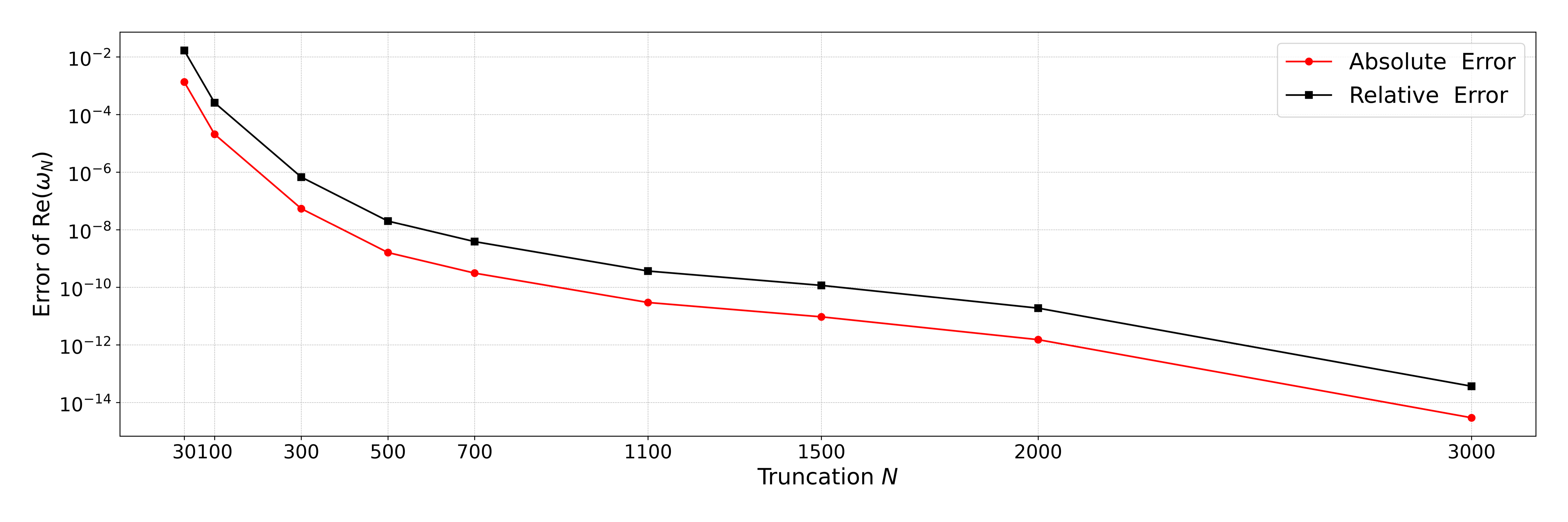}
\includegraphics[width=0.75\textwidth]{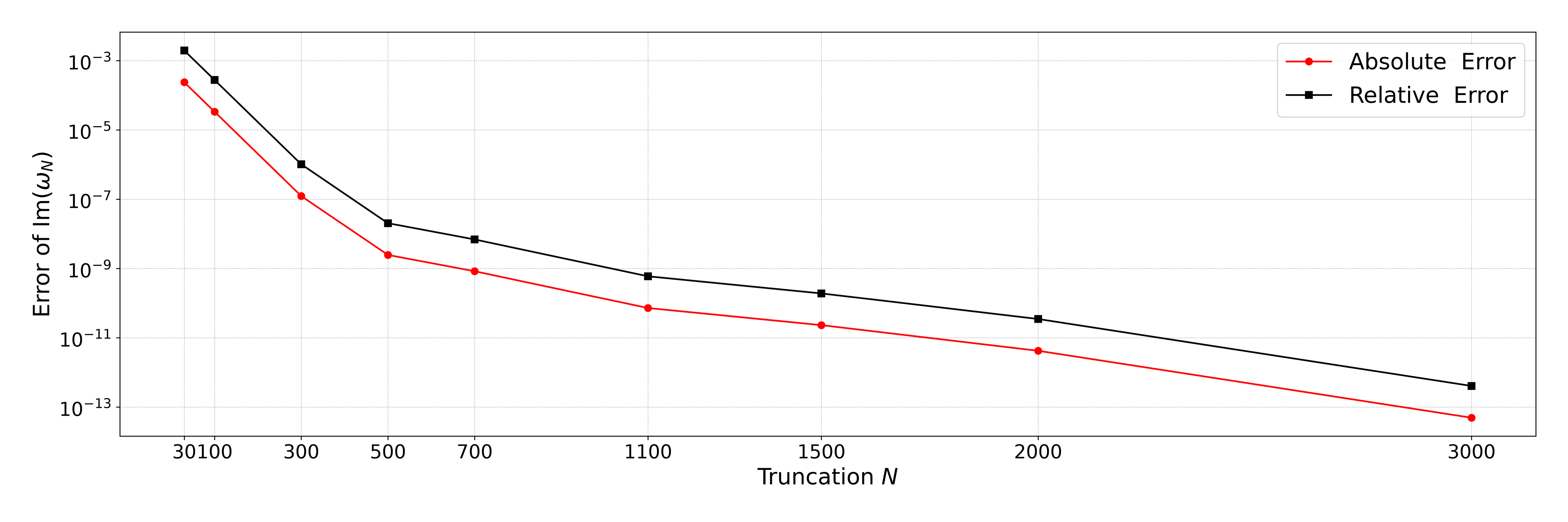}
\caption{
Absolute (red dashed lines) and relative (black solid lines) errors defined in Eqs.~\eqref{eq:error_abs} and~\eqref{eq:error_rel}, respectively,
of the real (top panel) and imaginary (bottom panel) parts of the fundamental \QNM{} frequency
of the $l=0$ mode,
as functions of the truncation level $N$.
The computation employs a conventional root-finding algorithm.
}
\label{fig:error_convergence}
\end{figure*}

In Fig.~\ref{fig:error_convergence}
we present the absolute and relative errors of the fundamental monopole \QNM{} frequencies
as functions of the truncation level $N$,
computed with conventional root-finding algorithms.
We show the errors in the real part of the frequency in the top panel, and the errors in the imaginary part in the bottom panel.
We see that the relative error in both quantities is $\lesssim 10^{-8}$ for truncation levels above $N\gtrsim 700$.

Based on this analysis, we use a truncation level of $N=700$ for most of the \QNM{} calculations presented in Sec.~\ref{sec:numerical_results} and up to $N=5000$ where convergence requires it.

\subsection{Error Analysis for the Automatic Differentiation Implementation}
\label{ssec:error_analysis_ad}

We repeat the truncation-error analysis for the implementation based on Leaver's continued fraction method combined with automatic differentiation,
introduced in Sec.~\ref{ssec:automaticdifferentiation}.
To allow a direct comparison with the conventional root-finding implementation, we use the same \bh{} and scalar-field parameters given in Eq.~\eqref{eq:error_analysis_parameters}, the same reference truncation level $N_{\rm ref}=6000$, and the same initial guesses of the frequencies.
For each truncation level, we compute the absolute and relative errors,
defined in Eqs.~\eqref{eq:error_abs} and~\eqref{eq:error_rel},
of the real and imaginary parts of the fundamental monopole \QNM{} frequencies.

\begin{figure*}[ht]
  \centering
\includegraphics[width=0.75\textwidth]{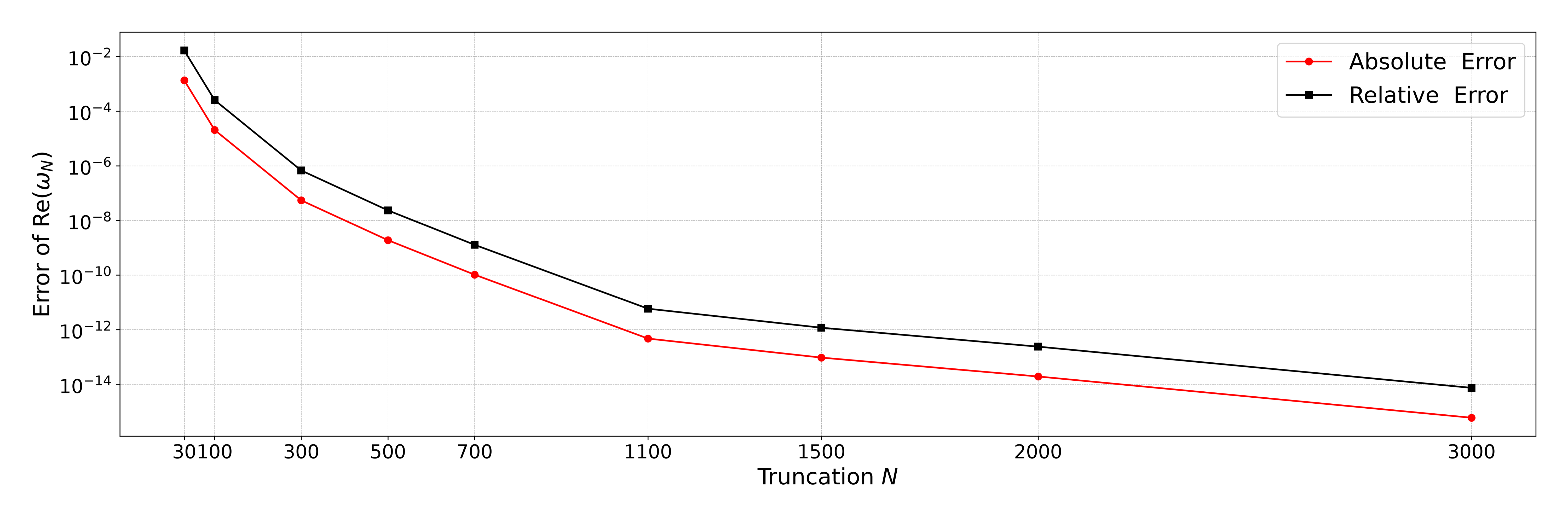}
\includegraphics[width=0.75\textwidth]{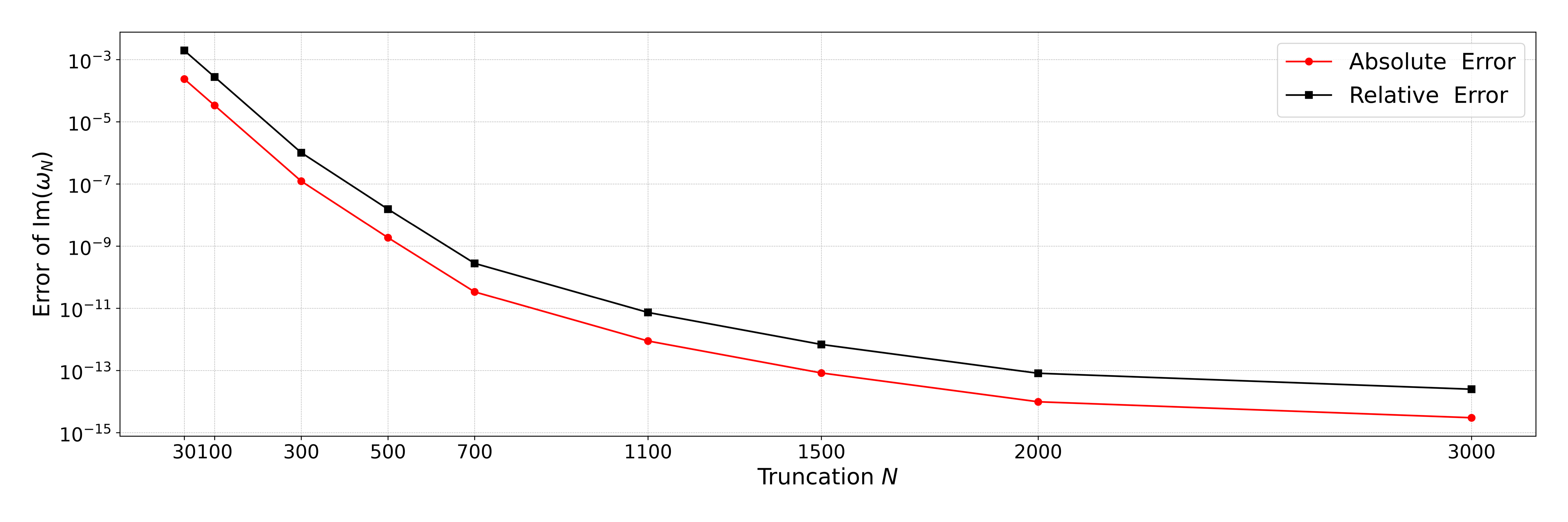}
\caption{
Absolute (dashed red lines) and relative (black solid lines) errors defined in Eqs.~\eqref{eq:error_abs} and~\eqref{eq:error_rel}, respectively,
of the real (top panel) and imaginary (bottom panel) parts of the fundamental \QNM{} frequency
of the $l=0$ mode,
as functions of the truncation level $N$.
The computation uses the automatic-differentiation implementation of Leaver's method.
}
\label{fig:error_convergence_ad}
\end{figure*}

The results are shown in Fig.~\ref{fig:error_convergence_ad},
where we plot the absolute and relative errors in the
real part (top panel) and imaginary part (bottom panel) of the frequency
as a function of the truncation level.
We see that both errors decrease exponentially with increasing truncation level.
All frequencies are stored in \texttt{float64} precision, which resolves approximately $15$--$17$ significant digits.

Compared with the conventional root-finding implementation shown in Fig.~\ref{fig:error_convergence}, the automatic-differentiation implementation in Fig.~\ref{fig:error_convergence_ad} produces smaller absolute and relative errors for both the real and imaginary parts at a sufficiently large truncation level $N$.
%

\section{Scaling of the quasinormal mode frequencies with the fluid parameter}
\label{app:leading_order_estimates}

In this section, we provide derivations that support the calculations in Sec.~\ref{sec:dependence_fluid} for the real and imaginary parts of the \QNM{} frequency,
and determine the dependence on the fluid parameter $K$.

We use the \WKB{} potential $U(r,\omega)$ in Eq.~\eqref{eq:WKBEffPotential}, together with the leading-order terms of the \WKB{} formula in Eq.~\eqref{eq:wkb3},
and the location of the real extremum, $r_0$, defined in Eq.~\eqref{eq:WKB-extremum-real}.
Quantities evaluated at the extremum are denoted by a subscript $0$.

Eqs.~\eqref{eq:wkb3} and~\eqref{eq:WKB-extremum-real} yield a set of coupled equations,
Eqs.~\eqref{eq:U-extremum-discussion}--~\eqref{eq:Eqs-balance-K},
relating the complex \QNM{} frequencies to the extremum of the \WKB{} potential.
For self-consistency in this section, we display these relations, to leading order, here
\begin{subequations}
\label{appeq:OmegaDepOnK}
\begin{align}
\label{appeq:U-extremum-discussion}
2 \left(\omegaR-\Phi_{0}\right) \Phi'(r_{0})
    & = -V'(r_{0})
\,,\\
\label{appeq:Re-balance-K}
\left(\omegaR - \Phi_0\right)^{2}
    & \simeq V_0+\omegaI^{2}
\,,\\
\label{appeq:Im-balance-K}
2 \left(\omegaR - \Phi_0\right) \omegaI
    & \simeq -\left(n+\tfrac{1}{2}\right) \sqrt{2\,U_{0}^{(2)}}
\,.
\end{align}
\end{subequations}

From Eqs.~\eqref{appeq:Re-balance-K} and~\eqref{appeq:Im-balance-K}, the leading-order contributions are
\begin{subequations}
\begin{align}
\label{appeq:Re-omega-K}
    \omegaR
    \simeq
    \Phi_0
    +
    \sqrt{V_0+\omegaI^2}
\,,\\
\label{appeq:Im-omega-K}
    \omegaI^{2}
    \simeq
    \frac{\left(n+\tfrac{1}{2}\right)^{2}U_{0}^{(2)}}
         {V_0+\sqrt{V_0^{2}+2\left(n+\tfrac{1}{2}\right)^{2}U_{0}^{(2)}}}
\,,
\end{align}
\end{subequations}
where we choose the branch with $\omegaR-\Phi_0>0$ in Eq.~\eqref{appeq:Re-omega-K}.
This branch is continuously connected to the standard solution with
positive real part of the frequency,
$\omegaR>0$, in the zero-coupling limit $qQ=0$.

Using Eqs.~\eqref{eq:2CompKiselev-Veff-polynomial}, \eqref{eq:WKBEffPotential}, and~\eqref{eq:WKB-eff-pot-deriv}, we now estimate the quantities entering Eqs.~\eqref{appeq:Re-omega-K} and~\eqref{appeq:Im-omega-K}.
We keep all terms needed to cover both the monopole case, $l=0$ with $\lambda=l(l+1)=0$, and the higher multipoles, $l\geq1$ with $\lambda\neq0$.

In the large-$r_{0}$ region,
the geometric potential $V_0$
evaluated at the location of the \WKB{} real extremum, $r_{0}$, has the leading-order form
\begin{equation}
\label{eq:V0-estimate-K}
    V_{0}
    \simeq
    f_{0}
    \left(
        \frac{\lambda}{r_{0}^{2}}
        +
        \frac{1}{r_{0}^{3}}
    \right),
    \quad
    f_{0}
    \simeq
    (1-K)-\frac{1}{r_{0}}+\frac{Q^2}{r_0^2}
\,.
\end{equation}
The corresponding first and second radial derivatives are
\begin{equation}
\label{eq:V0prime-estimate-K}
    V'_{0}
    \simeq
    -\frac{2\lambda(1-K)}{r_{0}^{3}}
    +
    \frac{3\lambda-3(1-K)}{r_{0}^{4}}
    +
    \frac{4(1-\lambda Q^2)}{r_{0}^{5}}
\,,
\end{equation}
and
\begin{equation}
\label{eq:V0second-estimate-K}
    V''_{0}
    \simeq
    \frac{6\lambda(1-K)}{r_{0}^{4}}
    +
    \frac{12\big[(1-K)-\lambda\big]}{r_{0}^{5}}
    -
    \frac{20(1-\lambda Q^2)}{r_{0}^{6}}
\,.
\end{equation}
The derivatives above are written through the orders needed for the
subsequent estimate. Terms of order $\mathcal O(r_0^{-6})$ in $V'_0$
and $\mathcal O(r_0^{-7})$ in $V''_0$ are omitted.

Then, the second derivative of the \WKB{} potential with respect to the tortoise coordinate at the real extremum reads
\begin{align}
\label{eq:U2-estimate-K}
    U_{0}^{(2)}
    &\simeq
    f_{0}^{2}
    \left[
        \frac{2(qQ)^{2}}{r_{0}^{4}}
        -
        \frac{2V'_{0}}{r_{0}}
        -
        V''_{0}
    \right]
\nonumber \\
    &\simeq
    f_{0}^{2}
    \biggl[
        \frac{2\big[(qQ)^{2}-\lambda(1-K)\big]}{r_{0}^{4}}
\\ & \quad
        +
        \frac{6\big[\lambda-(1-K)\big]}{r_{0}^{5}}
        +
        \frac{12(1-\lambda Q^{2})}{r_{0}^{6}}
    \biggr]. \nonumber
\end{align}

For fixed finite $\lambda$ and fixed coupling $qQ$, Eqs.~\eqref{eq:V0-estimate-K} and~\eqref{eq:U2-estimate-K} give the following order estimates
\begin{equation}
\label{eq:Im-omega-order-K-app}
    |\omegaI|
    =
    \mathcal{O}\!\left[
        \sqrt{f_0}
        \left(
        \frac{|qQ|}{r_0}
        +
        \frac{\sqrt{1-K}}{r_0}
        +
        \frac{1}{r_0^{3/2}}
        \right)
    \right],
\end{equation}
and
\begin{equation}
\label{eq:Re-omega-order-K-app}
    \omegaR
    =
    \frac{qQ}{r_0}
    +
    \mathcal{O}\!\left[
        \sqrt{f_0}
        \left(
        \frac{\sqrt{\lambda}}{r_0}
        +
        \frac{|qQ|}{r_0}
        +
        \frac{\sqrt{1-K}}{r_0}
        +
        \frac{1}{r_0^{3/2}}
        \right)
    \right].
\end{equation}

For an uncharged scalar field, Eq.~\eqref{appeq:U-extremum-discussion} reduces to $V'(r_{0})=0$, and the
location of the \WKB{} extremum to leading order
is given by
\begin{equation}
\label{eq:r0-estimate-K}
    r_0\simeq
    \begin{cases}
        \dfrac{4}{3(1-K)}, & \lambda=0,
        \\[1.2em]
        \dfrac{3}{2(1-K)}, & \lambda\neq0 .
    \end{cases}
\end{equation}
Thus, the location of the extremum and the metric function evaluated at this location scale as
\begin{equation}
\label{eq:r0-f0-scaling-K}
    r_0=\mathcal{O}\!\left((1-K)^{-1}\right)
\,,\quad
    f_0=\mathcal{O}(1-K)
\,,
\end{equation}
where we dropped the term $Q^2/r_0^2$ as a subleading contribution to $f_0$.
Consequently, the \QNM{} frequency
scales as
\begin{equation}
\label{eq:omegaI-scaling-K}
    |\omega_I|
    =
    \mathcal{O}\!\left((1-K)^2\right)
\,,
\end{equation}
and
\begin{equation}
\label{eq:omegaR-scaling-K}
    \omega_R
    =
    \begin{cases}
        \mathcal{O}\!\left((1-K)^2\right),
        & \lambda=0,
        \\[4pt]
        \mathcal{O}\!\left((1-K)^{3/2}\right),
        & \lambda\neq0.
    \end{cases}
\end{equation}

\section{Numerical Results and \WKB{} Comparison}\label{app:wkb_comparison}

\begin{figure*}[t]
\centering
\includegraphics[width=0.97\textwidth]{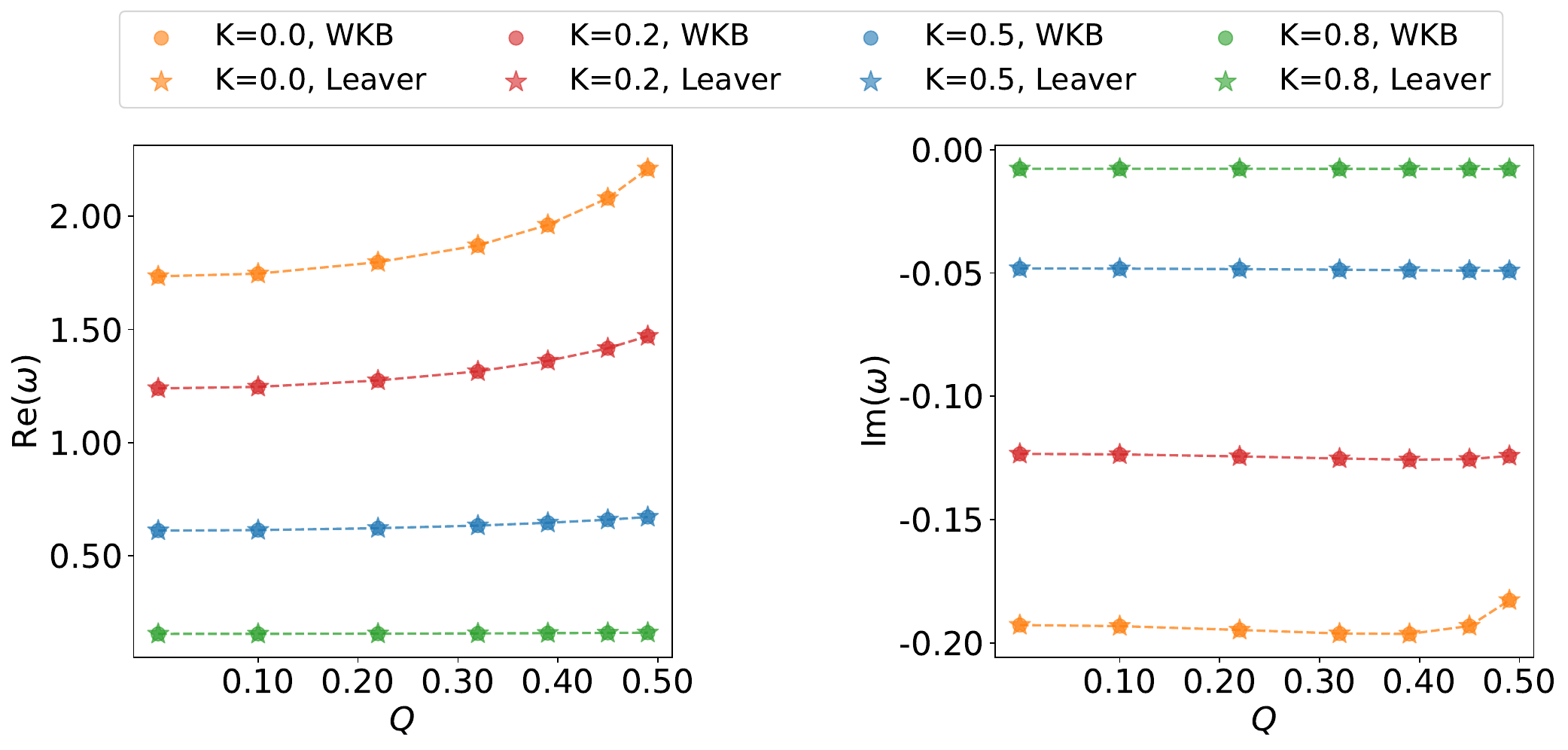}
\caption{Real (left panel) and imaginary (right panel) parts of the fundamental
\QNM{}
frequency for the $l = 4$ multipole of an uncharged scalar field ($q=0$)
as a function of the \bh{} charge $Q$.
The different lines correspond to different values of the parameter
$K \in \{0.0, 0.2, 0.5, 0.8\}$.
We compute the frequencies using Leaver's continued fraction method (stars) and the sixth-order \WKB{} approximation (filled circles).
The two methods are in good agreement, and the corresponding star and circle markers overlap in the parameter range.}
\label{fig:wkb_leaver_l=4qQ=0}
\end{figure*}

\begin{figure*}[htbp]
\centering
\includegraphics[width=0.97\textwidth]{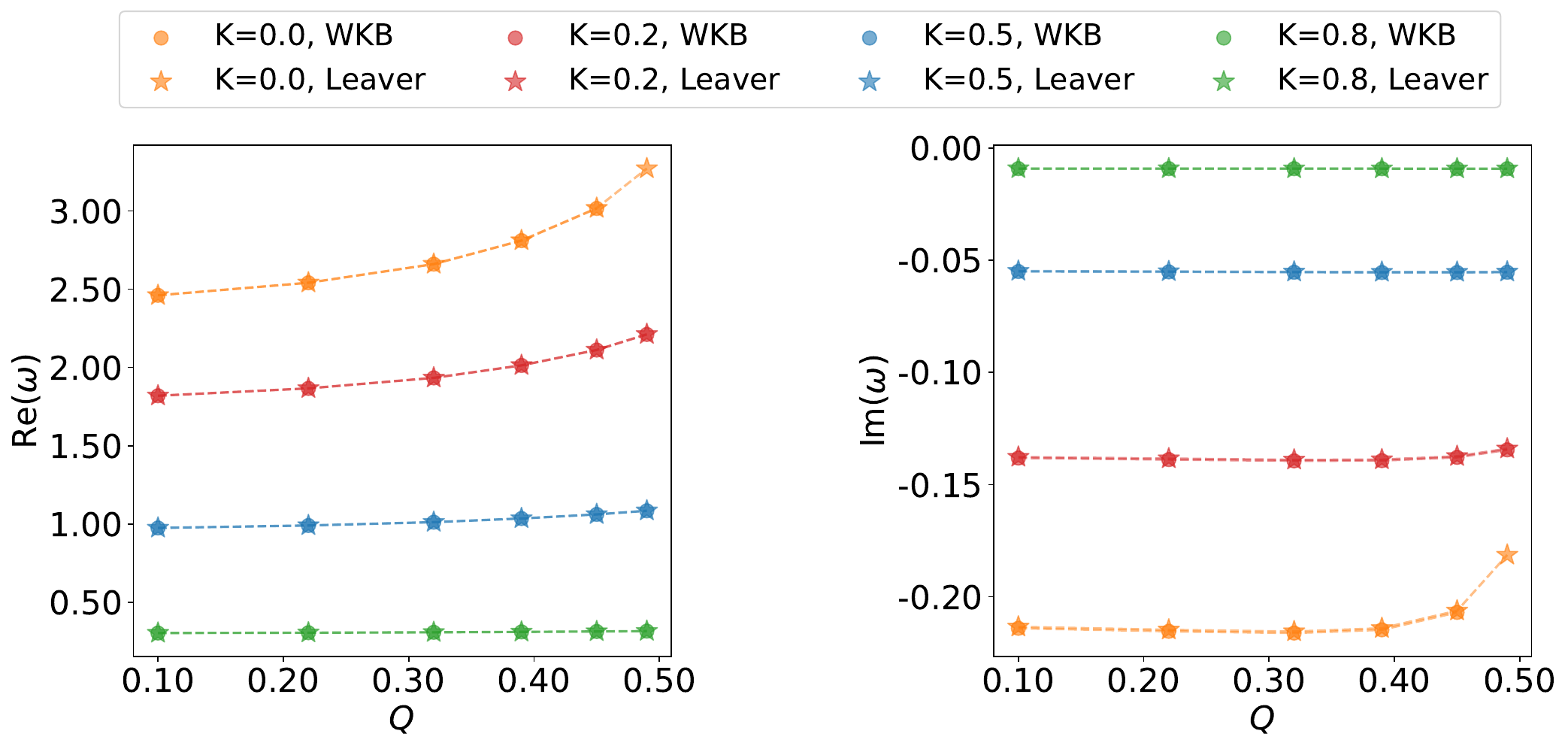}
\caption{Real (left panel) and imaginary (right panel) parts of the fundamental
\QNM{}
frequency of the $l = 4$ multipole of a charged scalar field with fixed electrostatic interaction $qQ=1$
as a function of the \bh{} charge $Q$.
The different lines correspond to different values of the parameter
$K \in \{0.0, 0.2, 0.5, 0.8\}$.
We compute the frequencies using Leaver's continued fraction method (stars) and the sixth-order \WKB{} approximation (filled circles).
The two methods are in good agreement, and the corresponding star and circle markers overlap in most of the parameter range.}
\label{fig:wkb_leaver_l=4qQ=1}
\end{figure*}

In this section, we present the numerical results for the \QNM{} frequencies obtained with Leaver's continued fraction method and the sixth-order \WKB{} approximation.
In Tables~\ref{tab:qnm_values_l1_qQ0_corrected} and \ref{tab:qnm_values_l1_qQ1_interaction}, we list the results of the scalar field $l=1$ mode, computed only with Leaver's method.
In Tables~\ref{tab:qnm_values_l4_qQ0} and \ref{tab:qnm_values_l4_qQ1} and
Figs.~\ref{fig:wkb_leaver_l=4qQ=0} and~\ref{fig:wkb_leaver_l=4qQ=1},
we present the results of the $l=4$ mode obtained with both Leaver's and the \WKB{} method.

For the uncharged scalar field, shown in Table~\ref{tab:qnm_values_l4_qQ0} and Fig.~\ref{fig:wkb_leaver_l=4qQ=0}, the sixth-order \WKB{} and Leaver's results agree extremely well, with relative differences below $1.7\times10^{-4}\%$ in the real part and $8.1\times10^{-4}\%$ in the imaginary part of the frequency.

For the charged scalar field with interaction $qQ=1$, shown in Table~\ref{tab:qnm_values_l4_qQ1} and Fig.~\ref{fig:wkb_leaver_l=4qQ=1}, the relative differences range from approximately $0.012\%$ to $0.065\%$ in the real part and from $0.30\%$ to $0.45\%$ in the imaginary part, excluding the parameter set for which the sixth-order \WKB{} approximation fails. The sixth-order \WKB{} approximation fails only for $K=0$ and $Q\approx0.49$, where the \bh{} charge parameter approaches its critical value.

It has been shown that
for frequency-independent \WKB{} potentials, Pad\'e resummations of the sixth-order and higher \WKB{} approximations agree very well with
accurate numerical \QNM{} frequencies
for Schwarzschild and \RN{} black holes~\cite{Matyjasek:2017psv}.
For Schwarzschild \bh{s},
the numerical
benchmark frequencies were taken from Ref.~\cite{Froeman:1992gp},
which used a modified Newman-Thorson method.
Ref.~\cite{Matyjasek:2017psv} found that, for low-lying modes, the Pad\'e-improved \WKB{} results are comparable to, and in some cases better than,
phase-integral results used as an independent benchmark.


\begin{table*}[t]
\centering
\caption{Real and imaginary parts of the fundamental \QNM{} frequency for the $l=1$ multipole of an uncharged, massless scalar field ($q=0$),
computed with Leaver's method.
Recall that, following the rescaling introduced in Eqs.~\eqref{eq:units}, the symbols $\omega$ and $Q$ denote the dimensionless frequency $\rg\omega$ and \bh{} charge $Q/\rg$, and we employ units with $\rg=1$.
}
\label{tab:qnm_values_l1_qQ0_corrected}
\begin{tabular}{ccccccccc}
\toprule
$Q$ & \multicolumn{2}{c}{$K = 0.0$} & \multicolumn{2}{c}{$K = 0.2$} & \multicolumn{2}{c}{$K = 0.5$} & \multicolumn{2}{c}{$K = 0.8$} \\
\cmidrule(lr){2-3} \cmidrule(lr){4-5} \cmidrule(lr){6-7} \cmidrule(lr){8-9}
& $\mathrm{Re}(\omega)$ & $\mathrm{Im}(\omega)$ & $\mathrm{Re}(\omega)$ & $\mathrm{Im}(\omega)$ & $\mathrm{Re}(\omega)$ & $\mathrm{Im}(\omega)$ & $\mathrm{Re}(\omega)$ & $\mathrm{Im}(\omega)$ \\
\midrule
0.000000 & 0.585872 & -0.195320 & 0.413416 & -0.124685 & 0.199907 & -0.048502 & 0.049448 & -0.007724 \\
0.100000 & 0.589886 & -0.195724 & 0.415670 & -0.124894 & 0.200582 & -0.048554 & 0.049514 & -0.007728 \\
0.223607 & 0.607227 & -0.197199 & 0.425253 & -0.125678 & 0.203389 & -0.048754 & 0.049784 & -0.007741 \\
0.316228 & 0.632486 & -0.198488 & 0.438734 & -0.126472 & 0.207152 & -0.048984 & 0.050129 & -0.007757 \\
0.387298 & 0.663347 & -0.198413 & 0.454375 & -0.126884 & 0.211244 & -0.049180 & 0.050486 & -0.007773 \\
0.447214 & 0.702870 & -0.194755 & 0.472989 & -0.126526 & 0.215728 & -0.049325 & 0.050853 & -0.007788 \\
0.489898 & 0.743864 & -0.183929 & 0.490903 & -0.125083 & 0.219650 & -0.049386 & 0.051156 & -0.007800 \\
\bottomrule
\end{tabular}
\end{table*}

\begin{table*}[htbp]
\centering
\caption{
Real and imaginary parts of the fundamental \QNM{} frequency for the $l=4$ multipole of an uncharged, massless scalar field ($q=0$).
For each value of $Q$, the first row gives the result from Leaver's method, and the second row gives the result from the sixth-order \WKB{} method.
Recall that, following the rescaling introduced in Eqs.~\eqref{eq:units}, the symbols $\omega$ and $Q$ denote the dimensionless frequency, $\rg\omega$, and \bh{} charge, $Q/\rg$, and we employ units with $\rg=1$.
}
\label{tab:qnm_values_l4_qQ0}
\begin{tabular}{ccccccccc}
\toprule
$Q$ & \multicolumn{2}{c}{$K = 0.0$}
& \multicolumn{2}{c}{$K = 0.2$}
& \multicolumn{2}{c}{$K = 0.5$}
& \multicolumn{2}{c}{$K = 0.8$} \\
\cmidrule(lr){2-3} \cmidrule(lr){4-5} \cmidrule(lr){6-7} \cmidrule(lr){8-9}
& $\mathrm{Re}(\omega)$ & $\mathrm{Im}(\omega)$ & $\mathrm{Re}(\omega)$ & $\mathrm{Im}(\omega)$
& $\mathrm{Re}(\omega)$ & $\mathrm{Im}(\omega)$ & $\mathrm{Re}(\omega)$ & $\mathrm{Im}(\omega)$ \\
\midrule
0.000000 & 1.734831 & -0.192783 & 1.239416 & -0.123339 & 0.610972 & -0.048155 & 0.154202 & -0.007701 \\
         & 1.734831 & -0.192784 & 1.239416 & -0.123339 & 0.610971 & -0.048155 & 0.154202 & -0.007701 \\
0.100000 & 1.746585 & -0.193204 & 1.246112 & -0.123555 & 0.613025 & -0.048208 & 0.154409 & -0.007704 \\
         & 1.746585 & -0.193205 & 1.246112 & -0.123556 & 0.613025 & -0.048208 & 0.154409 & -0.007704 \\
0.223607 & 1.797371 & -0.194761 & 1.274577 & -0.124372 & 0.621549 & -0.048413 & 0.155246 & -0.007718 \\
         & 1.797370 & -0.194762 & 1.274577 & -0.124372 & 0.621548 & -0.048413 & 0.155246 & -0.007718 \\
0.316228 & 1.871436 & -0.196198 & 1.314644 & -0.125222 & 0.632981 & -0.048651 & 0.156321 & -0.007734 \\
         & 1.871436 & -0.196198 & 1.314644 & -0.125222 & 0.632981 & -0.048651 & 0.156321 & -0.007734 \\
0.387298 & 1.962278 & -0.196362 & 1.361206 & -0.125713 & 0.645416 & -0.048856 & 0.157429 & -0.007750 \\
         & 1.962278 & -0.196362 & 1.361206 & -0.125713 & 0.645416 & -0.048856 & 0.157429 & -0.007750 \\
0.447214 & 2.080235 & -0.193149 & 1.416831 & -0.125476 & 0.659048 & -0.049012 & 0.158573 & -0.007766 \\
         & 2.080235 & -0.193150 & 1.416831 & -0.125476 & 0.659048 & -0.049012 & 0.158573 & -0.007766 \\
0.489898 & 2.210462 & -0.182717 & 1.470846 & -0.124186 & 0.670985 & -0.049085 & 0.159515 & -0.007777 \\
         & 2.210464 & -0.182718 & 1.470846 & -0.124186 & 0.670985 & -0.049085 & 0.159515 & -0.007777 \\
\bottomrule
\end{tabular}
\end{table*}


\begin{table*}[t]
\centering
\caption{
Real and imaginary parts of the fundamental \QNM{} frequency for the $l=1$ multipole
of a charged scalar field,
computed with Leaver's method.
We fix the electrostatic interaction to $qQ=1$,
yielding a scalar charge of $q=1/Q$.
Recall that, following the rescaling introduced in Eqs.~\eqref{eq:units}, the symbols $\omega$ and $Q$ denote the dimensionless  frequency, $\rg\omega$, and \bh{} charge, $Q/\rg$, in units with $\rg=1$.
}
\label{tab:qnm_values_l1_qQ1_interaction}
\begin{tabular}{ccccccccc}
\toprule
$Q$ & \multicolumn{2}{c}{$K = 0.0$} & \multicolumn{2}{c}{$K = 0.2$} & \multicolumn{2}{c}{$K = 0.5$} & \multicolumn{2}{c}{$K = 0.8$} \\
\cmidrule(lr){2-3} \cmidrule(lr){4-5} \cmidrule(lr){6-7} \cmidrule(lr){8-9}
& $\mathrm{Re}(\omega)$ & $\mathrm{Im}(\omega)$ & $\mathrm{Re}(\omega)$ & $\mathrm{Im}(\omega)$ & $\mathrm{Re}(\omega)$ & $\mathrm{Im}(\omega)$ & $\mathrm{Re}(\omega)$ & $\mathrm{Im}(\omega)$ \\
\midrule
0.000032 & 1.361688 & -0.235308 & 1.039938 & -0.152032 & 0.600475 & -0.060435 & 0.217840 & -0.009885 \\
0.022361 & 1.362258 & -0.235321 & 1.040292 & -0.152039 & 0.600607 & -0.060436 & 0.217860 & -0.009885 \\
0.054772 & 1.365119 & -0.235387 & 1.042069 & -0.152069 & 0.601270 & -0.060442 & 0.217962 & -0.009885 \\
0.100000 & 1.373269 & -0.235559 & 1.047114 & -0.152150 & 0.603144 & -0.060457 & 0.218248 & -0.009885 \\
0.223607 & 1.424232 & -0.236140 & 1.078095 & -0.152444 & 0.614345 & -0.060517 & 0.219911 & -0.009887 \\
0.316228 & 1.501672 & -0.235273 & 1.123155 & -0.152228 & 0.629685 & -0.060514 & 0.222066 & -0.009887 \\
0.387298 & 1.603250 & -0.230511 & 1.178098 & -0.150850 & 0.646831 & -0.060388 & 0.224311 & -0.009885 \\
0.447214 & 1.752177 & -0.214223 & 1.248511 & -0.147072 & 0.666252 & -0.060086 & 0.226655 & -0.009880 \\
0.489898 & 1.969469 & -0.156953 & 1.324084 & -0.140059 & 0.683876 & -0.059656 & 0.228607 & -0.009874 \\
\bottomrule
\end{tabular}
\end{table*}
\begin{table*}[t]
\centering
\caption{
Real and imaginary parts of the fundamental \QNM{} frequency for the $l=4$ multipole
of a charged scalar field.
We fix the electrostatic interaction to $qQ=1$,
yielding a scalar charge of $q=1/Q$.
For each value of $Q$, the first row gives the result from Leaver's method, and the second row gives the result from the sixth-order \WKB{} method.
Recall that, following the rescaling introduced in Eqs.~\eqref{eq:units}, the symbols $\omega$ and $Q$ denote the dimensionless  frequency, $\rg\omega$, and \bh{} charge, $Q/\rg$, in units with $\rg=1$.
Dashes indicate the parameter sets for which the sixth-order \WKB{} approximation fails.
}
\label{tab:qnm_values_l4_qQ1}
\begin{tabular}{ccccccccc}
\toprule
$Q$ & \multicolumn{2}{c}{$K = 0.0$} & \multicolumn{2}{c}{$K = 0.2$} & \multicolumn{2}{c}{$K = 0.5$} & \multicolumn{2}{c}{$K = 0.8$} \\
\cmidrule(lr){2-3} \cmidrule(lr){4-5} \cmidrule(lr){6-7} \cmidrule(lr){8-9}
& $\mathrm{Re}(\omega)$ & $\mathrm{Im}(\omega)$ & $\mathrm{Re}(\omega)$ & $\mathrm{Im}(\omega)$ & $\mathrm{Re}(\omega)$ & $\mathrm{Im}(\omega)$ & $\mathrm{Re}(\omega)$ & $\mathrm{Im}(\omega)$ \\
\midrule
0.100000 & 2.460978 & -0.213398 & 1.819677 & -0.137716 & 0.975376 & -0.054843 & 0.304004 & -0.009148 \\
         & 2.460674 & -0.214028 & 1.819245 & -0.138156 & 0.974978 & -0.055042 & 0.303847 & -0.009188 \\
0.223607 & 2.541222 & -0.214728 & 1.866723 & -0.138414 & 0.991072 & -0.055015 & 0.306015 & -0.009158 \\
         & 2.540764 & -0.215376 & 1.866210 & -0.138862 & 0.990653 & -0.055216 & 0.305857 & -0.009198 \\
0.316228 & 2.660462 & -0.215492 & 1.933960 & -0.138963 & 1.012343 & -0.055186 & 0.308611 & -0.009169 \\
         & 2.659737 & -0.216165 & 1.933319 & -0.139423 & 1.011894 & -0.055390 & 0.308451 & -0.009209 \\
0.387298 & 2.811193 & -0.214021 & 2.013846 & -0.138854 & 1.035792 & -0.055290 & 0.311304 & -0.009179 \\
         & 2.810048 & -0.214723 & 2.013030 & -0.139326 & 1.035308 & -0.055496 & 0.311142 & -0.009219 \\
0.447214 & 3.017484 & -0.206217 & 2.112365 & -0.137381 & 1.061911 & -0.055293 & 0.314102 & -0.009187 \\
         & 3.015538 & -0.206962 & 2.111288 & -0.137868 & 1.061385 & -0.055502 & 0.313937 & -0.009227 \\
0.489898 & 3.270555 & -0.181355 & 2.212338 & -0.134012 & 1.085183 & -0.055187 & 0.316421 & -0.009192 \\
         & --       & --        & 2.210929 & -0.134514 & 1.084617 & -0.055398 & 0.316254 & -0.009233 \\
\bottomrule
\end{tabular}
\end{table*}

 \clearpage
\bibliographystyle{apsrev4-2}
\bibliography{Refs_Kiselev.bib}

\end{document}